\documentclass[referee,useAMS,usenatbib]{mn2e}

\usepackage{amsmath,amssymb,mathrsfs}
\usepackage{subeqn, wasysym}
\usepackage{natbib}
\usepackage{wrapfig}
\usepackage{amsmath}
\usepackage{amssymb}
\usepackage{graphicx}
\usepackage{euscript}
\usepackage{units}
\usepackage{subcaption}
\usepackage{xcolor}
\usepackage{epstopdf}
\usepackage[export]{adjustbox}
\usepackage{placeins}
\usepackage{float}
\usepackage{morefloats}
\usepackage{color}
\usepackage{array}
\usepackage[]{multirow}
\usepackage{soul}

\newcommand{\beq}{\begin{equation}}
\newcommand{\eeq}{\end{equation}}

\newcommand{\bfe}{\mbox{\boldmath $e$}}
\newcommand{\bfr}{\mbox{\boldmath $r$}}

\newcommand{\bfu}{\mbox{\boldmath $u$}}

\newcommand{\p}{\mbox{$\partial$}}
\newcommand{\tk}{\mbox{$T_{\rm kep}$}}
\newcommand{\ts}{\mbox{$T_{\rm sec}$}}
\newcommand{\rpi}{{\rm \pi}}

\newcommand{\scrr}{{\cal R}}

\newcommand{\rmd}{{\rm d}}
\newcommand{\Msun}{\mbox{${\rm M}_\odot$}}

\begin{document}

\title[Secular Instabilities of Keplerian Stellar Discs]{Secular Instabilities of Keplerian Stellar Discs}

\author[Kaur, Kazandjian, Sridhar and Touma]
{Karamveer Kaur$^{1,4}$, Mher Kazandjian$^{2}$, S. Sridhar$^{1}$ and Jihad Touma$^{3}$\\
$^{1}$~Raman Research Institute, Sadashivanagar, Bangalore 560 080, India\\
$^{2}$~Leiden Observatory, Leiden University, PO Box 9513, 2300 RA Leiden, The Netherlands\\
$^{3}$~Department of Physics, American University of Beirut, PO Box 11-0236, Riad El-Solh, Beirut 11097 2020, Lebanon\\  
$^{4}$~karamveer@rri.res.in\\
}
\maketitle

\begin{abstract}
We present idealized models of a razor--thin, axisymmetric, Keplerian stellar disc around a massive black hole, and study non-axisymmetric secular instabilities in the absence of either counter-rotation or loss cones. These discs are prograde mono-energetic waterbags, whose phase space distribution functions are constant for orbits within a range of eccentricities ($e$) and zero outside this range. The linear normal modes of waterbags are composed of sinusoidal disturbances of the edges of distribution function in phase space. Waterbags which include circular orbits (\emph{polarcaps}) have one stable linear normal mode for each azimuthal wavenumber $m$. The $m=1$ mode always has positive pattern speed and, for polarcaps consisting of orbits with $e < 0.9428$, only the $m=1$ mode has positive pattern speed. Waterbags excluding circular orbits (\emph{bands}) have two linear normal modes for each $m$, which can be stable or unstable. We derive analytical expressions for the instability condition, pattern speeds, growth rates and normal mode structure. Narrow bands are unstable to modes with a wide range in $m$. Numerical simulations confirm linear theory and follow the non-linear evolution of instabilities. Long-time integration suggests that instabilities of different $m$ grow, interact non-linearly and relax collisionlessly to a coarse-grained equilibrium with a wide range of 
eccentricities.  
\end{abstract}


\section{Introduction}

Dense clusters of stars orbit massive black holes (MBH) in galactic nuclei. The best studied cases are the nuclear star clusters of the Milky Way and M31, each of which possesses a low mass (or Keplerian) stellar disc around the MBH. Since the black hole's gravity dominates the force on stars, Toomre $Q \gg 1$, so an axisymmetric Keplerian disc is expected to be linearly stable to axisymmetric perturbations on Keplerian orbital time scales. Even when a disc is stable to all modes on these short time scales, it may be unstable to modes that grow over the much longer \emph{secular} time scale of apse precession. Secular instabilities must necessarily be non-axisymmetric with the azimuthal wavenumber $m\neq 0$ \citep{st16a} --- hereafter ST1.  A good example is the $m=1$ instability of counter-rotating discs, which may be applicable to the nuclear disc of M31 \citep{tou02,kt13}. Stellar discs with distribution functions (DFs) even in the angular momentum and empty loss cones (i.e. DF is zero at zero angular momentum) may be unstable to $m=1$ modes \citep{tre05}. Mono-energetic discs dominated by nearly radial orbits, could be prone to loss cone instabilities of all $m$, if there is some amount of counter-rotating stars \citep{pps07}.

A natural question is the following: can prograde, axisymmetric discs support secular instabilities, even when counter-rotation and loss-cone are absent? The answers available in the literature pertain to the stability of razor-thin discs. \citet{tre01, jt12} proved that a Schwarzschild DF is stable to modes of all $m$ in the tight-winding limit.
This was generalised by ST1 who proved that a DF, which is a strictly monotonic function of the angular momentum at fixed semi--major axis (i.e. at fixed Keplerian energy), is stable to modes of all $m$.
However, these results are insufficient to address the general question, which could be relevant to the history of the clockwise disc of young stars at the centre of the Milky Way. If these stars formed in a fragmenting, circular gas disc around the MBH \citep{lb03}, then the initial  stellar orbits should have small eccentricities and the same sense of rotation (i.e. no counter-rotation) about the MBH. But \citet{yel14} found that the mean eccentricity of the stellar orbits is $\bar{e}\simeq 0.27$. Is this largish value the result of secular instabilities? The goal of this paper is to present the simplest models of stellar discs orbiting MBHs, whose instabilities can be studied explicitly. This is done by combining analytical methods from ST1 with numerical simulations derived from  \citet{ttk09}. 

In Section~\ref{sec_dyn_kep_discs} the problem is stated within the framework of ST1. Using their stability result as a guide we motivate the search for DFs that are either non-monotonic or not strictly monotonic in the angular momentum. This leads in Section~\ref{sec_mono-energetic discs intro} to mono-energetic discs, which are composed of stars with equal semi--major axes. The phase space of a mono-energetic disc is a sphere (see Figure~\ref{fig_phase_space}), and secular gravitational interactions between stars have an explicit logarithmic form.  
Drawing on earlier work in plasma physics we introduce the simplest of prograde, axisymmetric DFs, which correspond to `waterbags'. The phase space distribution function of a waterbag is constant for orbits whose eccentricities ($e$) lie within a certain range, and zero outside this range. These are of two types of waterbags: \emph{polarcaps}, which include  
circular orbits, and \emph{bands}, which exclude circular orbits --- see Figure~\ref{fig_sph_Pol_Band}. The linear stability analysis of these systems leads to normal modes which are composed of sinusoidal disturbances of the edges of distribution function in the phase space. For each $m\neq 0$, a polarcap has one stable  normal mode, whereas a band has two normal modes that may be stable or unstable. In Section~\ref{sec_simu_intro} we present numerical simulations of an unstable and a stable band; these give an immediate graphical picture, both in real space and phase space, of linear and non-linear evolution. The linear stability problem for a band is formulated and solved in Section~\ref{sec: instability-analytics}. Section~\ref{sec_simu_detailed} explores instabilities further, drawing detailed comparisons between linear theory and numerical simulations, as well as following the long-time evolution of an unstable band. We conclude in Section~\ref{sec_conclusions}.  

\section{Secular Dynamics of Keplerian Stellar Discs}
\label{sec_dyn_kep_discs}

Our model system is a razor-thin flat stellar disc of total mass $M$, 
composed of very many stars, orbiting a massive black hole (MBH) of mass $M_\bullet \gg M$. Since the mass ratio $\varepsilon = M/M_\bullet \ll 1\,$, 
the dominant gravitational force on the stars is the inverse-square Newtonian force of the MBH. The limiting case of negligible stellar self-gravity, $\,\varepsilon \to 0\,$, reduces to the problem of each star orbiting the MBH independently on a fixed Keplerian ellipse with period, $\tk = 2\rpi (a^3/GM_\bullet)^{1/2}$, where $a = \mbox{semi-major axis}$.  When $0 < \varepsilon \ll 1\,$, self-gravity is small but its effects build up over the long secular times, $\ts = \varepsilon^{-1}\tk \gg \tk\,$. ST1 describes the average behaviour of dynamical quantities over times $\ts\,$, by systematically averaging over the fast Keplerian orbital phase --- a method that goes back to Gauss. The secular orbit of each star in the disc is represented by a Gaussian ring, which is a Keplerian ellipse with the MBH at one focus, of fixed semi-major axis, whose eccentricity and apsidal longitude can evolve over times $\ts$. Hence the natural measure of time in secular theory is $\,\tau = \varepsilon \times  \mbox{time}$, the `slow' time variable. The state of a Gaussian ring at any time $\tau$ can be specified by giving its three-dimensional Delaunay coordinates, $\scrr = \{I, L, g\}$, where $I\,=\,\sqrt{GM_\bullet a\,} = \mbox{constant}$ which is a measure of the Keplerian energy, $\,L$ is the specific angular momentum which 
is restricted to the range $-I\leq L\leq I$, and $\,0\leq g < 2\rpi$ is the longitude of the periapse. Ring space (or $\scrr$-space) is topologically equivalent to $\mathbb{R}^3$, with $I$ the `radial coordinate', $\arccos{(L/I)}$ the `colatitude', and $g$ the `azimuthal angle'. A disc composed of $N \gg 1$ stars, each of mass $m_\star = M/N$, is a collection of $N$ points in $\scrr$-space. The simplest description of a stellar disc uses the single-ring probability DF, $F(\scrr, \tau) = F(I, L, g, \tau)$, which is normalized as,
\beq
\int {\rm d}\scrr\,\,F(\scrr, \tau) \;=\;
\int {\rm d}I\,{\rm d}L\,{\rm d}g\,\,F(I, L, g, \tau) \;=\; 1\,.
\label{eqn_norm-disc}
\eeq

Over times much shorter than the resonant relaxation times, $T_{\rm res} = N\ts$, the graininess of the ring-ring interactions has negligible effects and the stellar system can be thought of as collisionless. Formally, the collisionless limit corresponds to assuming that the system is composed of an infinite number of stars, each of infinitesimal mass, the whole having a mass $M$ equal to the total stellar mass: $N\to\infty\,$, $\,m_\star\to 0\,$ with $\,M = Nm_\star\,$ held constant. Then each star is like a test-ring, whose 
motion is governed by the secular Hamiltonian, $\Phi(I, L, g, \tau)$, which is equal to the (scaled) self-gravitational disc potential:\footnote{ST1 include relativistic effects of the MBH and tidal forces due to external gravitational fields, but these are not considered in this paper.}
\beq
\Phi(I, L, g, \tau) \;=\; \int {\rm d}I'\,{\rm d}L'\,{\rm d}g'\,\,\Psi(I, L, g, I', L', g')\,F(I', L', g', \tau)\,,
\label{eqn_phi-def}
\eeq
where
\beq
\Psi(I, L, g, I', L', g') \;=\; -GM_\bullet\oint\oint\frac{{\rm d}w}{2\rpi}\,
\frac{{\rm d}w'}{2\rpi}\,\frac{1}{\left|\bfr - \bfr'\right|}
\label{eqn_psi-def}
\eeq
\noindent
is the (scaled) interaction potential between two rings.\footnote{Here $\bfr = (x, y)$ and $\bfr' = (x', y')$ are the position vectors of the two stars with respect to the MBH --- see \S~4.1 of ST1 for details of the transformation from $\bfr$ and $\bfr'$ to the corresponding Delaunay variables. Here $w$ and $w'$ are the mean anomalies of the stars representing the Keplerian orbital phase on their respective Gaussian rings. } Ring orbits are determined by the Hamiltonian equations of motion:
\beq
I \;=\; \sqrt{GM_\bullet a} \;=\; \mbox{constant}\,,\qquad\quad
\frac{{\rm d}L}{{\rm d}\tau} \;=\; -\,\frac{\p \Phi}{\p g}\,,\qquad\quad  
\frac{{\rm d}g}{{\rm d}\tau} \;=\; \frac{\p \Phi}{\p L}\,.
\label{eqn_eom}
\eeq
\noindent
This is a Hamiltonian flow in $\scrr$-space which is restricted to the 
$I = \mbox{constant}$ two-sphere. The flow carries with it the DF, whose evolution is governed by the secular collisionless Boltzmann equation (CBE):
\beq  
\frac{\p F}{\p \tau} \;+\; \left[F\,,\,\Phi\right]_{Lg} \;=\; 0\,,
\quad\qquad\mbox{where}\quad\qquad\left[F\,,\,\Phi\right]_{Lg} \;=\;
\frac{\p F}{\p g}\frac{\p \Phi}{\p L} \,-\, \frac{\p F}{\p L}\frac{\p \Phi}{\p g} 
\label{eqn_cbe}
\eeq
\noindent
is the two-dimensional Poisson Bracket in $(L, g)$-space. $\Phi$ itself
depends on $F$ through the $\scrr'$-space integral of equation~(\ref{eqn_phi-def}). Therefore equation~(\ref{eqn_cbe}), together with the secular Hamiltonian of equation~(\ref{eqn_phi-def}), defines the self-consistent initial value problem of the secular time evolution of the DF, given an arbitrarily specified initial DF $F(I, L, g, 0)$. 
A general property of this time evolution is the following: since
the $I$ of any ring is constant in time, the probability for a ring to be 
in $(I,\, I + {\rm d}I)$ is a conserved quantity. In other words the probability distribution function in one-dimensional $I$-space, defined by
\beq
P(I) \;=\; \int\, dL\,dg\,\, F(I,\, L,\, g, \,\tau)\,, 
\label{eqn_conserved}
\eeq
\noindent
is independent of $\tau$, as can be verified directly using the CBE 
of equation~(\ref{eqn_cbe}). 

\subsection{Axisymmetric equilibria and linear stability}
\label{sec: axisymmetric_LCBE}

Secular equilibria are DFs that are time-independent and self-consistent solutions of the CBE. They can be constructed using the secular Jeans theorem of ST1, which states that $F$ must be function of the isolating integrals of motion of the secular Hamiltonian. An axisymmetric equilibrium DF is independent of $g$ and can be written as $F = (2\rpi)^{-1}F_0(I, L)\,$, because $I$ and $L$ are two isolating integrals of motion of the axisymmetric Hamiltonian, $\Phi_0(I, L)$. Equation~(\ref{eqn_phi-def}) gives $\Phi_0$ self-consistently in terms of $F_0\,$:\footnote{$\Psi(I, L, g, I', L', g')$ depends on the apses only in the combination $\vert g - g'\vert$, so the integral over $g'$ is independent 
of $g\,$.}  
\beq
\Phi_0(I, L) \;=\; \int {\rm d}I'\,{\rm d}L'\,F_0(I', L')
\oint\frac{{\rm d}g'}{2\rpi}\,\Psi(I, L, g, I', L', g')\,.
\label{eqn_phi0-def}
\eeq
The equations of motion (\ref{eqn_eom}) for a ring become very simple in an axisymmetric disc: 
\beq
I \;=\; \mbox{constant}\,,\qquad L \;=\; \mbox{constant}\,,\qquad
\frac{{\rm d}g}{{\rm d}\tau} \;\equiv\; \Omega_0(I, L) \;=\; \frac{\p \Phi_0}{\p L}\,. 
\label{eqn_orbaxi}
\eeq
The semi-major axis and eccentricity of a ring are constant, with the 
apsidal longitude precessing at the constant angular frequency 
$\Omega_0(I, L)$.

The time evolution of perturbations to an axisymmetric equilibrium DF can 
be studied by considering the total DF to be $F = (2\rpi)^{-1}F_0(I, L) + 
F_1(I, L, g, \tau)$, where the perturbation $F_1$ contains no net mass: 
\beq
\int {\rm d}I\,{\rm d}L\,{\rm d}g\,\,F_1(I, L, g, \tau) \;=\; 0\,.
\label{eqn_nomass}
\eeq
If $\Phi_1(I, L, g, \tau)$ is the self-gravitational potential due to $F_1$, 
then the total Hamiltonian is $\Phi = \Phi_0(I, L) + \Phi_1(I, L, g, \tau)$. By substituting for $F$ and $\Phi$ in the CBE (\ref{eqn_cbe}), and using 
$\left[F_0\,, \Phi_0\right]_{Lg} = 0$, we can derive the equation governing the time evolution of $F_1$. For small perturbations $\vert F_1 \vert \ll F_0\,$ this is the linearized collisionless Boltzmann equation (LCBE):
\begin{subequations}
\begin{align}  
&\frac{\p F_1}{\p \tau} \;+\; \Omega_0\frac{\p F_1}{\p g} \;=\;
\frac{1}{2\rpi}\frac{\p F_0}{\p L}\frac{\p \Phi_1}{\p g} \,,
\label{eqn_lcbe}\\[1ex] 
&\Phi_1(I, L, g, \tau) \;=\; \int {\rm d}I'\,{\rm d}L'\,{\rm d}g'\,\,\Psi(I, L, g, I', L', g')\,F_1(I', L', g', \tau)\,.
\label{eqn_phi1-def}
\end{align}
\end{subequations}
\noindent
The LCBE is a linear (partial) integro-differential equation for $F_1$, 
and determines the linear stability of the axisymmetric DF, 
$F_0(I, L)$. 

An axisymmetric perturbation $F_1(I, L, \tau)$ gives rise to a 
$\Phi_1(I, L, \tau)$ that is also independent of $g$. Then the LCBE (\ref{eqn_lcbe}) implies $\p F_1/\p \tau = 0$, whose physical solution is $F_1 = 0$, because an axisymmetric perturbation cannot change the angular momentum of a star. Hence it is only non-axisymmetric, or 
$g$-dependent, perturbations that are of interest in secular theory. 
Since $\tau$ and $g$ appear in the LCBE only as $(\p/\p \tau)$ and $(\p/\p g)$ we can look for linear modes of the form $F_1 \propto\exp{[{\rm i}(m g - \omega\tau)]}$, where $m \neq 0$ is the azimuthal wavenumber. Using only the general symmetric properties of $\Psi(\scrr, \scrr')$, the following result was  proved in ST1 for DFs that are strictly monotonic functions of $L$:

\medskip 
\noindent
$\bullet\,$ Stationary, axisymmetric discs with DFs $F_0(I, L)$ 
are neutrally stable (i.e. $\omega$ is real) to secular perturbations 
of all $m$ when $\,\p F_0/\p L\,$ is of the same sign (either positive 
or negative) everywhere in its domain of support, $\,-I \leq L \leq I\,$ and $I_{\rm min} \leq I \leq I_{\rm max}\,$.

\medskip
\noindent
As noted in ST1 these secularly stable DFs can have both prograde and retrograde populations of stars because $-I \leq L \leq I\,$. The discs
have net rotation and include physically interesting cases, such as a 
secular analogue of the well-known Schwarzschild DF. To investigate secular instabilities, the above stability result motivates us to look at axisymmetric discs with DFs, $F_0(I, L)$, that are either non-monotonic or not strictly monotonic functions of $L$ at fixed $I$. 

A general way to proceed would be to develop stability theory, using only the symmetry properties of $\Psi(\scrr, \scrr')$, as ST1 did. But the goal of this paper is more specific: We wish to construct the simplest class of disc models that permits quantitative study of the onset and growth of linear non-axisymmetric instabilities. In order to do this we must be able to calculate physical quantities such as the apse precession frequency 
$\Omega_0(I, L)$, using equations~(\ref{eqn_phi0-def}) and (\ref{eqn_orbaxi}). Hence we need to use explicit forms for $\Psi$, for a physically motivated model of a stellar disc.

\section{Mono-energetic discs}
\label{sec_mono-energetic discs intro}

\subsection{Collisionless Boltzmann equation}
 
$\Psi(I, L, g, I', L', g')$ depends on the apses only in the combination 
$\vert g - g'\vert$, and can be developed in a Fourier series in $(g - g')$. 
When the spread in the semi--major axes of the disc stars is comparable to the mean disc radius, the Fourier coefficients are, in general, complicated functions of $(I, L, I', L')$ --- although for numerical calculations it is straightforward to calculate them on any grid in this four dimensional 
space. Analytical forms are readily available if restrictions are placed on $L$ and $L'$, such as both the rings being near-circular and well-separated (the `Laplace--Lagrange' limit of planetary dynamics) or both rings being very eccentric, corresponding to the `spoke' limit of
\citet{pps07}. But secular dynamics and statistical mechanics are really about the exchange of angular momentum of stars at fixed semi--major axes, so it seems preferable if we do not place such severe restrictions on $L$ or $L'$. Let us consider discs with a small spread in semi--major axes; since this is equivalent to a small spread in Keplerian orbital energies, the disc may be called nearly mono-energetic. Having nearly the same semi--major axes, any two rings either cross each other or come very close to each other, so $\Psi(\scrr, \scrr')$ can be large, even infinite, in magnitude. For nearly-circular rings the dominant contribution, which is a logarithmic singularity, was worked out by \citet{bgt83}. 

In a nearly mono-energetic disc most pairs of rings intersect each other. It is useful to consider the strictly mono-energetic limit, $I = I_0 = \sqrt{GM_\bullet a_0}$, when every ring intersects every other ring. Since all rings have the same semi-major axis $a_0$, they also have the same Keplerian orbital period, $\tk = 2\rpi (a_0^3/GM_\bullet)^{1/2}\,$. Hence it is convenient to use a dimensionless slow time variable, $\,t = \tau/\tk = \mbox{time}/\ts\,$, to study the dynamics of mono-energetic discs. The state of a ring at time $t$ can be specified by giving its periapse, $g$, and the dimensionless specific angular momentum $\ell = L/I_0\,$. Since $-1 \leq \ell \leq 1$, the motion of any ring is restricted to the unit sphere (Figure~{\ref{fig_phase_space}) on which $\ell =\cos{(\mbox{colatitude})}$ and $g = \mbox{azimuthal angle}$ are canonical coordinates. For a mono-energetic disc $F$ takes the form:
\beq
F(I, L, g, \tau) \;=\; \frac{\delta(I - I_0)}{I_0}\,f(\ell, g, t)\,.
\label{eqn_df-mono}
\eeq
Then equation~(\ref{eqn_norm-disc}) implies the following normalization for $f\,$: 
\beq
\int {\rm d}\ell\,{\rm d}g\,\,f(\ell, g, t) \;=\; 1\,. 
\label{eqn_norm-mono}
\eeq
Hence $f(\ell, g, t)$ is the (dimensionless) DF for mono-energetic discs on the $(\ell, g)$ phase space of Figure~{\ref{fig_phase_space}}. The eccentricity of a ring, $e = \sqrt{1 - \ell^2}$, is equal to the length of the projection of the corresponding position vector on the sphere's equatorial plane. The eccentricity vector (or Lenz vector) is defined as $\bfe = (e_x, e_y)$ with $e_x = e\cos{g}$ and $e_y = e\sin{g}$. We can think of $(e_x, e_y, \ell)$ as a right-handed Cartesian coordinate system, with the ring phase space realized as the unit sphere, $e_x^2 + e_y^2 + \ell^2 = 1\,$. 

\begin{figure}
\centering
\includegraphics[width=0.9 \textwidth,trim={4cm 4cm 4cm 2cm}]{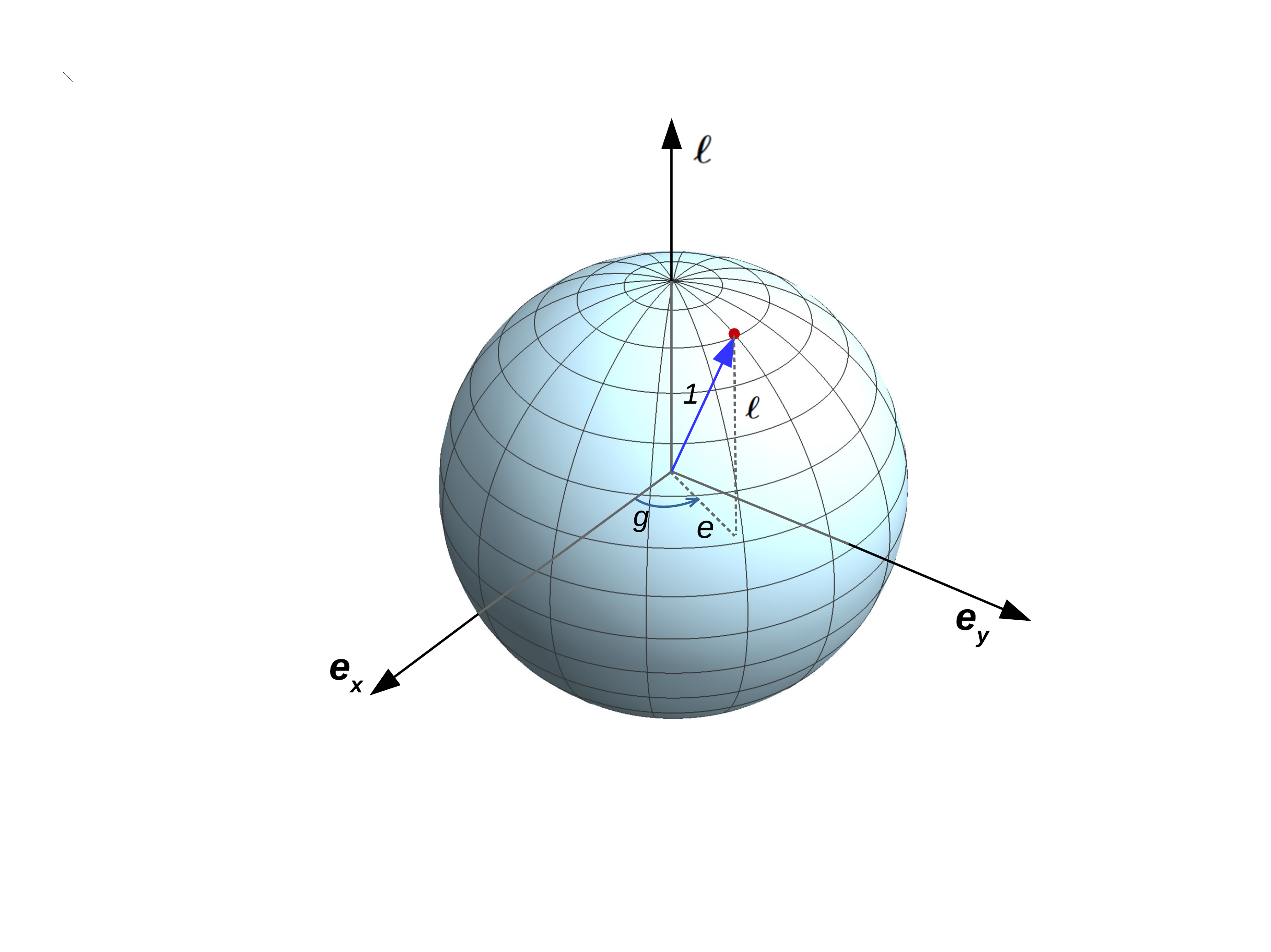}
\caption{\emph{Phase space of a mono-energetic disc}. Each star in the disc is represented by point on the unit sphere (shown in red), with canonical coordinates $(\ell, g)$. The latitudes are lines of constant $\ell$, and longitudes are lines of constant $g$. The projection of $(\ell, g)$ onto the equatorial plane gives the eccentricity vector $\bfe = (e_x, e_y)$.}
\label{fig_phase_space}
\end{figure}

The formula of \citet{bgt83} for the ring-ring interaction potential, 
$\psi(\ell, \ell', g-g') = \Psi(I_0, I_0\ell, g, I_0, I_0\ell', g')$, 
takes the following attractive form given in \citet{tt14}:    
\beq
\psi(\ell, \ell', g-g') \;=\; 
\frac{GM_\bullet}{a_0}\,\left\{-\frac{4}{\rpi}\,\log 2 \;+\; 
\frac{1}{2\rpi}\,\log\left\vert\bfe - \bfe'\right\vert^2\,\right\}\,.
\label{eqn_psi-mono-def}
\eeq
This expression for $\psi$ is, strictly speaking, valid only when $e, e' \ll 1\,$. But \citet{tt14} have shown that this formula for $\psi$ serves as a good approximation for all values of $e$ and $e'$, and used this fact to study axisymmetric and non-axisymmetric secular thermodynamic equilibria; they also provide an improved fitting formula but we do not use this. 
Henceforth we take equation~(\ref{eqn_psi-mono-def}) as the basic `law of interaction', between any two rings in a mono--energetic disc. Using equation~(\ref{eqn_df-mono}) in (\ref{eqn_phi-def}) we see that the 
mean-field self-gravitational potential, $\varphi(\ell, g, t) = \Phi(I_0, I_0\ell, g, \tau)$ is given in explicit form as:
\begin{align}
\varphi(\ell, g, t) &\;=\;
\int {\rm d}\ell'\,{\rm d}g'\,\,\psi(\ell, \ell', g-g')f(\ell', g', t)
\nonumber\\[1ex]
&\;=\; -\frac{4GM_\bullet}{\rpi a_0}\,\log 2 \;+\;
\frac{GM_\bullet}{2\rpi a_0}\int {\rm d}\ell'\,{\rm d}g'\,
\log\left\vert\bfe - \bfe'\right\vert^2 f(\ell', g', t)\,. 
\label{eqn_phi-mono}
\end{align}

We have already cast the independent variables $(\ell, g, t)$ in dimensionless form. Equations~(\ref{eqn_eom}), governing the dynamics of a ring, can now be written in the following dimensionless form: 
\beq 
\frac{{\rm d}\ell}{{\rm d}t} \;=\; -\,\frac{\p H}{\p g}\,,\qquad\quad  
\frac{{\rm d}g}{{\rm d}t} \;=\; \frac{\p H}{\p \ell}\,,
\label{eqn_eom-mono}
\eeq
where
\beq
H(\ell, g, t) \;=\; \frac{\tk}{I_0}\,\varphi(\ell, g, t) \;\;=\;
\int {\rm d}\ell'\,{\rm d}g'\,
\log\left\vert\bfe - \bfe'\right\vert^2 f(\ell', g', t)
\;\;+\; \mbox{constant}  
\label{eqn_ham-mono}
\eeq 
is the dimensionless secular Hamiltonian. These equations of motion imply the natural Poisson Bracket on the $(\ell, g)$ unit sphere:
\beq
\left[\,f\,,\,H\,\right] \;=\;
\frac{\p f}{\p g}\frac{\p H}{\p \ell} \;-\; \frac{\p f}{\p \ell}
\frac{\p H}{\p g}\,. 
\label{eqn_pb-mono}
\eeq
Substituting equation~(\ref{eqn_df-mono}) in (\ref{eqn_cbe}) we obtain the following CBE governing the self-consistent evolution of the DF:
\beq  
\frac{\p f}{\p t} \;+\; \left[\,f\,,\,H\,\right] \;=\; 0\,.
\label{eqn_cbe-mono}
\eeq
Equations~(\ref{eqn_ham-mono})---(\ref{eqn_cbe-mono}) provide a complete, dimensionless description of the collisionless dynamics of mono-energetic Keplerian discs.

\subsection{Linear stability of axisymmetric equilibria}
\label{sec_lin_axi}

In the study of axisymmetric equilibria and their linear, non-axisymmetric 
perturbations it is useful to have at hand the Fourier expansion of the ring--ring interaction potential, $\log\left\vert\bfe - \bfe'\right\vert^2$, that appears in the definition of the Hamiltonian in equation~(\ref{eqn_ham-mono}). From equation~(C.2) of \citet{tt14} we have,
\begin{align}
\log\left\vert\bfe - \bfe'\right\vert^2 &\;=\;
\log\left[e^2 -2ee'\cos(g-g') + e'^2\,\right]\nonumber\\[1ex]
&\;=\; \log\left(e_>^2\right) \;-\; 2\,\sum_{m=1}^{\infty}\,\frac{1}{m}
\left(\frac{e_<}{e_>}\right)^m\,\cos{[m(g-g')]}\,,
\label{eqn_log-fou}
\end{align}
where $e_< = \mbox{min}\left(e, e'\right)$ and $e_> = \mbox{max}\left(e, e'\right)$.

Any DF of the form $\,f = (2\rpi)^{-1}f_0(\ell)\,$, which is normalised as
$\int_{-1}^{1}{\rm d}\ell\,f_0(\ell) = 1\,$, represents an axisymmetric equilibrium. Using equation~(\ref{eqn_log-fou}) in (\ref{eqn_ham-mono}), we have the corresponding axisymmetric Hamiltonian: 
\begin{align}
H_0(\ell) &\;=\; \int_{-1}^{1}\,{\rm d}\ell'\,\log\left(e_>^2\right)\,f_0(\ell')
\nonumber\\[1ex]
&\;=\; \int_{0}^{\vert\ell\vert}{\rm d}\ell'\,\log\left(1-\ell'^2\right)\left\{f_0(\ell') + f_0(-\ell')\right\}
\;+\; \log\left(1-\ell^2\right)\int_{\vert\ell\vert}^{1} 
{\rm d}\ell'\left\{f_0(\ell') + f_0(-\ell')\right\}\,, 
\label{eqn_h0-def}
\end{align}
where we have dropped a constant term. The apse precession frequency is given:
\beq
\Omega_0(\ell) \;=\; \frac{{\rm d} H_0}{{\rm d} \ell} \;=\; - \frac{2 \, \ell}{1-\ell^2} \int_{\vert \ell \vert}^{1}\;{\rm d}\ell' \, \left\{ f_0(\ell') + f_0(-\ell')  \right\} .
\label{eqn_omega0-def-mono}
\eeq
Some general properties of $\Omega_0$ are: (i) Since the product $\ell.
\Omega_0(\ell) \leq 0$, the apse precession of a ring is always opposite
to the faster Keplerian orbital motion; (ii) As $\ell \to 0$ we have 
$\Omega_0(\ell) \to -2\ell$, so highly eccentric rings precess very 
slowly; (iii) In the limit of circular rings $\ell \to \pm 1$, and  
$\Omega_0(\ell) \to \mp \left\{f_0(1) + f_0(-1)\right\}$ goes to a finite 
limit. 

When the axisymmetric equilibrium is perturbed the total DF is 
$f(\ell,g,t) = (2\rpi)^{-1}f_0(\ell) + f_1(\ell,g,t)$, and the corresponding self-consistent Hamiltonian is $H_0(\ell) + H_1(\ell,g,t)$. Substituting 
these in the mono-energetic CBE (\ref{eqn_cbe-mono}) and linearizing, we
obtain the LCBE governing the evolution of $f_1\,$:
\begin{align}
&\frac{\p f_1}{\p t} + \Omega_0 (\ell) \frac{\p f_1}{\p g} \;=\; \frac{1}{2 \rpi} \frac{{\rm d} f_0}{ {\rm d} \ell} \frac{\p H_1}{\p g}\,, \label{eqn_lcbe-mono}\\[1 ex]
\mbox{where}\qquad &H_1(\ell,g,t)  \;=\; \int {\rm d}\ell'\,{\rm d}g'\, \log\left\vert\bfe - \bfe'\right\vert^2 f_1(\ell', g', t)\;.  
\label{eqn_H1-def-mono}
\end{align}
We seek solutions of the form $f_1(\ell,g,t; m) = {\rm Re}\left\{f_{1m}(\ell) \exp{[{\rm i}(m g - \omega_m t)]}\right\}$ and $H_1(\ell,g,t) = {\rm Re}\left\{H_{1m}(\ell) \exp{[{\rm i}(m g - \omega_m t)]}\right\}$ where, without 
loss of generality, we take $m$ to be a positive integer. Equation~(\ref{eqn_H1-def-mono}) gives $H_{1m} = -2\rpi/m\int_{-1}^{1}{\rm d} \ell'(e_</e_>)^m f_{1m}(\ell')$. Then the LCBE reduces to the following equation,
\beq
\left[\,\omega_m - m\Omega_0(\ell)\,\right]\,f_{1m}(\ell) \;=\;  \frac{ {\rm d} f_0 }{ {\rm d} \ell  } \int_{-1}^{1}{\rm d} \ell'\,\left(\frac{e_<}{e_>}\right)^m  f_{1m}(\ell')\;,
\label{eqn_integral-eigen-eqn-mono}
\eeq
which is an integral eigenvalue problem, for the eigenvalues $\omega_m$ 
and corresponding eigenfunctions $f_{1m}(\ell)$. This equation is a special case of equation~(75) of ST1, which is valid for a general disc. Proceeding in a manner similar to ST1, it is straightforward to prove the stability result: \emph{all DFs $f_0(\ell)$ that are strictly monotonic functions of 
$\ell$ are linearly stable}. This raises again the question of the stability of DFs that are not strictly monotonic in $\ell$. Since this question is now posed in the context of equation~(\ref{eqn_integral-eigen-eqn-mono}) --- which is given in explicit form --- we can proceed to explore it quantitatively. 
Among all the DFs that are not strictly monotonic functions of $\ell$, the simplest are probably the `waterbag' DFs which are discussed below.

\subsection{Waterbags and the linear stability problem}
\label{sec_waterbags_intro}

A mono-energetic waterbag is a region of the unit sphere phase space of Figure~{\ref{fig_phase_space}} within which the DF takes a constant positive value and is zero outside this region.\footnote{The ``waterbag'' model was originally developed for the Vlasov equation by \citet{br70}.} Time evolution that is governed by the CBE of equations~(\ref{eqn_ham-mono})---(\ref{eqn_cbe-mono}) conserves both the area of the region as well as the value of the DF. Hence the dynamical problem reduces to following the evolution of the contour(s) bounding the region. Analogous to the contour dynamics of fluid vortices on a sphere \citep{Dri88}, the deformation of the contour(s) defining a waterbag stellar disc can be very complicated. 

\subsubsection{Axisymmetric equilibria}
An axisymmetric mono-energetic waterbag has a DF, $f_0(\ell)$, that takes a constant positive value for $\ell \in [\ell_1,\ell_2]$, and is zero outside this interval. Since our primary interest in this paper concerns the stability of discs in which stars orbit the MBH in the same sense, we assume that $0 \le \ell_1 <\ell_2 \le 1$. The normalized DF for such a `prograde waterbag' is: 
\beq
f_0(\ell)\;=\;
\begin{cases}
\displaystyle{\;\frac{1}{\ell_2-\ell_1}}\qquad\mbox{for}\quad\ell_1 \le \ell \le \ell_2\,,\\
\qquad0\qquad\quad\mbox{otherwise.} 
\end{cases}                 
\label{eqn_waterbag-def}
\eeq
There are two different cases, corresponding to $\ell_2 =1$ (Polarcap) 
and $\ell_2 < 1$ (Band) --- see Figure~{\ref{fig_sph_Pol_Band}}. It can be seen that bands have DFs that are non-monotonic in $\ell$, whereas polarcaps have DFs that are not strictly monotonic 
in $\ell$. Hence the stability result, stated below equation~(\ref{eqn_integral-eigen-eqn-mono}), does not apply to either of these systems. But their stability properties can be determined completely,
as we show below.

\begin{figure}
\begin{subfigure}{0.5\linewidth}
\centering
\includegraphics[width=1.25 \textwidth,trim={6cm 4cm 0cm 0cm}]{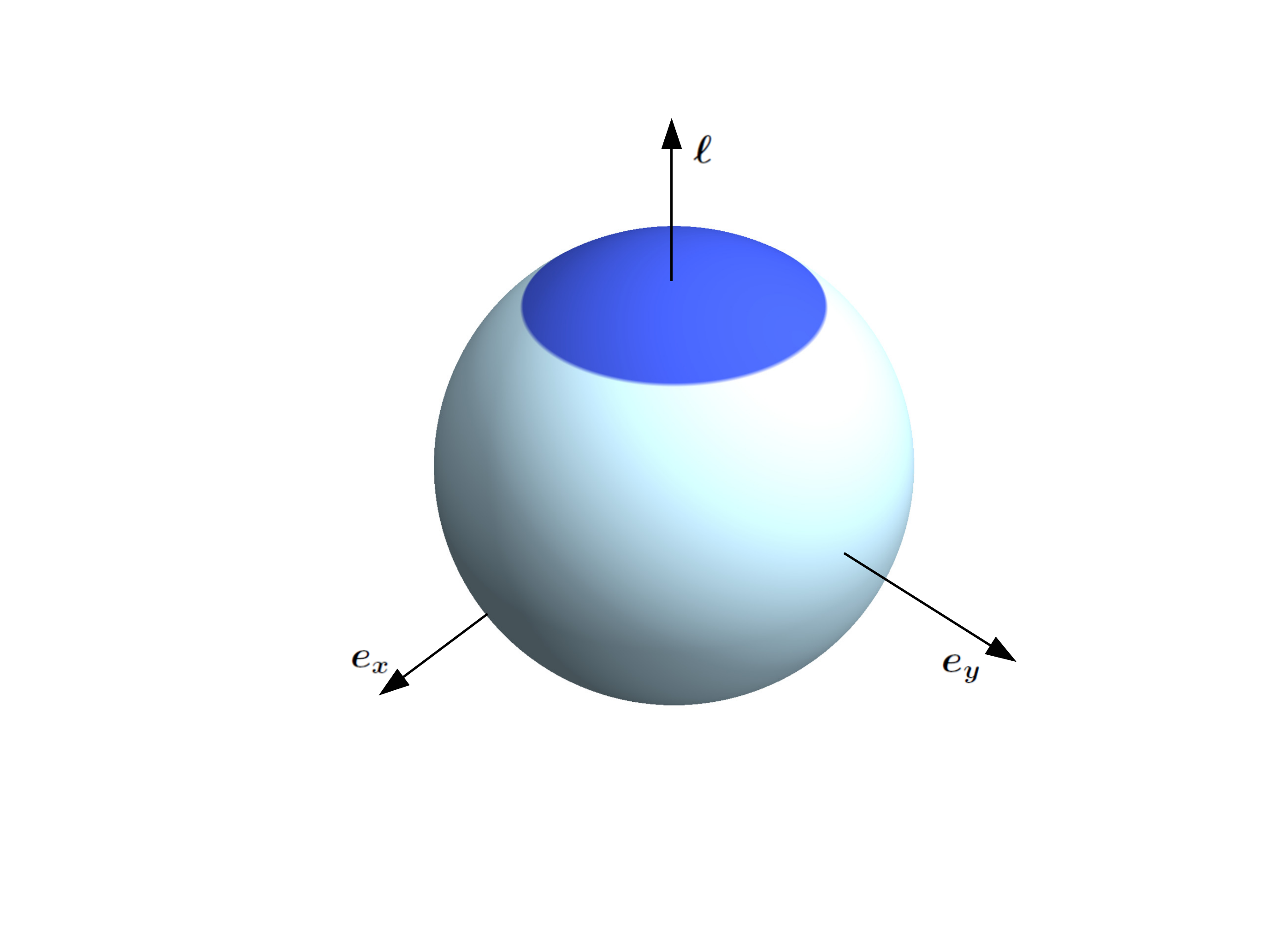}
\caption{Polarcap with $\ell_1 = 0.8$ and $\ell_2 = 1$}
\end{subfigure}
\begin{subfigure}{0.5\linewidth}
\centering
\includegraphics[width=1.25 \textwidth,trim={6cm 4cm 0cm 0cm}]{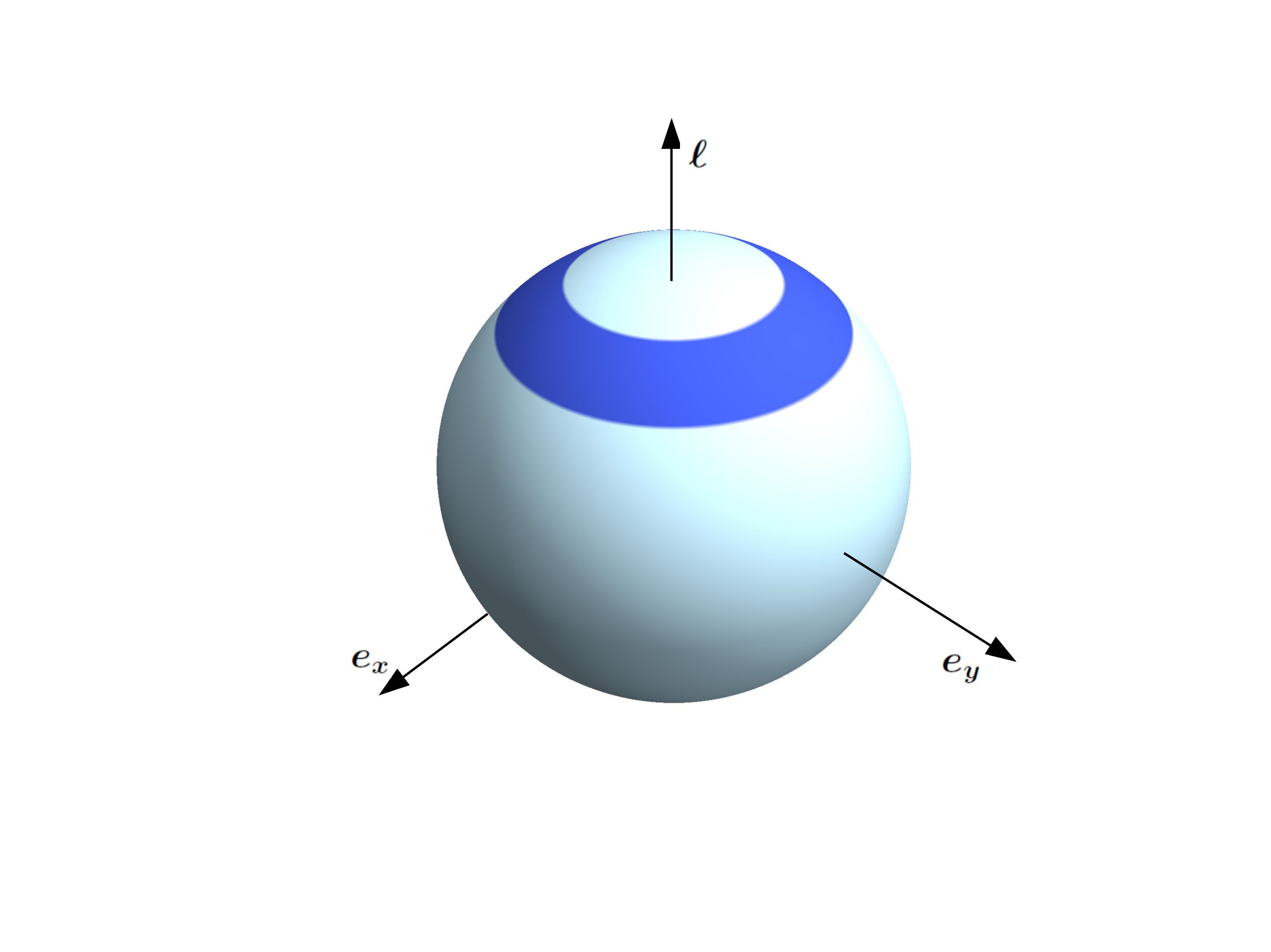}
\caption{Band with $\ell_1 = 0.7$ and $\ell_2 = 0.9\,$}
\end{subfigure}
\caption{\emph{Two types of prograde waterbags}}
\label{fig_sph_Pol_Band}
\end{figure}

The waterbag DF describes a circular annular disc composed of stars with eccentricities $e = \sqrt{1-\ell^2} \,\in [e_2,e_1]$, where $e_i =
\sqrt{1-\ell_i^2}\,$ for $i = 1,2\,$. The inner and outer radii of the disc are $r_{\rm min} = a_0(1-e_1)$ and $r_{\rm max} = a_0 (1+e_1)$ are determined by the most eccentric rings in the disc. The normalized surface density profile, $\Sigma_0(r)$, is obtained by integrating $f_0(\ell)$ over the velocities, as is done in appendix \ref{app:sec: surface_density_waterbags}. This gives
\beq
\Sigma_0(r) \;=\; 
\begin{cases}
\displaystyle{
\;\frac{\sin^{-1}\left[\ell_2/\ell_0(r)\right] \,-\,\sin^{-1}\left[\ell_1/\ell_0(r)\right]}{2 \pi^2 a_0^2(\ell_2 - \ell_1)}\,,\,\qquad\qquad|r-a_0| \,\le\, a_0 e_2}\\[2em]
\displaystyle{
\;\qquad \frac{\cos^{-1}\left[\ell_1/\ell_0(r)\right]}{2 \pi^2 a_0^2(\ell_2 - \ell_1)}\,\,,\qquad\qquad \qquad\;\,   a_0 e_2 \,<\, |r-a_0|\,\le\, a_0  e_1}\\[2em]   
\qquad\qquad\quad\,  0\,,\quad\qquad\qquad\qquad\qquad\quad 
a_0 e_1 \,<\, |r-a_0|  
\end{cases}
\label{eqn_surface density-final}
\eeq
where $\ell_0(r) = \sqrt{2r/a_0 \,-\, r^2/a_0^2\,}\,$.
Surface density profiles are plotted in Figure~{\ref{fig_sig_om}}a 
for the polarcap and band of Figure~{\ref{fig_sph_Pol_Band}}, and also
a broad band ($\ell_1 = 0.1\,,\ell_2 =0.9\,$), whose stability is studied later. We note that the $\Sigma_0(r)$ profiles of a polarcap and a band are very different: the former has a single maximum at the centre of the disc, whereas the latter has a characteristic double-horned shape. 

\begin{figure}
\hspace{-1cm}
\begin{subfigure}{0.45\linewidth}
\centering
\includegraphics[width=1.2 \textwidth,trim={0cm 0cm 0cm 0cm}]{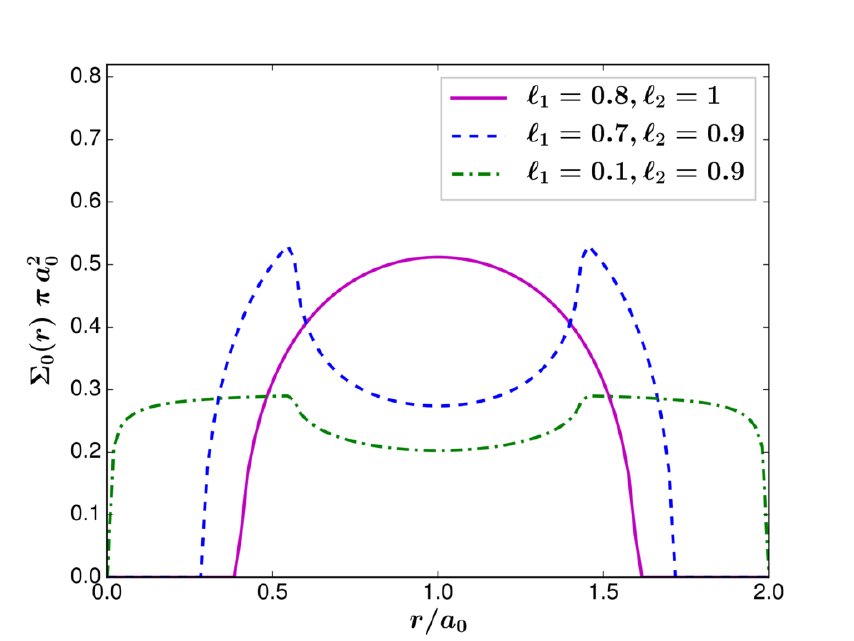}
\caption{Surface probability density  }
\end{subfigure}
\hspace{1.6cm}
\begin{subfigure}{0.45\linewidth}
\centering
\includegraphics[width=1.2 \textwidth,trim={0cm 0cm 0cm 0cm}]{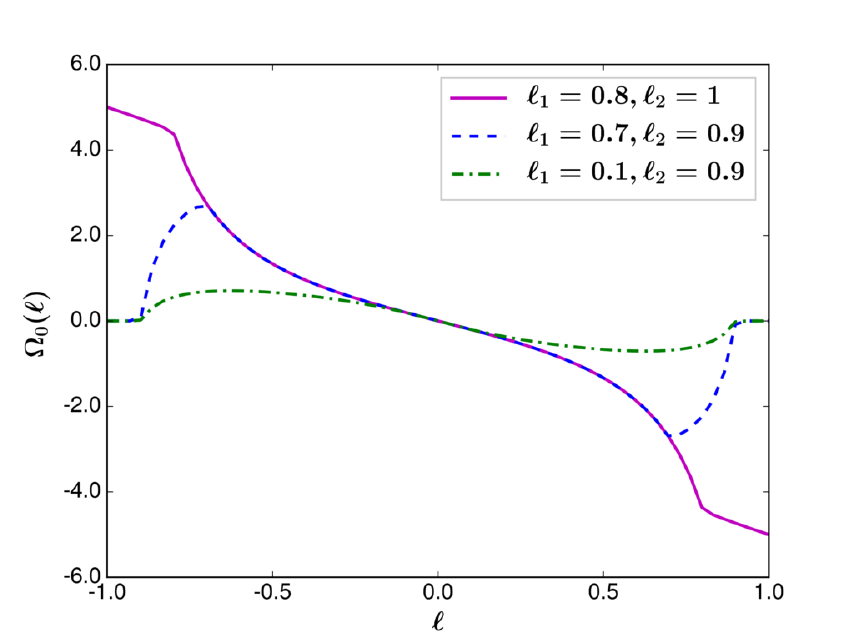}
\caption{Apse precession rate }
\end{subfigure}
\caption{\emph{ Physical features of waterbags}: Solid and dashed lines are for the polarcap and band of Figure~{\ref{fig_sph_Pol_Band}}, respectively. The broken dashed line is for a broad band, to be studied later.}
\label{fig_sig_om}
\end{figure}

The apse precession frequency $\Omega_0(\ell)$ can be determined by using equation~(\ref{eqn_waterbag-def}) in (\ref{eqn_omega0-def-mono}). For a 
polarcap, 
\beq
\Omega_0(\ell) \;=\; 
\begin{cases} 
\displaystyle{- \frac{2  \, \ell}{(1 -\ell^2)}}\,,\qquad\qquad\qquad\qquad\qquad  0 \leq \left|\ell \right| \leq \ell_1 \\[1 em] 
\displaystyle{ - \frac{2  \, \ell}{(1 + \left|\ell\right|)(1-\ell_1)}}\,, 
\qquad\qquad\qquad\;\,\ell_1 < \left| \ell \right| \le  1\,,               
\end{cases}                                                          
\label{eqn_omega0-waterbag-l2-unity} 
\eeq
and for a band,
\beq
\Omega_0(\ell) \;=\; 
\begin{cases} 
\displaystyle{- \frac{2  \, \ell}{(1 -\ell^2)}}\,,\qquad\qquad\qquad\qquad\quad  0 \leq \left|\ell \right| \leq \ell_1 \\[1 em] 
\displaystyle{ - \frac{2  \, \ell}{(1 -\ell^2)} \, \left( \frac{\ell_2 - \left| \ell \right|}{\ell_2-\ell_1} \right) }\,,\qquad\qquad 
\ell_1 < \left| \ell \right| \leq  \ell_2 \\[1 em]
\qquad \qquad 0\,,\qquad\qquad\qquad\qquad\quad\ell_2 < \left| \ell \right| \le 1\,.                                                                   \end{cases}                                                                 \label{eqn_omega0-waterbag-l2-general}
\eeq
Even though the waterbag itself occupies only the interval $[\ell_1,\ell_2]$ 
we calculate $\Omega_0(\ell)$ for all $\ell \in [-1, 1]$, because it gives the apse precession frequency of any test-ring that may be introduced
into the system. $\Omega_0$ is an antisymmetric function of $\ell$, as can 
be seen in Figure~{\ref{fig_sig_om}}b. For a polarcap $\Omega_0$ is
non zero when $\ell = \pm 1$, whereas for a band $\Omega_0(\ell)$ vanishes for all $\left|\ell \right| > \ell_2$. 

\subsubsection{Stability to non-axisymmetric modes}

An arbitrary collisionless perturbation of a waterbag can be described as a deformation of its boundaries. From Figure~{\ref{fig_sph_Pol_Band}} we
see that a polarcap has just one boundary at $\ell = \ell_1$ whereas a band has two boundaries, at $\ell = \ell_1$ and $\ell = \ell_2$. Non-axisymmetric
perturbations of the boundaries can be resolved as a Fourier series in the apsidal longitude $g$.  Figure~{\ref{fig_edge_modes}} shows a $m = 3$ deformation of the polarcap
and band of Figure~{\ref{fig_sph_Pol_Band}} where $m$ is the azimuthal wavenumber of perturbation. 

\begin{figure}
\begin{subfigure}{0.5\linewidth}
\centering
 \includegraphics[width=1.25 \textwidth ,trim={5cm 1cm 0cm 2cm}]{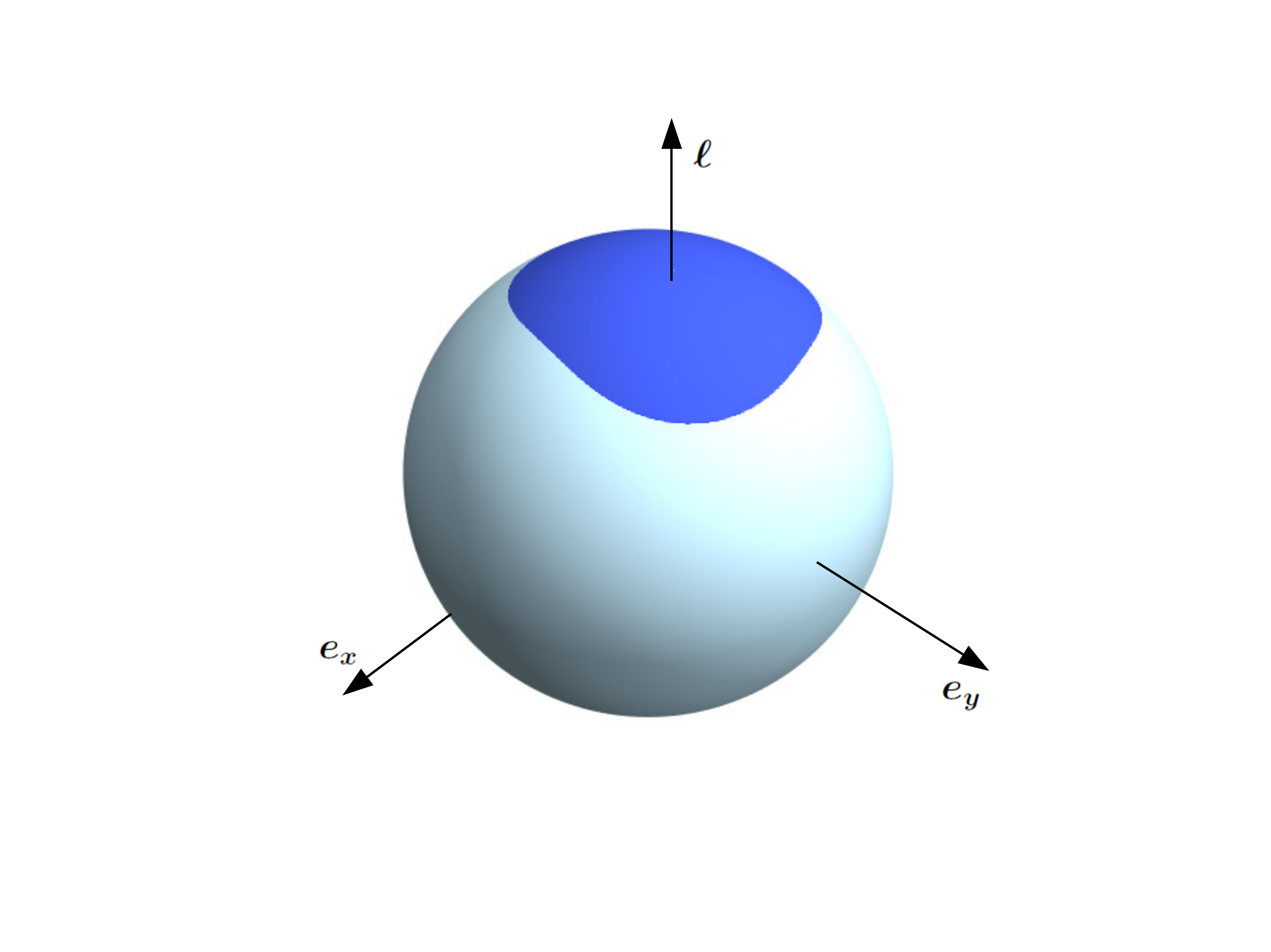}
\end{subfigure}
\hfill
\begin{subfigure}{0.45\linewidth}
\centering
\includegraphics[width=1.25 \textwidth ,trim={5cm 1cm 0cm 2cm}]{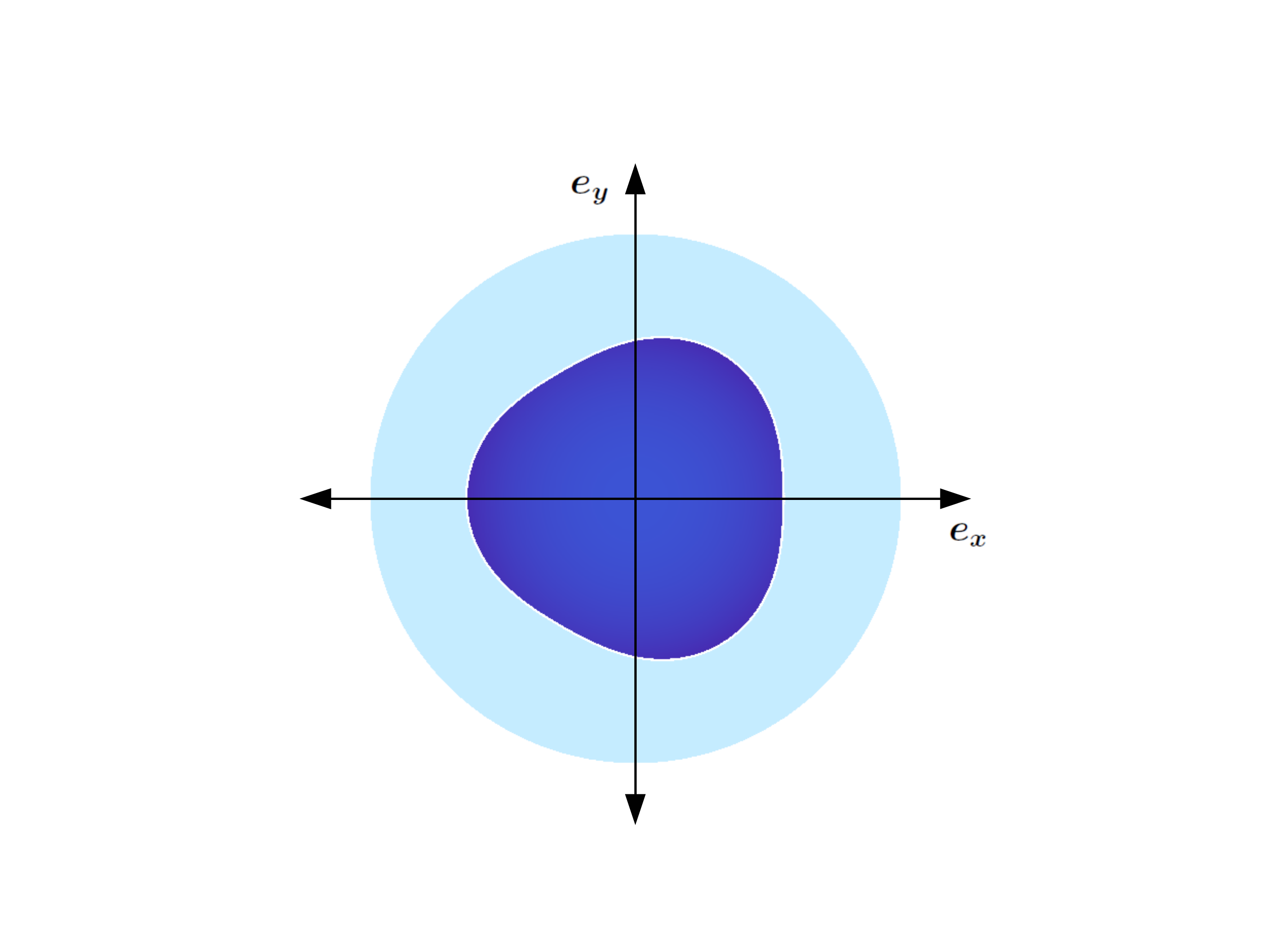}
\end{subfigure}
\hfill
\begin{subfigure}{0.5\linewidth}
\centering
\includegraphics[width=1.25 \textwidth ,trim={5cm 1cm 0cm 2cm}]{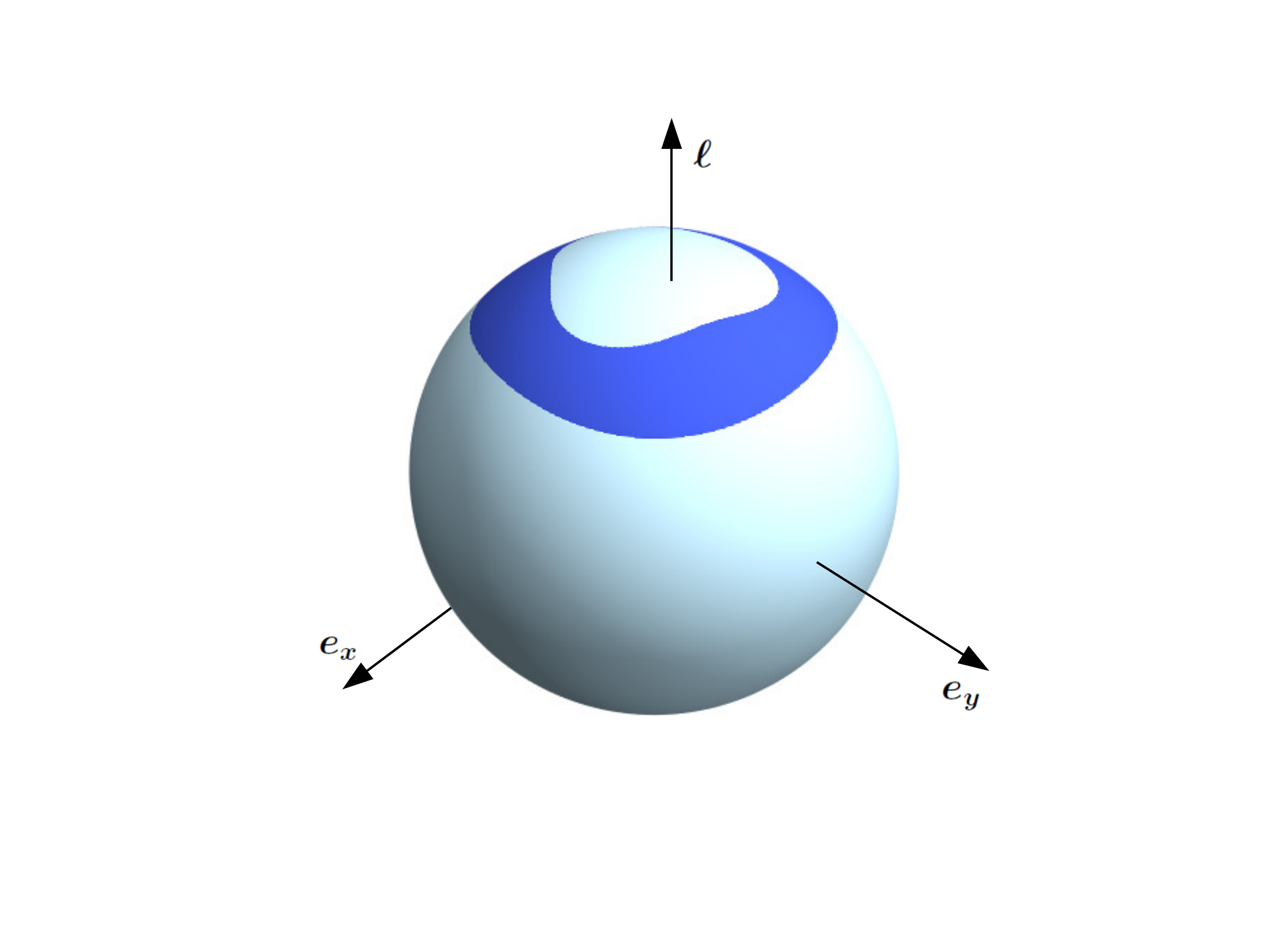}
\end{subfigure}
\hfill
\begin{subfigure}{0.45\linewidth}
\centering
\includegraphics[width=1.25 \textwidth ,trim={5cm 1cm 0cm 2.5cm}]{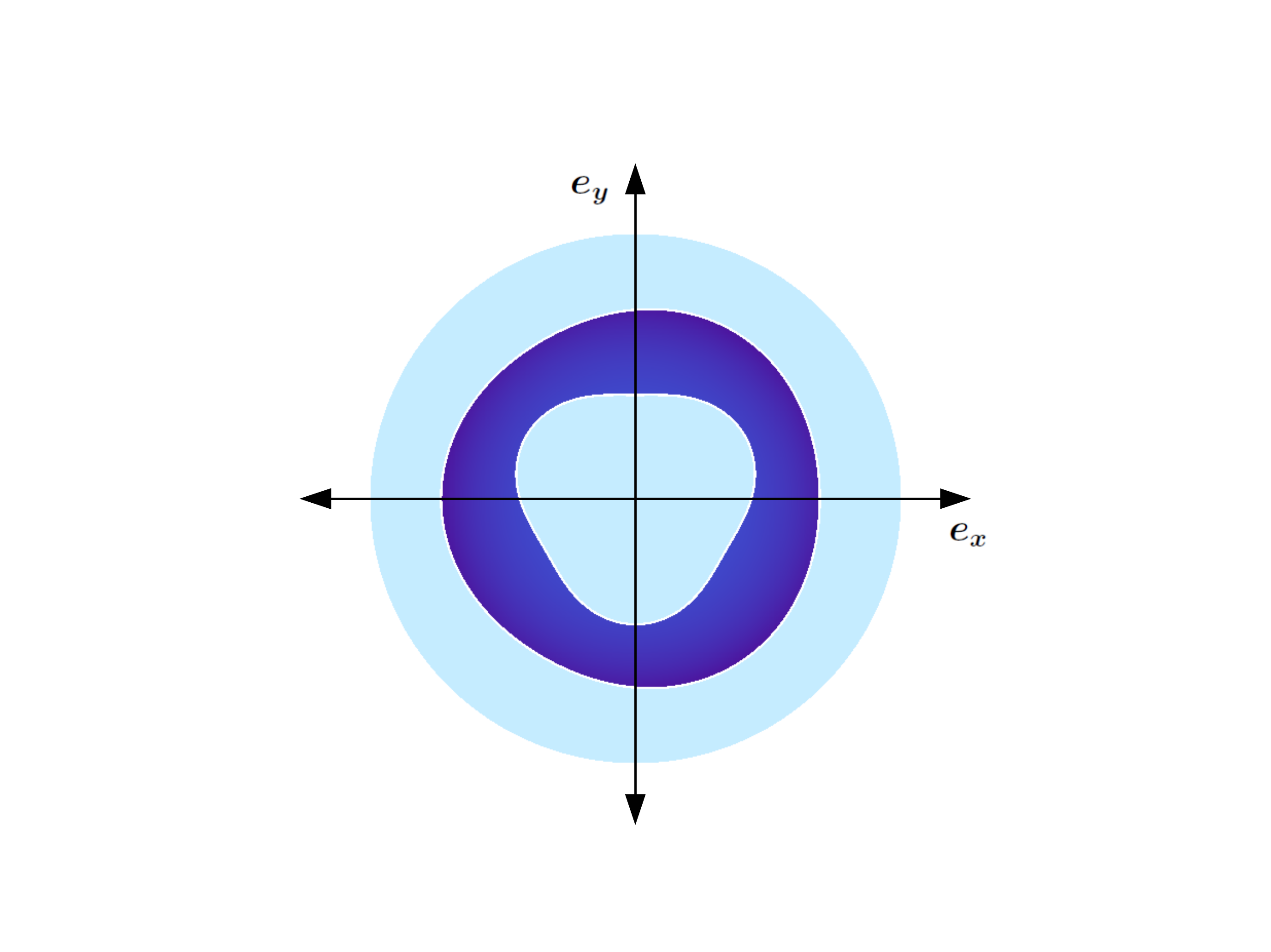}
\end{subfigure}

\hspace{10.2cm}
\begin{subfigure}{0.25\linewidth}
\centering
\includegraphics[width=1.25 \textwidth ]{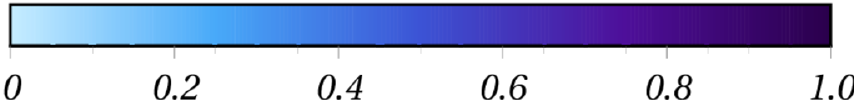}
\end{subfigure}
\caption{\emph{m=3 normal mode for Polarcap and Band}. The panels on the left show the deformed polarcap (Upper panel) and band (Lower panel) DFs. The panels on the right are for the corresponding probability densities, $n(e_x, e_y) = \ell^{-1}\times DF$, in the $(e_x, e_y)$ plane. Since the DF is constant within the deformed boundaries, $n\propto 1/\sqrt{1-e^2}\,$.}
\label{fig_edge_modes}
\end{figure}

\begin{figure}
\centering
\includegraphics[width=0.8 \textwidth,trim={0cm 0cm 0cm 0cm}]{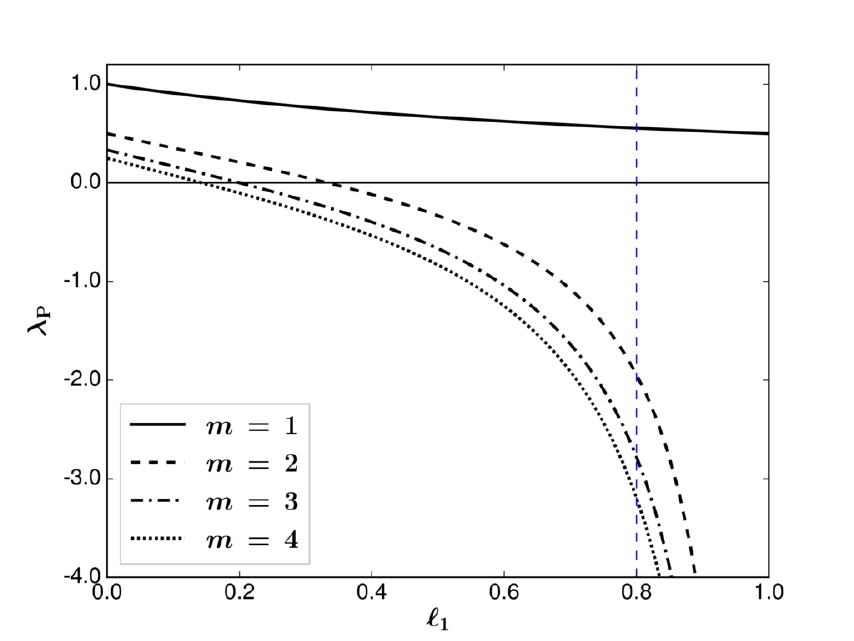}
\caption{\emph{Precession frequency of normal modes of Polarcaps}. The 
intersections of the vertical dashed line with the $\lambda_{\rm P}$ curves gives the spectrum of the normal modes of the polarcap of Figure~{\ref{fig_sph_Pol_Band}}. Only the $m=1$ normal mode has positive precession for all values of $\ell_1$.} 
\label{fig_prec_freq_polar}
\end{figure}

\medskip\noindent
{\bf Polarcaps} are linearly stable to all non-axisymmetric modes. 
In order to prove this we note that, for a polarcap, $\rmd f_0/\rmd \ell = (1 -\ell_1)^{-1} \delta(\ell-\ell_1)$. Substituting this in the integral equation~(\ref{eqn_integral-eigen-eqn-mono}) we obtain:
\beq
\left[\,\omega_m - m\Omega_0(\ell)\,\right]\,f_{1m}(\ell) \;=\;  
\frac{\delta(\ell-\ell_1)}{1 -\ell_1} 
\int_{-1}^{1}{\rm d} \ell'\,\left(\frac{e_<}{e_>}\right)^m  f_{1m}(\ell')\;,
\label{eqn_integral-eigen-polar}
\eeq
where $\Omega_0(\ell)$ is given by equation~{\ref{eqn_omega0-waterbag-l2-unity}}. The physical solution is $f_{1m}(\ell) = A_m\,\delta(\ell-\ell_1)$, where $A_m$ is a complex amplitude. Using this in equation~(\ref{eqn_integral-eigen-polar}) we obtain the eigenvalue, 
\beq
\omega_m \;=\; m\, \Omega_0(\ell_1) \;+\; \frac{1}{1 - \ell_1}\,. 
\label{eqn_l2_1_eigenvalue}
\eeq
Since $\omega_m$ is real for all $m = 1, 2, \ldots$ and $0\leq \ell_1 < 1$, all normal modes are stable and purely oscillatory. For each $m$ there is a 
normal mode with
\beq
f_1(\ell,g,t; m) \;=\; {\rm Re}\left\{A_m\delta(\ell-\ell_1)\, \exp{\left[{\rm i}m(g - \lambda_{\rm P} t)\right]}\right\},
\eeq
where 
\beq 
\lambda_{\rm P}(m, \ell_1) \;=\; \frac{\omega_m}{m} \;=\; 
- \frac{2  \, \ell_1}{(1 -\ell_1^2)} \;+\; \frac{1}{m(1 - \ell_1)}
\eeq
is the precession frequency of the $m$-lobed, sinusoidal deformation of the polarcap boundary. The first term on the right side is just the apse
precession frequency in the unperturbed polarcap, and is negative. The 
second term comes from the self-gravity of the deformation, which is positive. The competition between these two terms results in the following interesting features of $\lambda_{\rm P}(m, \ell_1)\,$, as can be seen 
in Figure~{\ref{fig_prec_freq_polar}}: 
\begin{itemize}
\item For a polarcap with given $\ell_1$, $\,\lambda_{\rm P}$ is a decreasing function of $m\,$. This is because the self-gravity of the deformed edge
is smaller for bigger $m$, due to mutual cancellation from its lobes and dips. In the limit $m\to\infty\,$ this vanishes altogether and $\,\lambda_{\rm P}\to\Omega_0(\ell_1)$.  

\item
The $m=1$ mode always has prograde precession, with $\lambda_{\rm P} = 1/(1+ \ell_1)\,$.  

\item
Modes with $m=2,3,\ldots$ precess in a prograde sense for $0\leq \ell_1 < 1/(2m -1)\,$, and in a retrograde sense for $1/(2m -1) < \ell_1 \leq 1\,$.
$\lambda_{\rm P}$ vanishes when a polarcap is such that $\ell_1 = 1/(2m -1)$ for some $m$; then it has a stationary time-independent deformation with 
$m$ lobes.

\item
For $\ell_1 > 1/3$, only the $m=1$ mode has positive pattern speed. 
\end{itemize}

\medskip\noindent 
{\bf Bands} have richer stability properties because, for each $m$, there 
are two normal modes (as shown in Section~5). Each of these is composed of sinusoidal disturbances of the two edges of phase space DF --- see the lower panels of Figure~{\ref{fig_edge_modes}} for a representation of a $m=3$ mode. For bands $\rmd f_0/\rmd \ell = \{\delta(\ell-\ell_1) - \delta(\ell-\ell_2)\}/\Delta \ell\,$, where $\Delta\ell = (\ell_2 - \ell_1)\,$. Substituting this in  equation~(\ref{eqn_integral-eigen-eqn-mono}) we obtain the following integral 
equation:
\beq
\left[\,\omega_m - m\Omega_0(\ell)\,\right]\,f_{1m}(\ell) \;=\;  
\frac{\delta(\ell-\ell_1) - \delta(\ell-\ell_2)}{\Delta \ell}
 \int_{-1}^{1}{\rm d} \ell'\,\left(\frac{e_<}{e_>}\right)^m  f_{1m}(\ell')\;,
\label{eqn_integral-eigen-band}
\eeq
where $\Omega_0(\ell)$ is given by equation~{(\ref{eqn_omega0-waterbag-l2-general})}. Hence the eigenfunctions are of the form:
\beq
f_{1m}(\ell) = A_{m1}\,\delta(\ell - \ell_1) + A_{m2} \, \delta(\ell - \ell_2).
\label{eqn_edgemode-form}
\eeq
where $A_{m1}$ and $A_{m2}$ are complex amplitudes. When equation~(\ref{eqn_edgemode-form}) for $f_{1m}(\ell)$ is substituted in  
equation~(\ref{eqn_integral-eigen-band}) the integral equation reduces to
a $2 \times 2$ matrix eigenvalue problem. This is the simplest linear stability problem in secular dynamics that can be studied analytically in detail --- see Section~\ref{sec: instability-analytics}. Before doing this 
we present numerical simulations of an unstable band and a stable band, so the reader may have an immediate picture of the time evolution going beyond the linear evolution of small disturbances.

\section{Numerical exploration of waterbag stability}
\label{sec_simu_intro}

\begin{figure}
\begin{subfigure}{0.3\linewidth}
\includegraphics[width=1.25 \textwidth ]{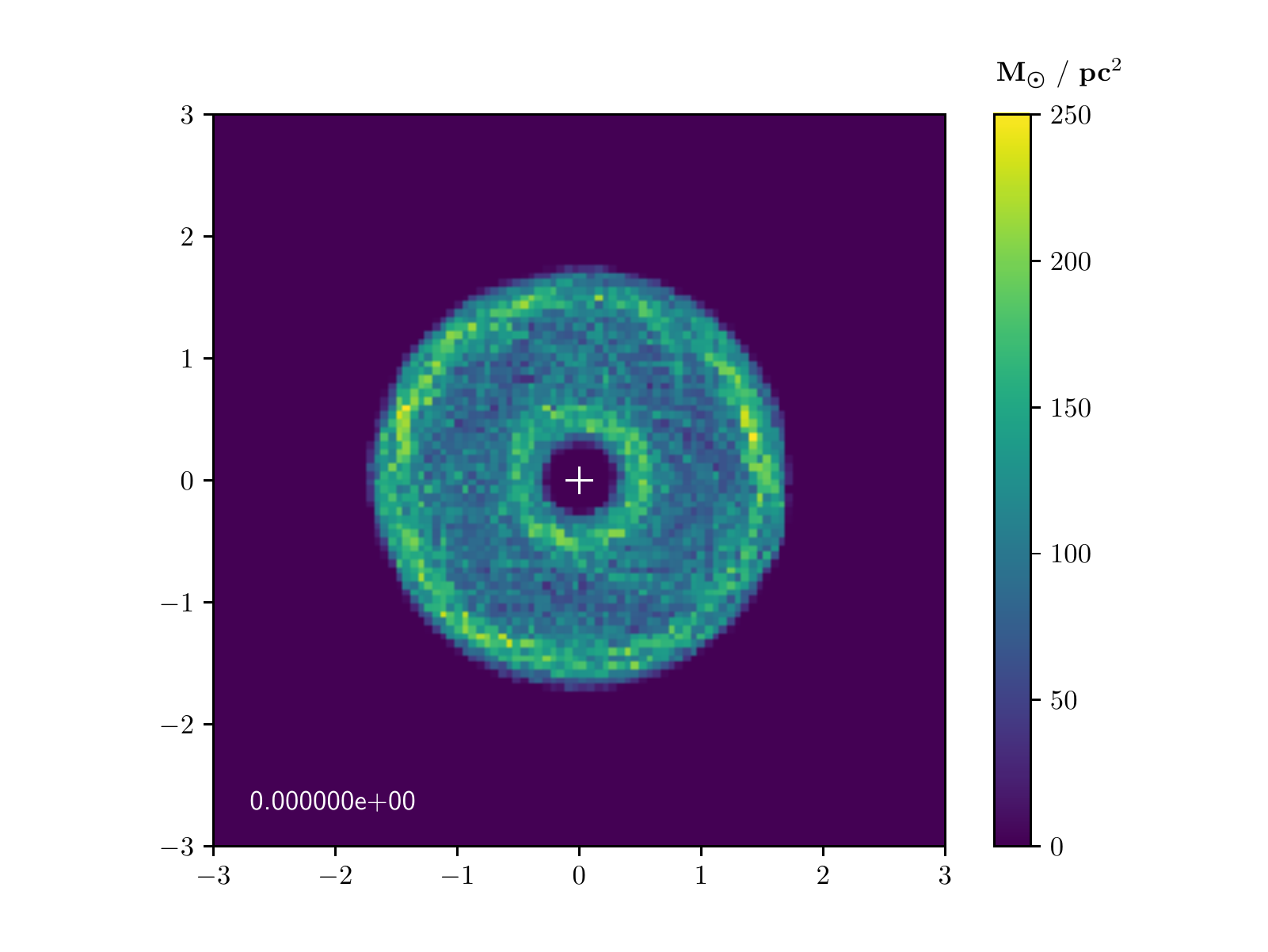}
\end{subfigure}
\hfill
\begin{subfigure}{0.3\linewidth}
\includegraphics[width=1.25 \textwidth ]{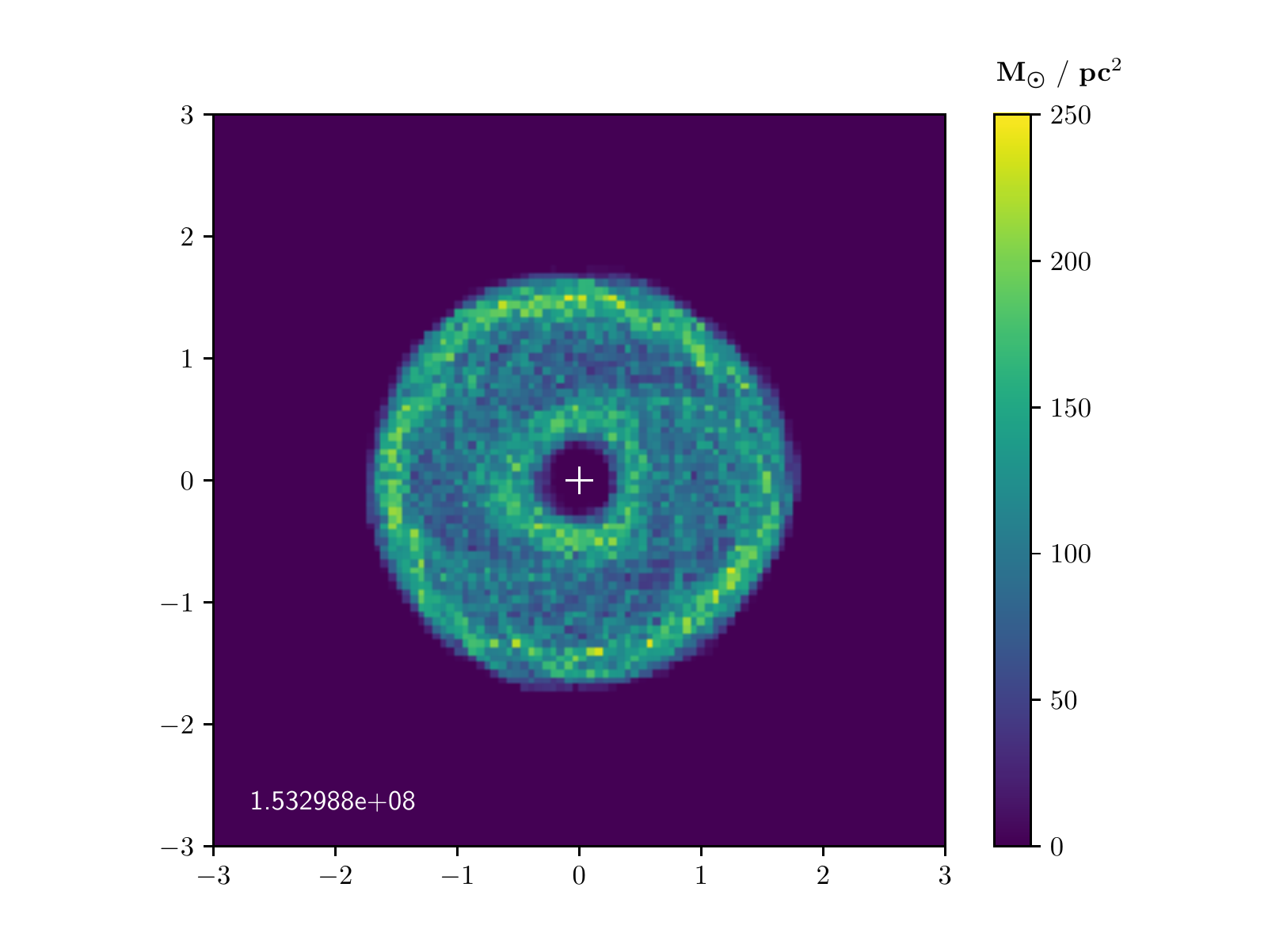}
\end{subfigure}
\hfill
\begin{subfigure}{0.3\linewidth}
\includegraphics[width=1.25 \textwidth ]{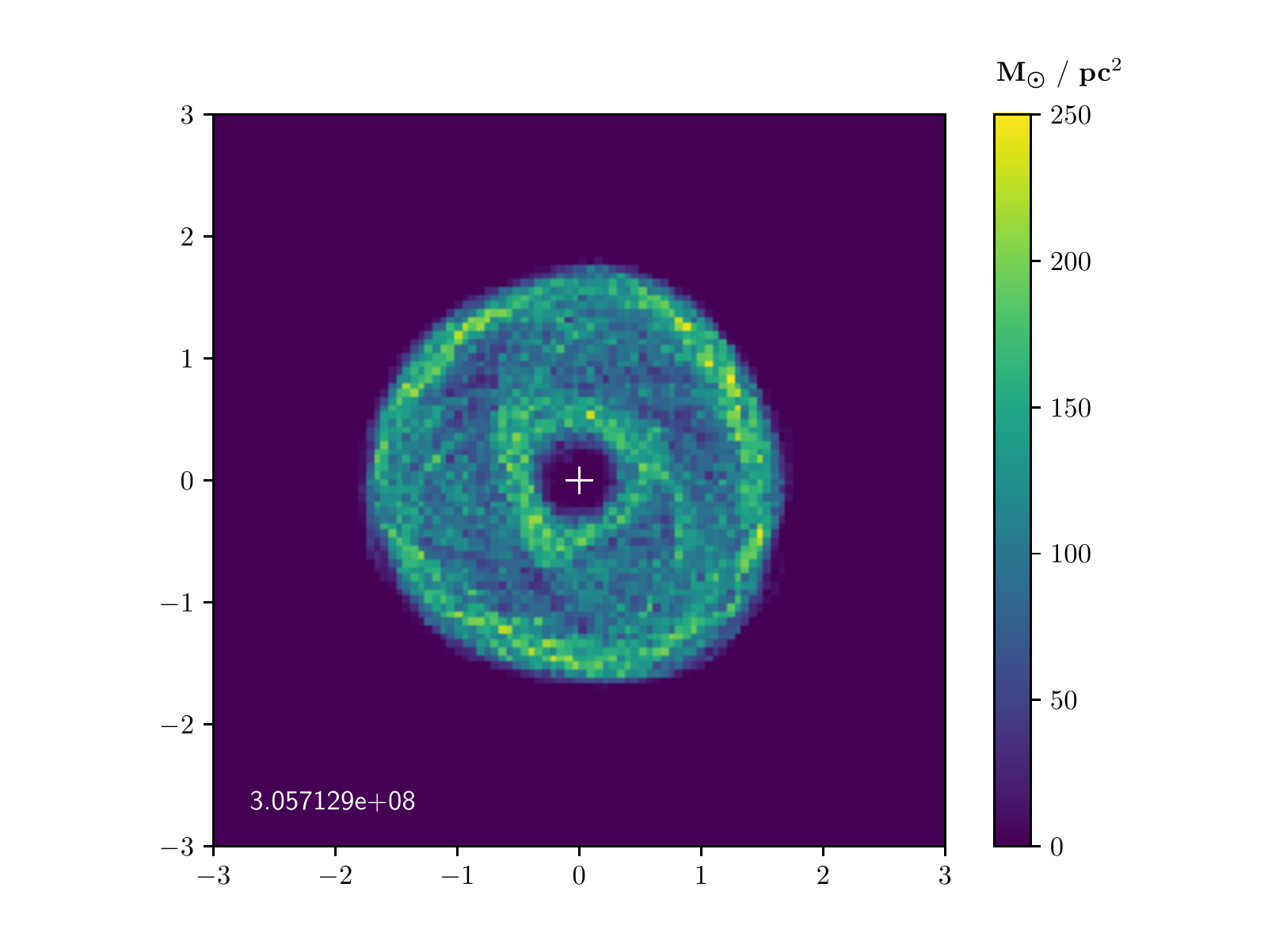}
\end{subfigure}
\hfill

\begin{subfigure}{0.3\linewidth}
\includegraphics[width=1.25 \textwidth ]{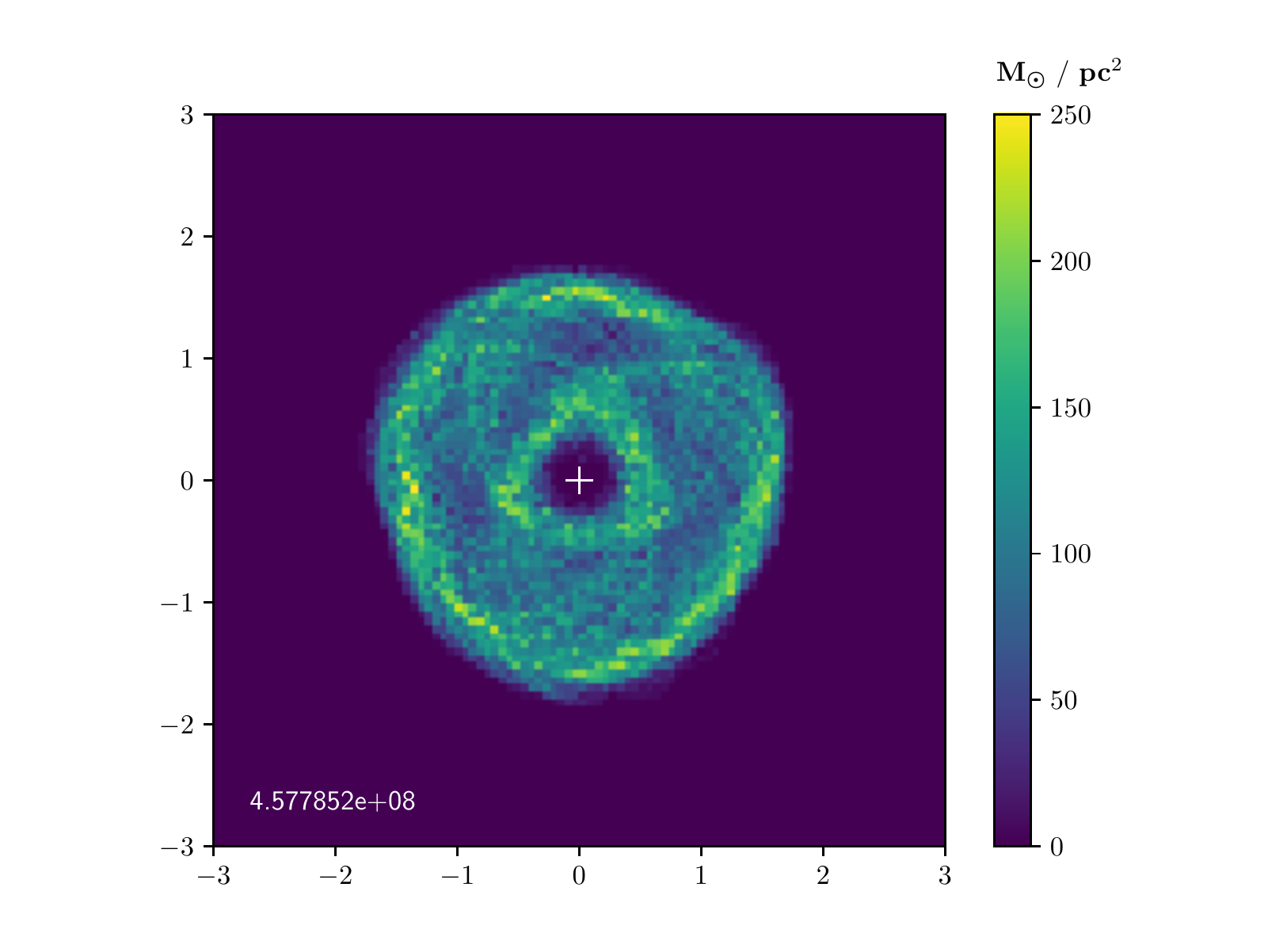}
\end{subfigure}
\hfill
\begin{subfigure}{0.3\linewidth}
\includegraphics[width=1.25 \textwidth ]{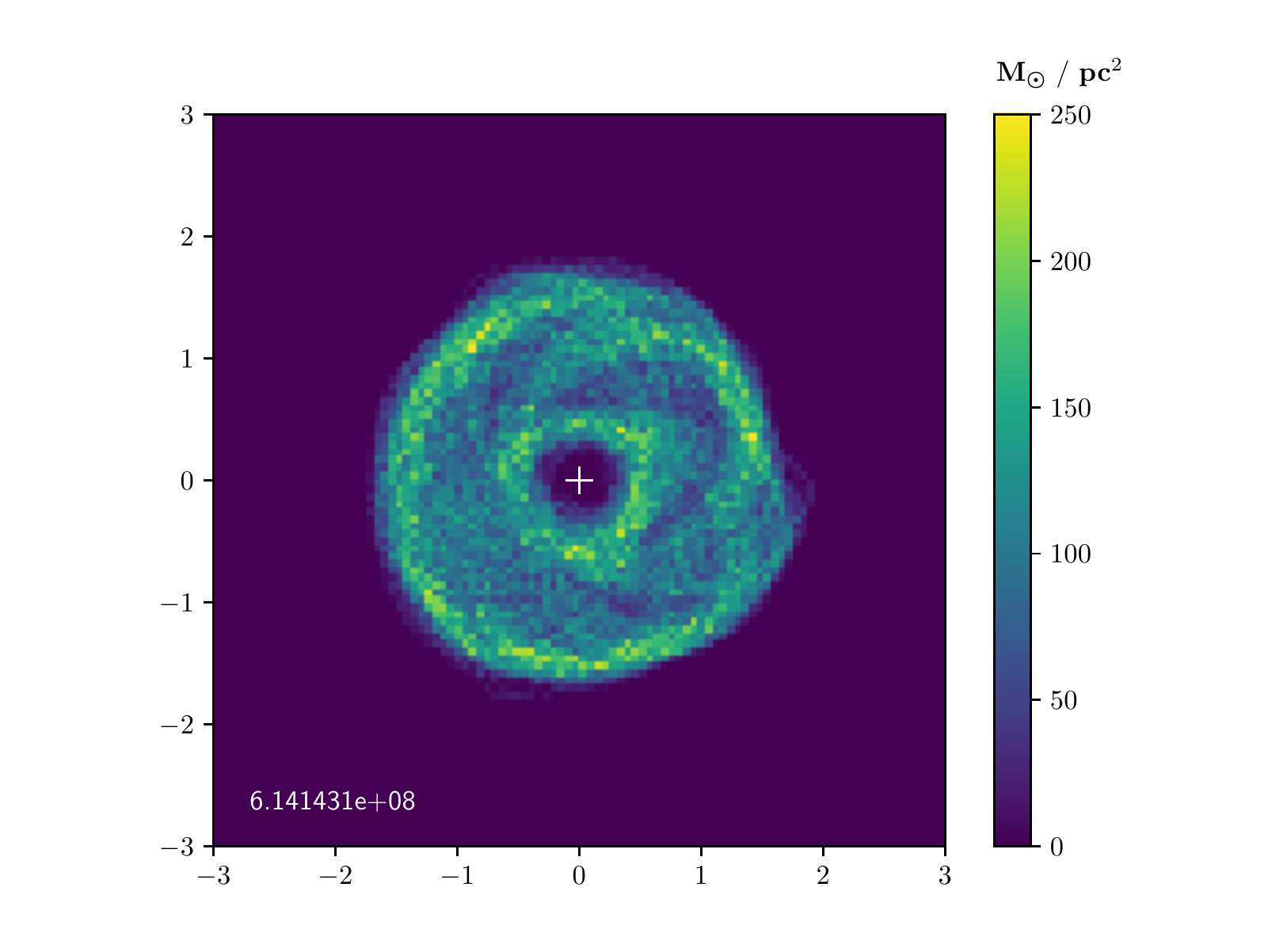}
\end{subfigure}
\hfill
\begin{subfigure}{0.3\linewidth}
\includegraphics[width=1.25 \textwidth ]{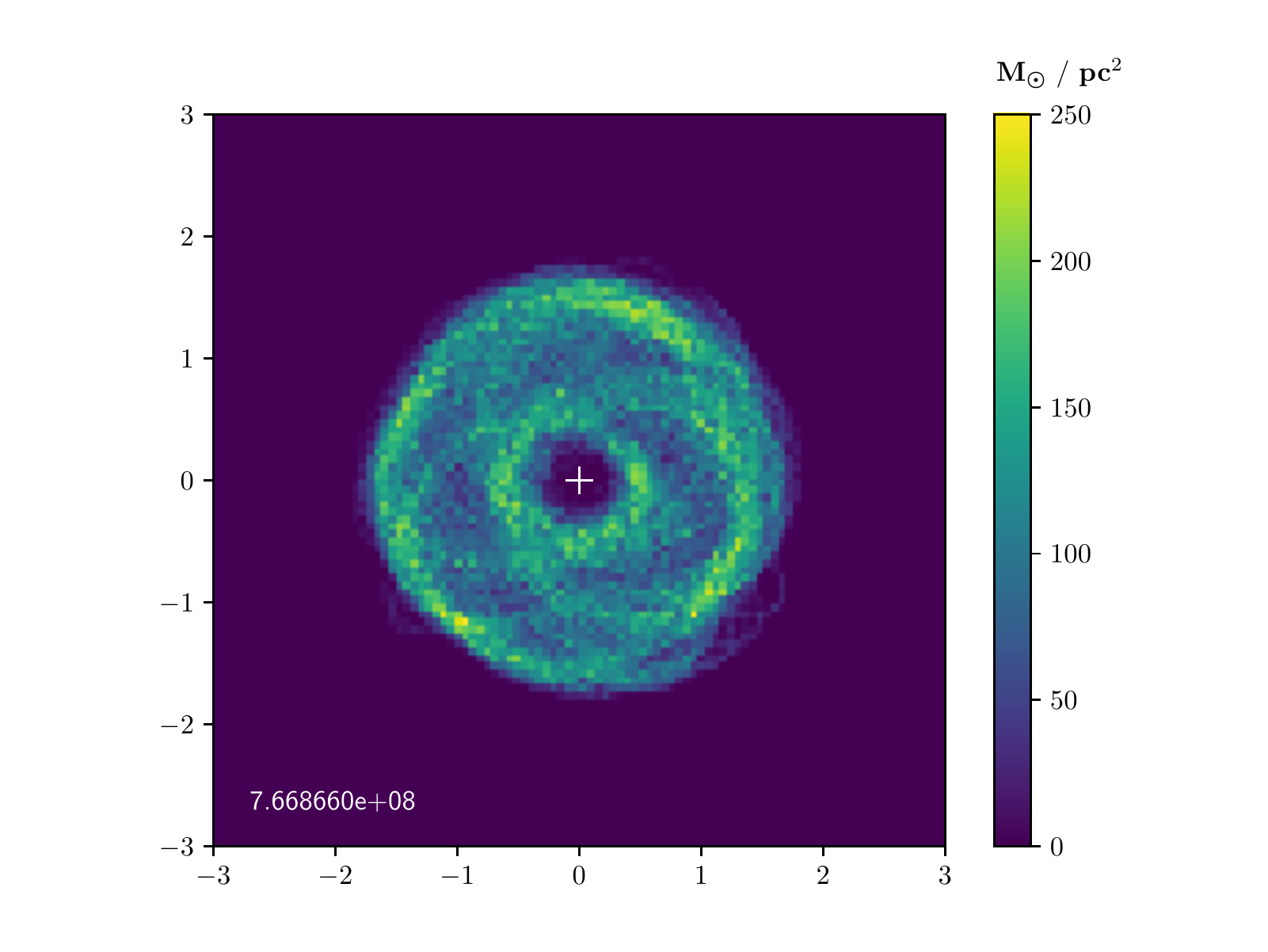}
\end{subfigure}

\begin{subfigure}{0.3\linewidth}
\includegraphics[width=1.0 \textwidth ]{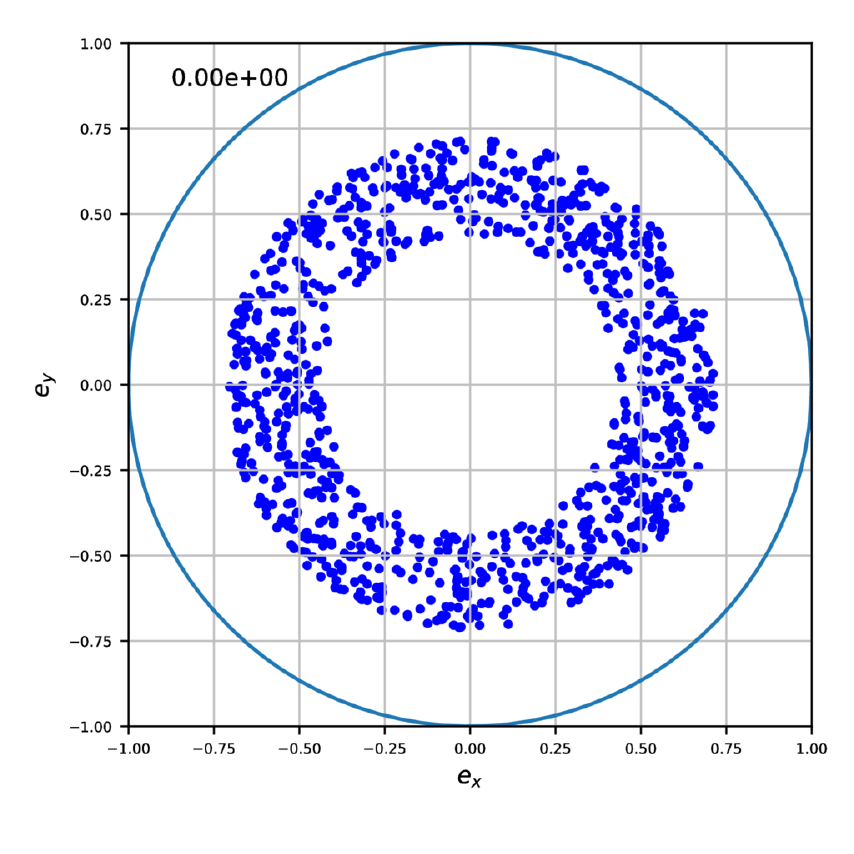}
\end{subfigure}
\hfill
\begin{subfigure}{0.3\linewidth}
\includegraphics[width=1.0 \textwidth ]{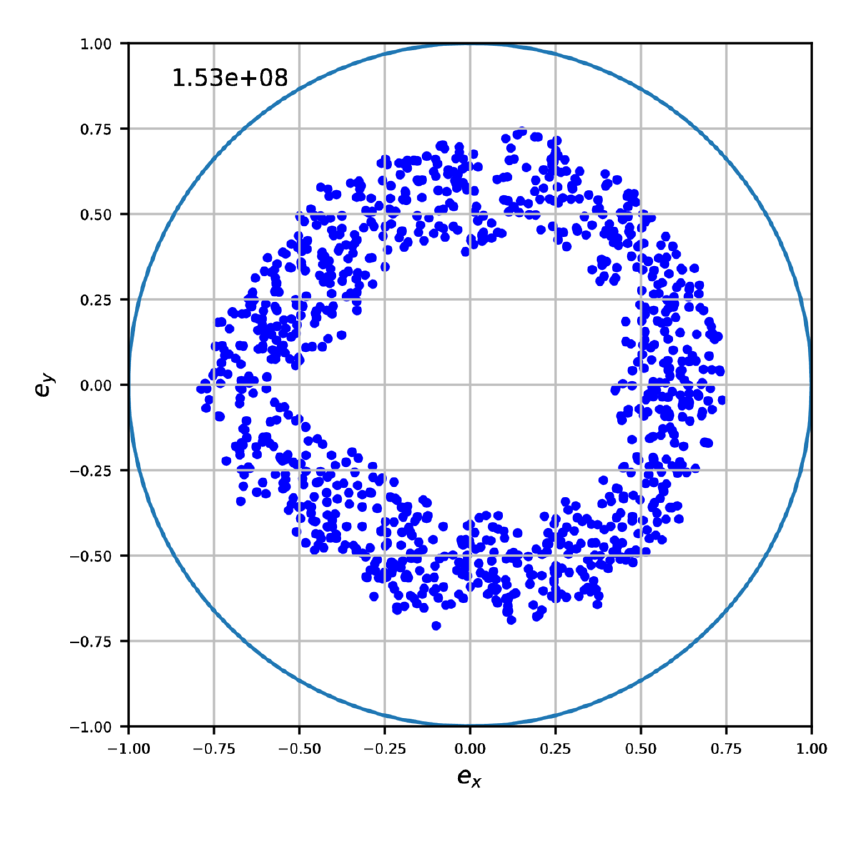}
\end{subfigure}
\hfill
\begin{subfigure}{0.3\linewidth}
\includegraphics[width=1.0 \textwidth ]{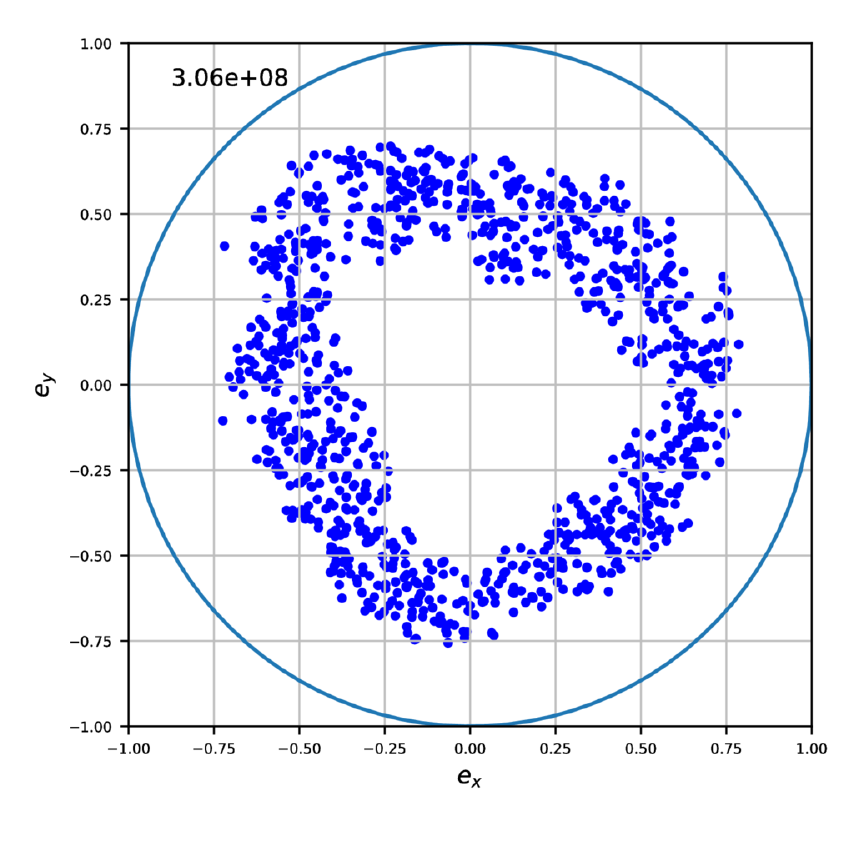}
\end{subfigure}
\hfill

\begin{subfigure}{0.3\linewidth}
\includegraphics[width=1.0 \textwidth ]{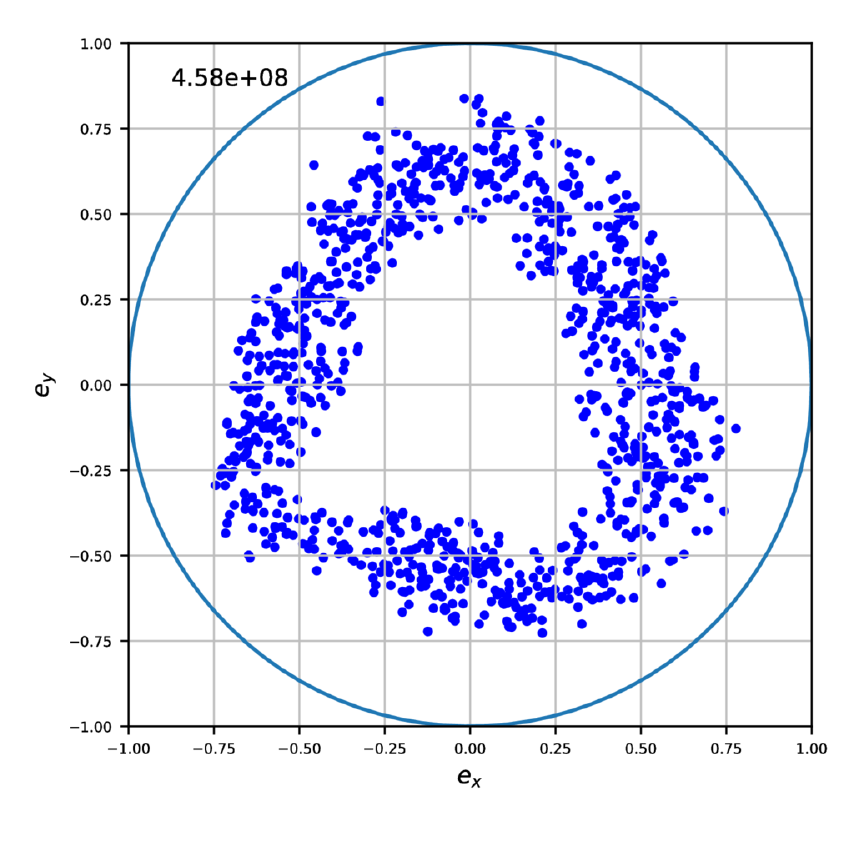}
\end{subfigure}
\hfill
\begin{subfigure}{0.3\linewidth}
\includegraphics[width=1.0 \textwidth ]{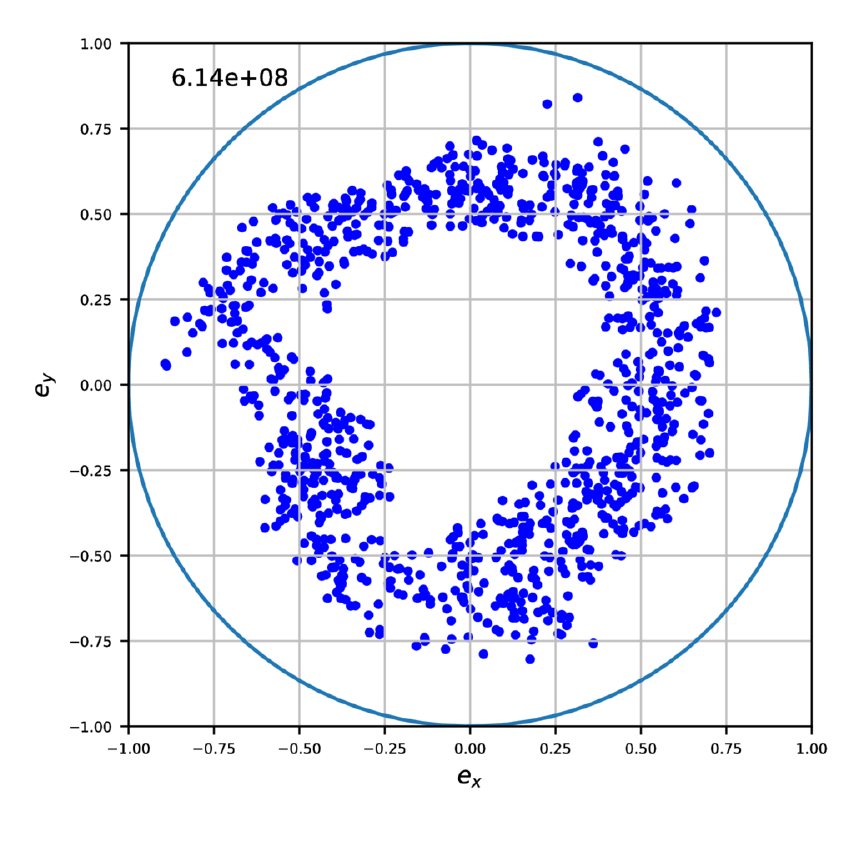}
\end{subfigure}
\hfill
\begin{subfigure}{0.3\linewidth}
\includegraphics[width=1.0 \textwidth ]{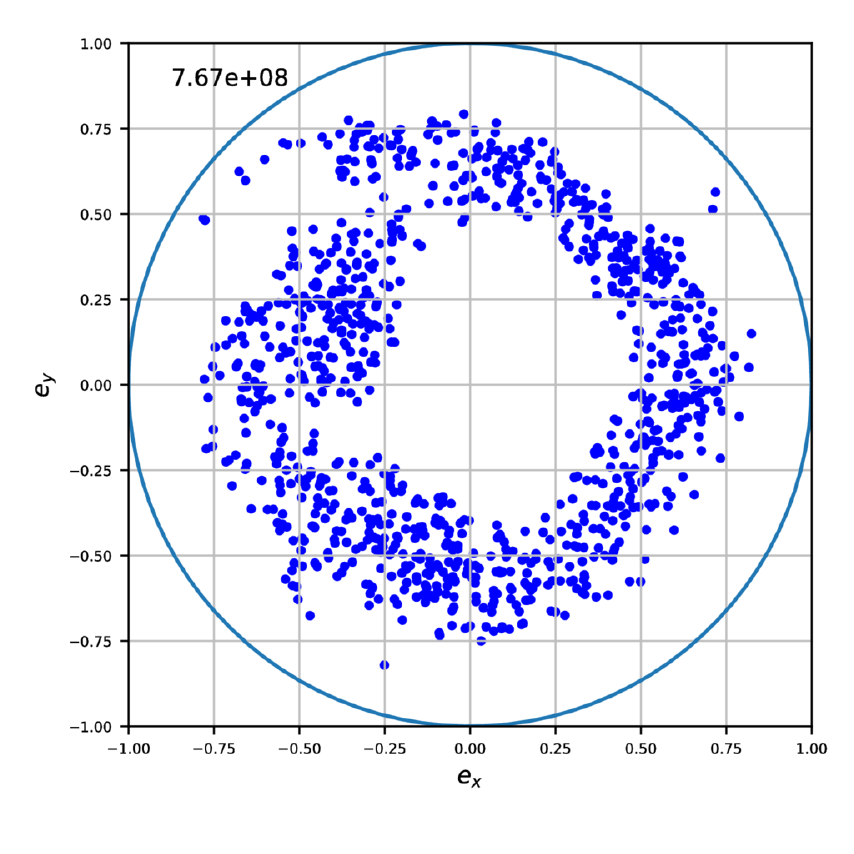}
\end{subfigure}
 \caption{\emph{Evolution of the unstable band} \texttt{waterbag\_1\_s0}. Upper two rows show the surface density in real space (with distances measured in parsec), and the lower two rows show the distribution in the eccentricity plane at the same respective time. The $m=3$ mode is clealy visible as three overdensity lumps in the surface density plots and as a triangular feature in the eccentricity plane. Note that the time (in years) is indicated within the subfigures. }
\label{fig_wb1_DF_plane}
\end{figure}

\begin{figure}
\begin{subfigure}{0.3\linewidth}
\includegraphics[width=1.25 \textwidth ]{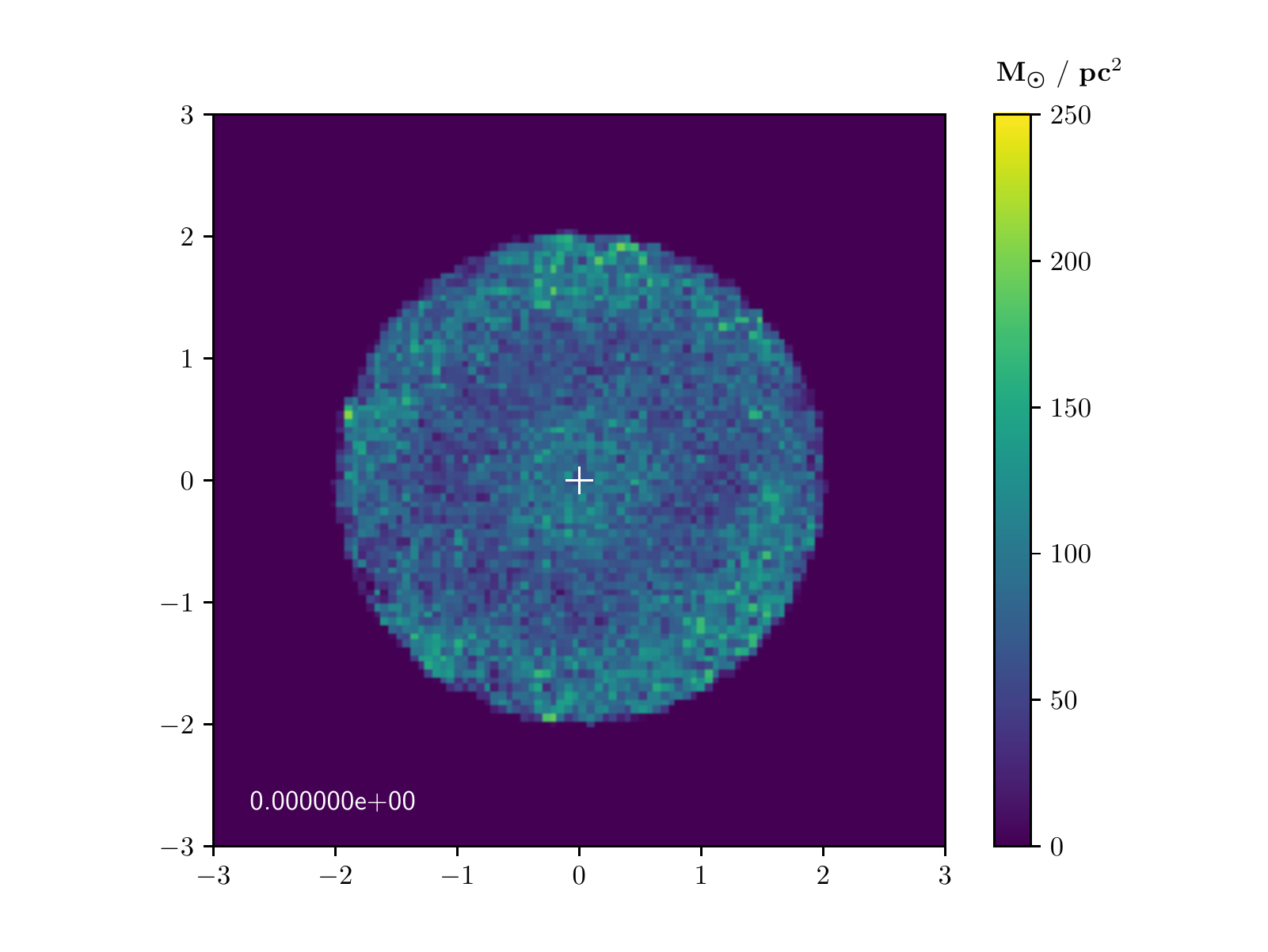}
\end{subfigure}
\hfill
\begin{subfigure}{0.3\linewidth}
\includegraphics[width=1.25 \textwidth ]{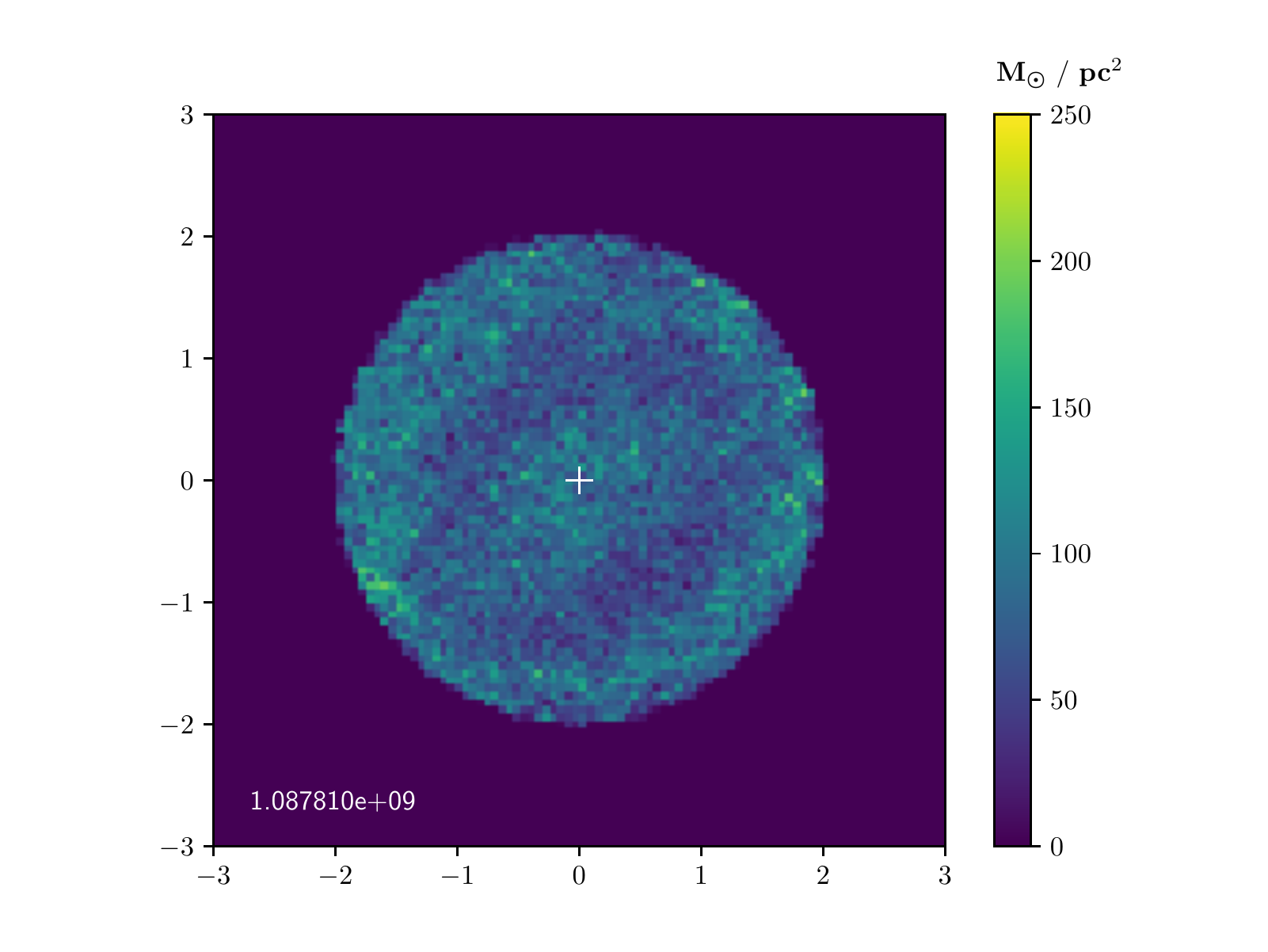}
\end{subfigure}
\hfill
\begin{subfigure}{0.3\linewidth}
\includegraphics[width=1.25 \textwidth ]{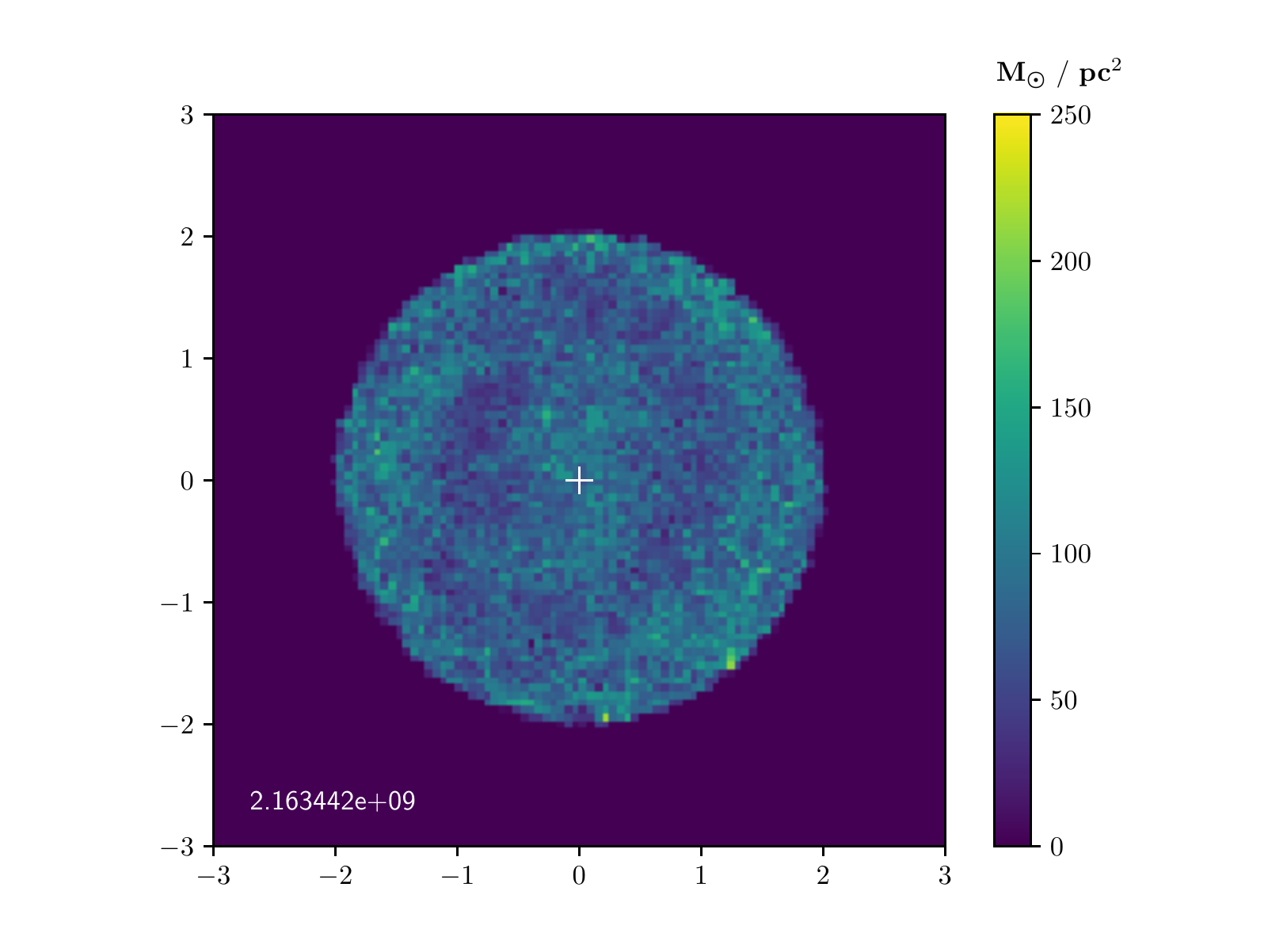}
\end{subfigure}
\hfill

\begin{subfigure}{0.3\linewidth}
\includegraphics[width=1.25 \textwidth ]{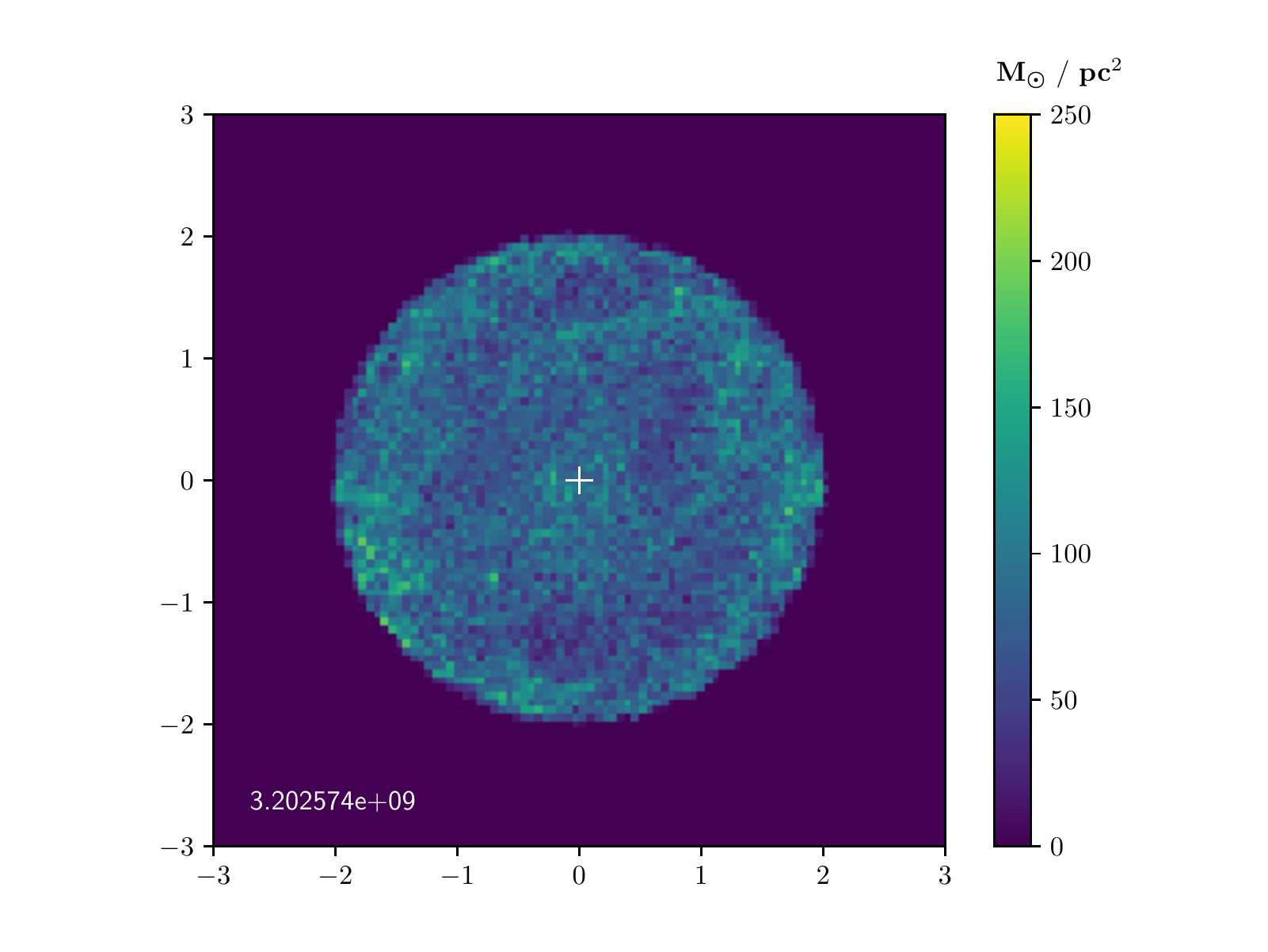}
\end{subfigure}
\hfill
\begin{subfigure}{0.3\linewidth}
\includegraphics[width=1.25 \textwidth ]{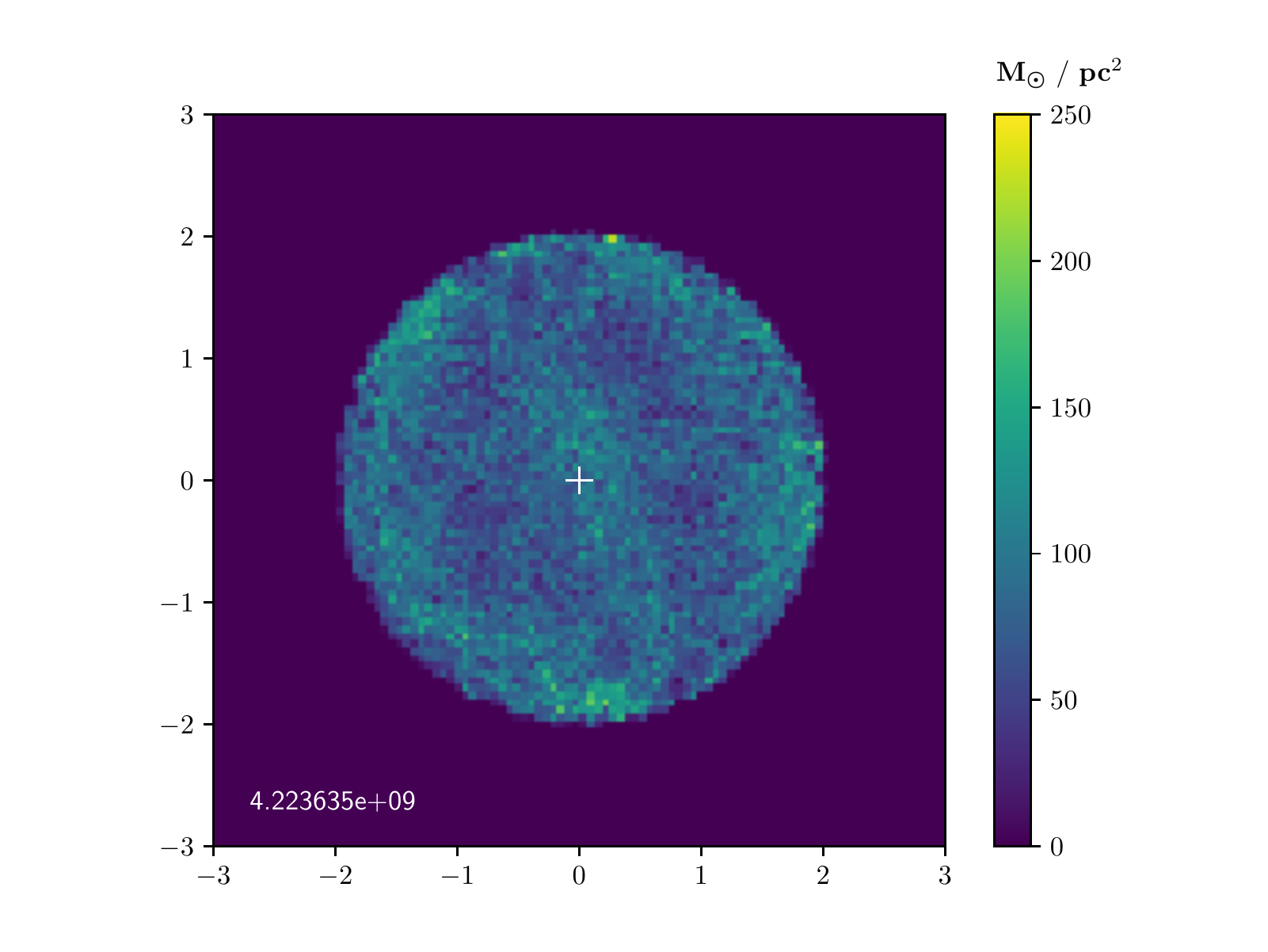}
\end{subfigure}
\hfill
\begin{subfigure}{0.3\linewidth}
\includegraphics[width=1.25 \textwidth ]{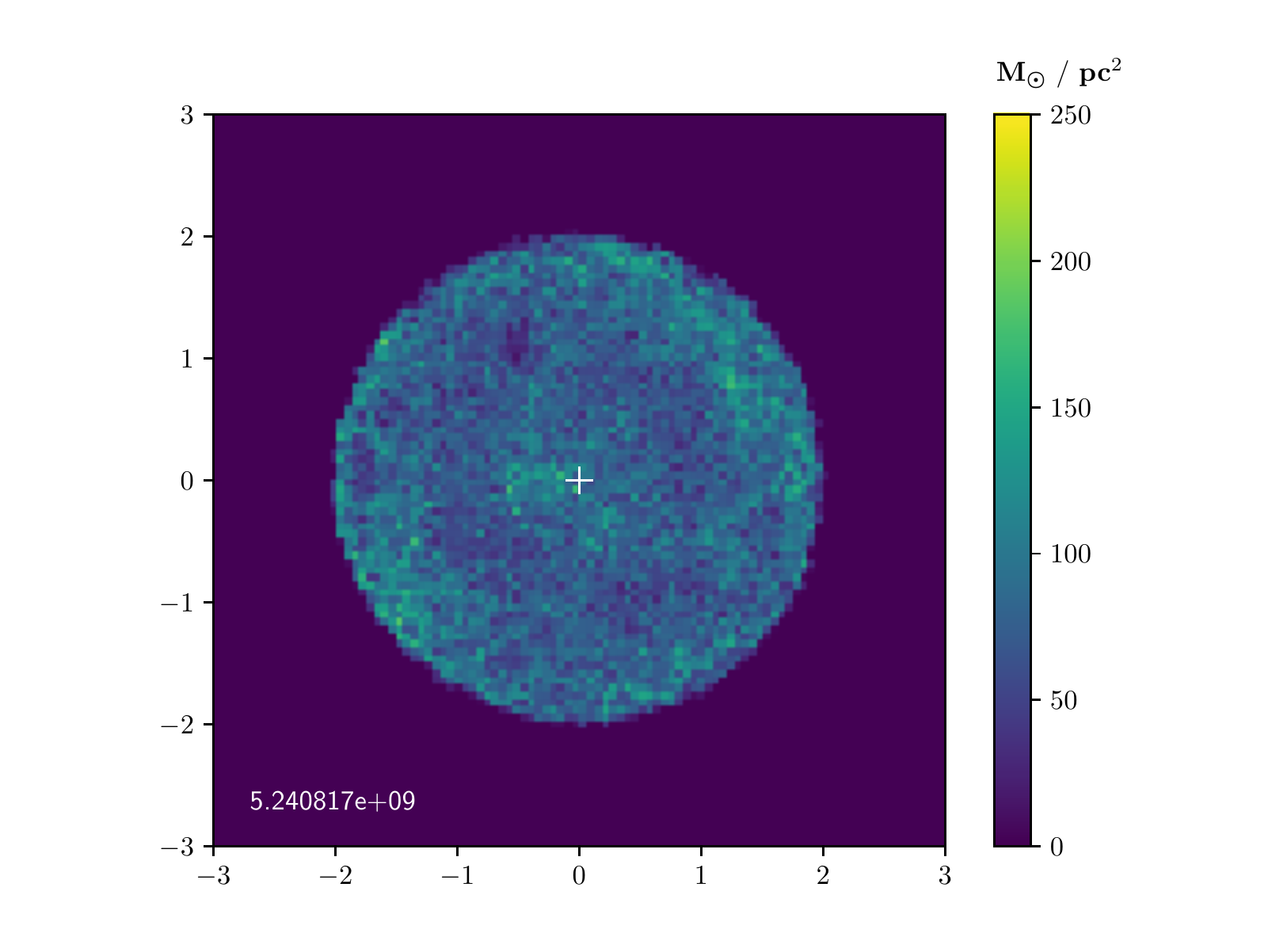}
\end{subfigure}

\begin{subfigure}{0.3\linewidth}
\includegraphics[width=1.0 \textwidth ]{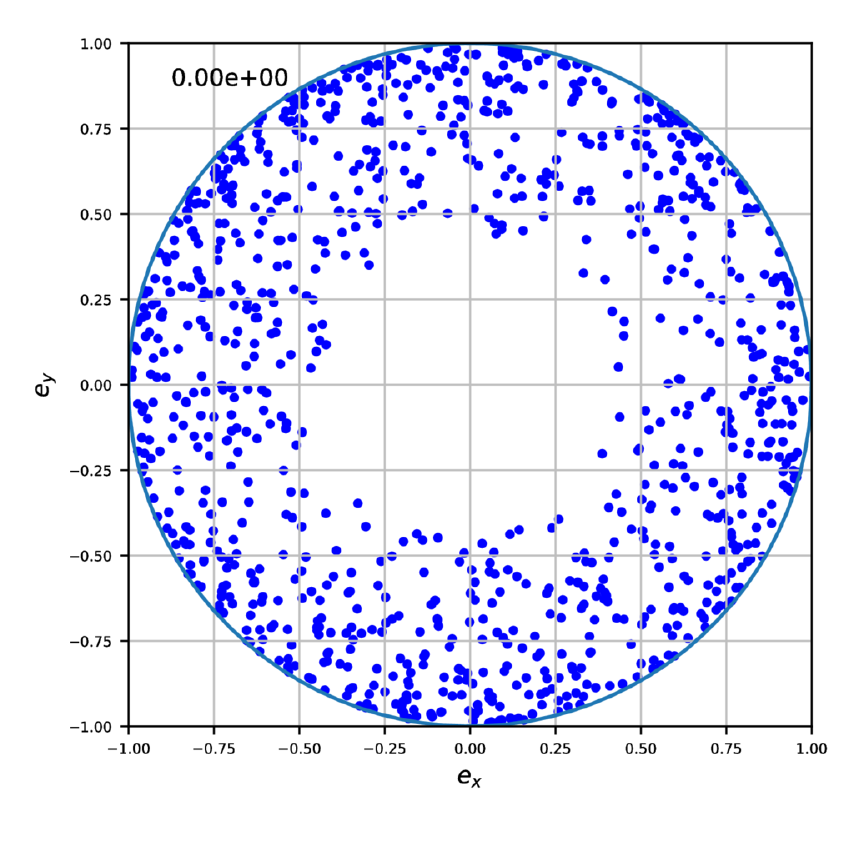}
\end{subfigure}
\hfill
\begin{subfigure}{0.3\linewidth}
\includegraphics[width=1.0 \textwidth ]{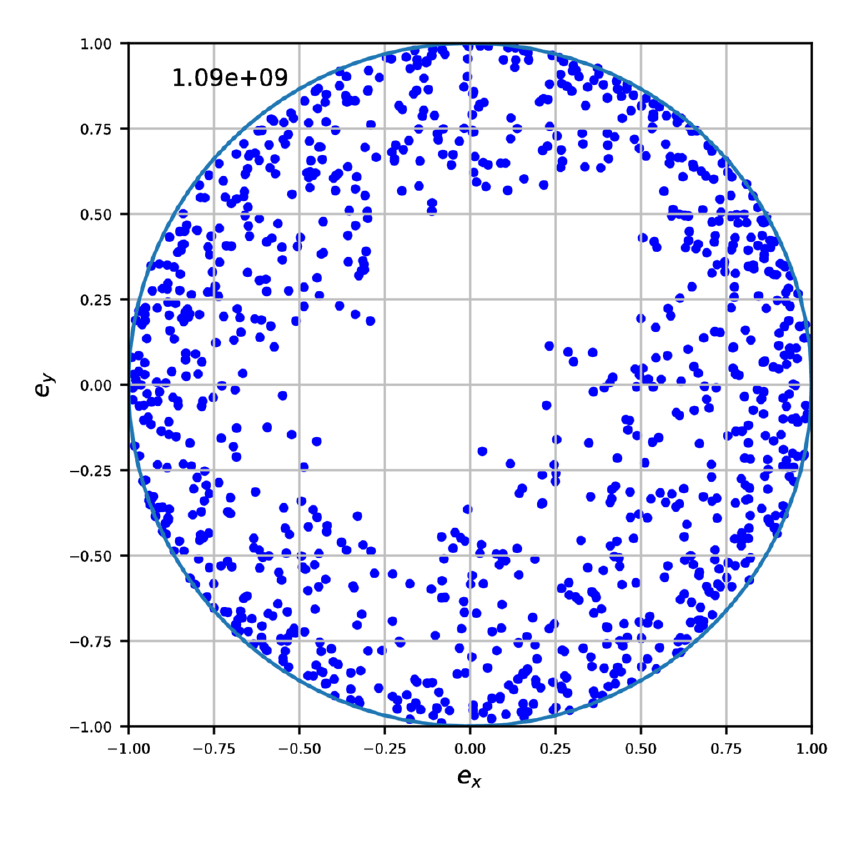}
\end{subfigure}
\hfill
\begin{subfigure}{0.3\linewidth}
\includegraphics[width=1.0 \textwidth ]{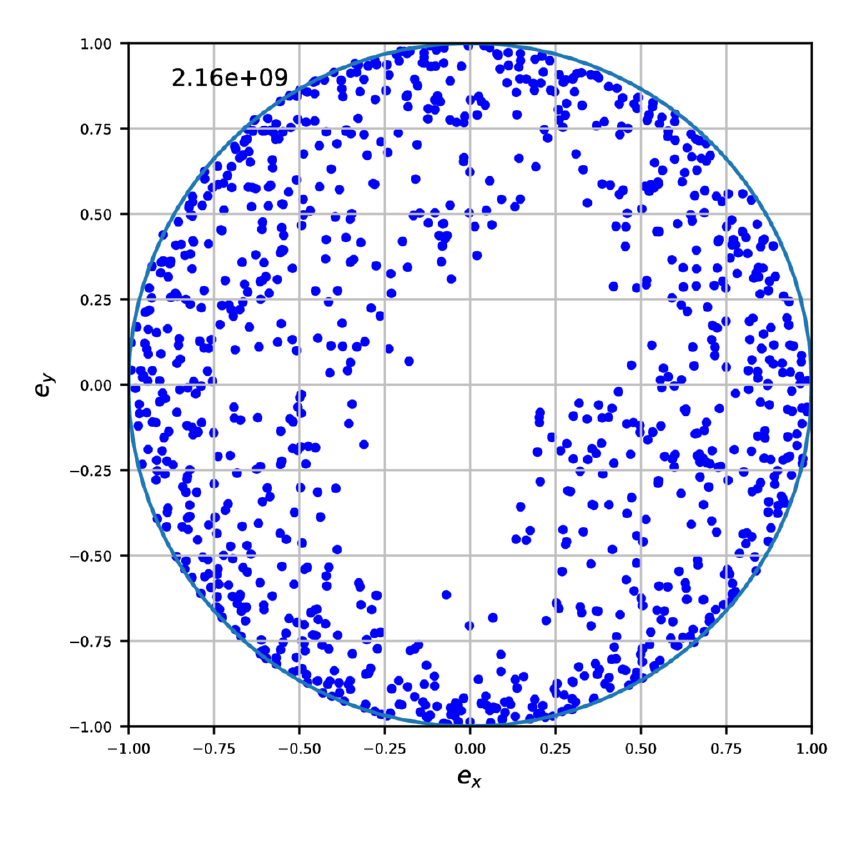}
\end{subfigure}
\hfill

\begin{subfigure}{0.3\linewidth}
\includegraphics[width=1.0 \textwidth ]{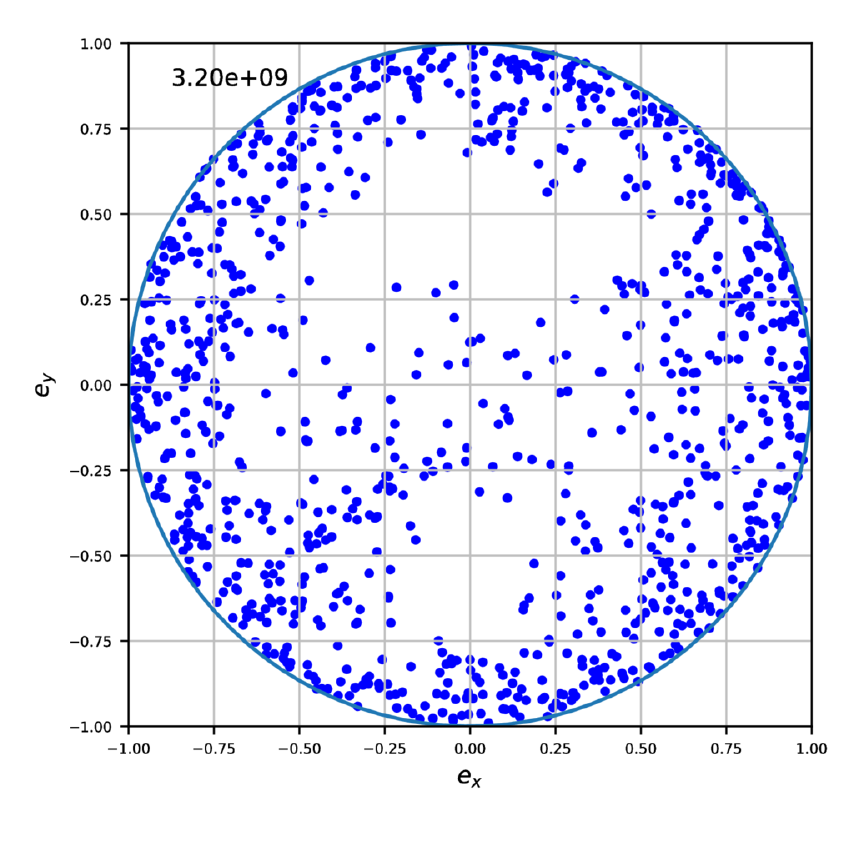}
\end{subfigure}
\hfill
\begin{subfigure}{0.3\linewidth}
\includegraphics[width=1.0 \textwidth ]{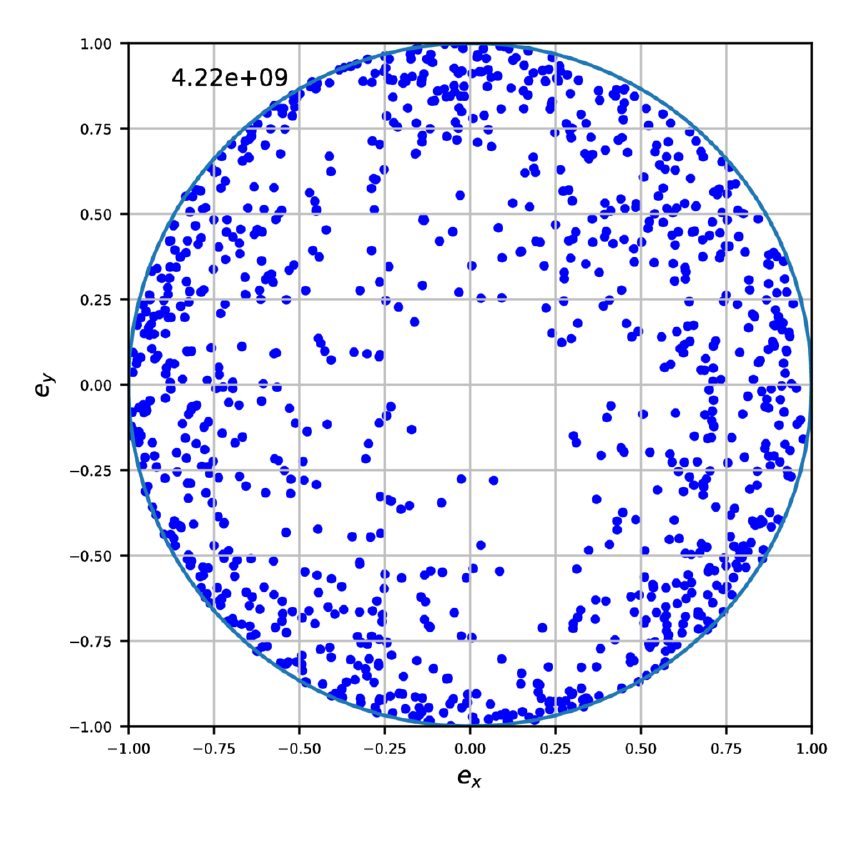}
\end{subfigure}
\hfill
\begin{subfigure}{0.3\linewidth}
\includegraphics[width=1.0 \textwidth ]{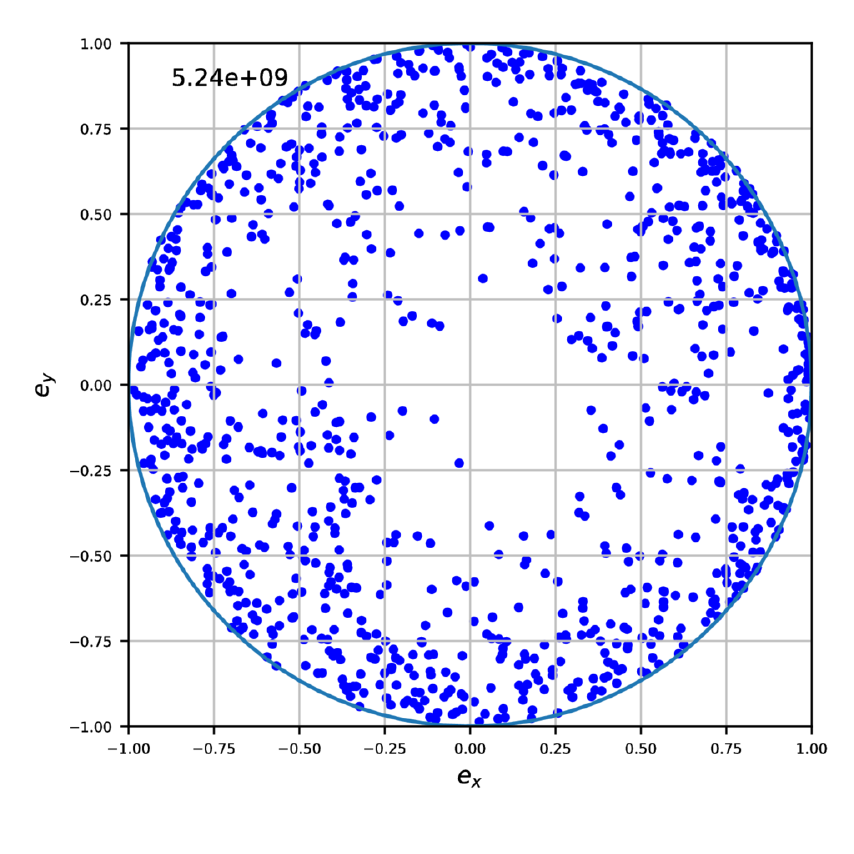}
\end{subfigure}
  \caption{\emph{Evolution of the stable broad band} \texttt{waterbag\_2\_s0}. Upper two rows show the surface density in real space (with distances measured in parsec), and the lower two rows show the distribution in the eccentricity plane at the same respective time. Note that the time (in years) is indicated within the subfigures.
}
\label{fig_wb2_DF_plane}
\end{figure}

We performed $N$-ring numerical simulations of waterbag bands, for a range of system parameters $(\ell_1,\,\ell_2)$. The full list is given in Table~{\ref{tbl:simulations}} of Section~\ref{sec_simu_detailed}. 
The last entry has $\ell_2 = 1$, so is a polarcap and not a band. It is included in the table as a limiting case of a class of broad bands.
Here we discuss the stability of the two bands whose $\Sigma_0(r)$ and 
$\Omega_0(\ell)$ profiles feature in Figure~\ref{fig_sig_om}: one is the band 
\texttt{waterbag\_1\_s0} with $(\ell_1 = 0.7, \,\ell_2 = 0.9)$, and the other 
is the broad band \texttt{waterbag\_2\_s0} with $(\ell_1 = 0.1,\,\ell_2 = 0.9)$.

We simulate a planar system of $N$ rings, each of which has the same semi-major axis $a_0$ and mass $m_\star\,$, orbiting a MBH of mass $M_\bullet\,$.
The total disc mass $M = Nm_\star$ is chosen to be much smaller than $M_\bullet\,$, so $\varepsilon = M/M_\bullet \ll 1$ and the secular time scale, $T_{\rm sec} = \varepsilon^{-1}T_{\rm kep}\,$, is much longer than the Kepler orbital period. Each ring can be thought of as a point on the unit sphere 
phase space of Figure~\ref{fig_phase_space}, with coordinates $(\ell^i, \,g^i)$ for $i = 1,2,\ldots,N$. The projection of the points onto the equatorial plane gives $N$ eccentricity vectors, $\bfe^i = e^i(\cos{g^i}\,\hat{x} + \sin{g^i}\,\hat{y})$,  where $e^i = \sqrt{1-(\ell^i)^2}$
is the eccentricity. Then the normalised secular energy of the whole system is:
\beq
{\cal{H}} \;=\; \frac{1}{N}\sum_{\substack{i,j \\ j > i}} \log{\left| \bfe^i - \bfe^j \right|^2}\,, 
\eeq
which serves as the $N$-ring Hamiltonian for secular dynamics on the sphere:
\beq
\frac{\rmd g^i}{\rmd t} \;=\; \frac{\partial \cal{H}}{\partial \ell^i}\,, 
\qquad\frac{\rmd \ell^i}{\rmd t} \;=\; - \frac{\partial \cal{H}}{\partial g^i}\qquad\qquad\mbox{(for $i = 1,2,\ldots,N$)}\,,  
\eeq
where $t = {\rm time}/T_{\rm sec}\,$ is, as earlier, the dimensionless time 
variable. The Hamiltonian equations can be rewritten compactly as: 
\beq
\frac{\rmd \bfe^i}{\rmd t} \;=\; \frac{2}{N} \sum_{\substack{j=1 \\ j \neq i}}^{N} \frac{ (\bfe^i -\bfe^j) \times {\boldsymbol{\ell}^i}}  { \left| \bfe^i -\bfe^j \right|^2}
\label{eqn_ecc_evo_n-ring}
\eeq
where $\boldsymbol{\ell}^i = \ell^i\,\hat{z}$. These vectorial equations
are similar to those presented in \citet{ttk09}, with the difference 
that our interaction Hamiltonian is unsoftened and logarithmic. The
equations have been solved using a Bulirsch-Stoer integrator, with relative and absolute tolerances equal to $10^{-8}$. Our fiducial system has the following parameters:
\begin{itemize}
\item The disc is composed of $N = 1000$ rings.
\item Semi-major axis of each ring is $a_0 = 1~{\rm pc}$. 
\item Black hole mass $M_{\bullet} = 10^7\,\Msun$, giving a Kepler
orbital period $\tk = 0.03~{\rm Myr}$.
\item Disc mass $M = 10^3~\Msun$, so $\varepsilon = 10^{-4}$ and the
secular time scale $\ts =  0.3~{\rm Gyr}$.
\end{itemize}
The typical relative energy and angular momentum errors for the simulations listed in Table~{\ref{tbl:simulations}} of Section~\ref{sec_simu_detailed}
are $\sim 10^{-6}$.

The evolution of the two bands, \texttt{waterbag\_1\_s0} and 
\texttt{waterbag\_2\_s0}, is shown in Figure~\ref{fig_wb1_DF_plane} and Figure~\ref{fig_wb2_DF_plane}, respectively. The upper two panels are for the surface mass density in the the $x$-$y$ plane, and the lower two panels show the rings represented as $1000$ points on the $(e_x, e_y)$ plane.\footnote{Since we are dealing with prograde discs, all the points have positive $\ell^i\,$.} We begin with initial conditions corresponding to the two bands of Figure~\ref{fig_sig_om}. The following overall features
can be noticed:
\begin{itemize}
\item For \texttt{waterbag\_1\_s0} a non-axisymmetric $m=3$ instability grows; it is seen very clearly around $0.3$~Gyr and, by $\sim 0.6$~Gyr, there are distinct signs of nonlinear evolution.

\item  In contrast the broad band \texttt{waterbag\_2\_s0} is seen to be stable
over a time scale of $5$~Gyr. 
\end{itemize}

\begin{figure}
 \begin{subfigure}{0.47 \linewidth}
\centering
\includegraphics[width=1.1\textwidth,trim={1cm 0cm 0cm 0cm}]{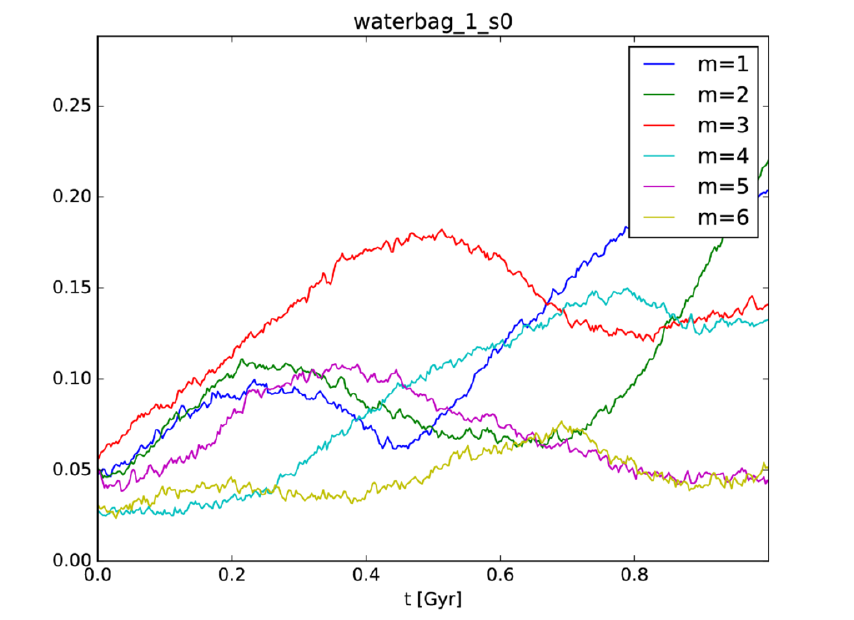}
  \caption{ \texttt{waterbag\_1\_s0}}
\label{fig_wb1_mode_amp_grates}
\end{subfigure}
\hfill
\begin{subfigure}{0.47 \linewidth}%
\centering
\includegraphics[width=0.95 \textwidth,trim={0cm 0cm 1cm 0cm}]{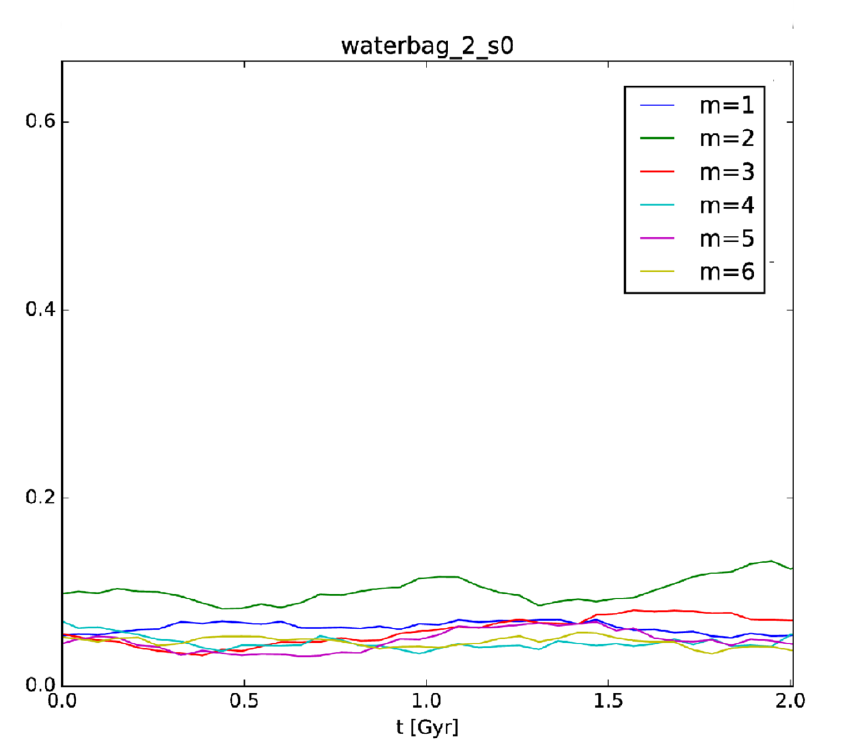}
  \caption{ \texttt{waterbag\_2\_s0}}
\label{fig_wb2_mode_amp_grates}
\end{subfigure}
  \caption{\emph{Evolution of mode amplitudes $a_m(t)$}.}
\end{figure}

Dynamical behaviour can be characterized in more detail by looking at mode 
amplitudes, $a_m(t)\,$, which were evaluated by computing Fast Fourier Transforms over annuli of the projected mass density. These are plotted in  
Figure~\ref{fig_wb1_mode_amp_grates} for \texttt{waterbag\_1\_s0} and Figure~\ref{fig_wb2_mode_amp_grates} for \texttt{waterbag\_2\_s0}. The main features are:
\begin{itemize}
\item 
For \texttt{waterbag\_1\_s0} the initially unstable mode has $m=3$, and 
this remains dominant until about $0.6$~Gyr. Later there is growth of other modes, especially, $m=1$ and $m=2$. 

\item 
Modes of all $m$ maintain a low amplitude for \texttt{waterbag\_2\_s0}.
We note that sampling noise, which is unavoidable in the 
initial conditions, was such that a $m=2$ mode had a greater initial 
amplitude than the other modes (see Figures~\ref{fig_wb2_mode_amp_grates}).
The $m=2$ mode is seen to be stable and precessing in 
Figure~\ref{fig_wb2_DF_plane}. Interactions of some stars with the
$m=2$ mode has, presumably, scattered them in phase space. Whereas 
a study of this mode-particle scattering is beyond the scope of this paper, 
simulations with a larger number of particles will help clarify the
nature of this process.

\end{itemize}
In the next section we present a detailed account of the linear stability of bands. We will also discuss how linear theory accounts for the behaviour of \texttt{waterbag\_1\_s0} and \texttt{waterbag\_2\_s0}.

\section{Linear stability of bands}
\label{sec: instability-analytics}

A normal mode of a waterbag band 
has the form $f_1(\ell, g, t; m) = {\rm Re}\left\{f_{1m}(\ell)\exp{[{\rm i}(m g - \omega_m t)]}\right\}$, where $\omega_m$ is a complex eigenfrequency. Since a normal mode is composed of sinusoidal disturbances of the two edges of the phase space DF, the corresponding eigenfunction is of the form, $f_{1m}(\ell) = A_{m1}\,\delta(\ell - \ell_1) + A_{m2}\,\delta(\ell - \ell_2)$, where $A_{m1}$ and $A_{m2}$ are complex amplitudes --- see equation~(\ref{eqn_edgemode-form}). When this is substituted in the integral equation~(\ref{eqn_integral-eigen-band}), it reduces to the following 
$2 \times 2$ matrix eigenvalue problem: 
\beq
\left(\begin{array}{cc} \displaystyle{\frac{1}{\Delta \ell}  }+m\, \Omega_0(\ell_1) & \displaystyle{\frac{1}{\Delta \ell}\left( \frac{e_2}{e_1} \right)^m }\\[4 ex] - \displaystyle{\frac{1}{\Delta \ell}\left( \frac{e_2}{e_1} \right)^m} & - \displaystyle{\frac{1}{\Delta \ell}}+m\, \Omega_0(\ell_2) \end{array}\right)  
\left(\begin{array}{c} A_{m1} \\[6ex] A_{m2} \end{array}\right) \;=\; \omega_m \, \left( \begin{array}{c} A_{m1} \\[6ex] A_{m2}   
\end{array} \right)\,.
\label{eqn_matrix_eqns}
\eeq
Here $\Delta\ell = (\ell_2 - \ell_1)\,$,  and equation~(\ref{eqn_omega0-waterbag-l2-general}) gives $\Omega_0(\ell_1) \equiv \Omega_1 = -2\ell_1/(1 - \ell_1^2)$ and $\Omega_0 (\ell_2) = 0$. The solutions for the eigenfrequency and the ratio of edge disturbance amplitudes are, 
\begin{subequations}
\begin{align}
\omega_m^\pm &\;=\; \frac{m\Omega_1}{2}  \;\pm\; \frac{1}{\Delta\ell}\,
\sqrt{\;\left[ 1 \,+\, \frac{m \, \Delta \ell \, \Omega_1}{2} \right]^2 
\,-\, \left( \frac{e_2}{e_1} \right)^{2m}\;}\quad,  
\label{eqn_eigenvalue} \\[3ex]
\left(\frac{A_{m2}}{A_{m1}}\right)^{\pm} &\;=\; - \left[1 \,+\, \frac{m \, \Delta \ell \,\Omega_1}{2}   \right]  \left( \frac{e_1}{e_2} \right)^m \;\pm\; \sqrt{\;\left[ 1 \,+\, \frac{m \, \Delta \ell \, \Omega_1}{2} \right]^2  \left( \frac{e_1}{e_2} \right)^{2m} \,-\,1\;}\quad. 
\label{eqn_eigenfunc}
\end{align}
\end{subequations}
A number of properties of linear modes follow:
\begin{itemize}
\item For each $m=1,2,\ldots$ there are two normal modes denoted by `$\pm$'. Each normal mode is made up of two edge disturbances corresponding to the DF boundaries $\ell = \ell_1$ and $\ell=\ell_2$.  

\item The eigenfrequencies, $\omega_m^\pm\,$, are either real or complex 
conjugates of each other. If they are both real then both the normal modes
are stable with pattern speed $\lambda_{\rm P}^\pm = \omega_m^\pm/m$. When the eigenfrequencies are complex conjugates, then one normal mode grows exponentially (an instability) and the other decays exponentially, with both modes having the same pattern precession frequency.

\item
From equation~(\ref{eqn_eigenvalue}) we see that the condition for 
instability is: 
\beq
\left( \frac{1 \,-\, \ell_2^2}{1 \,-\,\ell_1^2} \right)^{m/2} \;>\;
\left|\;1- \frac{m \, (\ell_2 - \ell_1) \, \ell_1}{1-\ell_1^2}\;\right|  \,.
\label{eqn_instability_rel}
\eeq

\item 
It can be verified that the above inequality cannot be satisfied 
for any $0\leq\ell_1 < \ell_2 < 1\,$, when $m=1,2\,$. So all bands
have stable $m=1$ and $m=2$ modes, and only modes with $m = 3,4,\ldots$
can be unstable.

\item
 The unstable band \texttt{waterbag\_1\_s0} has $\ell_1 = 0.7$ and $\ell_2 = 0.9$. The stable broad band \texttt{waterbag\_2\_s0} has $\ell_1 = 0.1$ and 
 $\ell_2 = 0.9$. Using these values of $(\ell_1,\,\ell_2)$ in equation~(\ref{eqn_instability_rel}) it can be verified that 
(i) \texttt{waterbag\_1\_s0} has precisely two unstable modes, for $m=3$ and $m=4\,$; (ii) For \texttt{waterbag\_2\_s0} modes of all $m$ are stable.
This is in agreement with the numerical simulations discussed in 
Section~\ref{sec_simu_intro}.

\item
The inequality condition~(\ref{eqn_instability_rel}) defines a region of instability in the $(\ell_1,\,\ell_2)$ parameter plane, for each value of $m$. These are displayed in Figure~\ref{fig_inst_reg} for $m=3,4,5,6\,$. 
As $m$ increases the cresecent-like region of instability expands. 
\end{itemize}
  
\begin{figure}
\centering
\begin{subfigure}{ 0.48 \linewidth}
\centering
\includegraphics[width=0.55 \textwidth ,trim={3cm 0cm 2cm 0cm}]{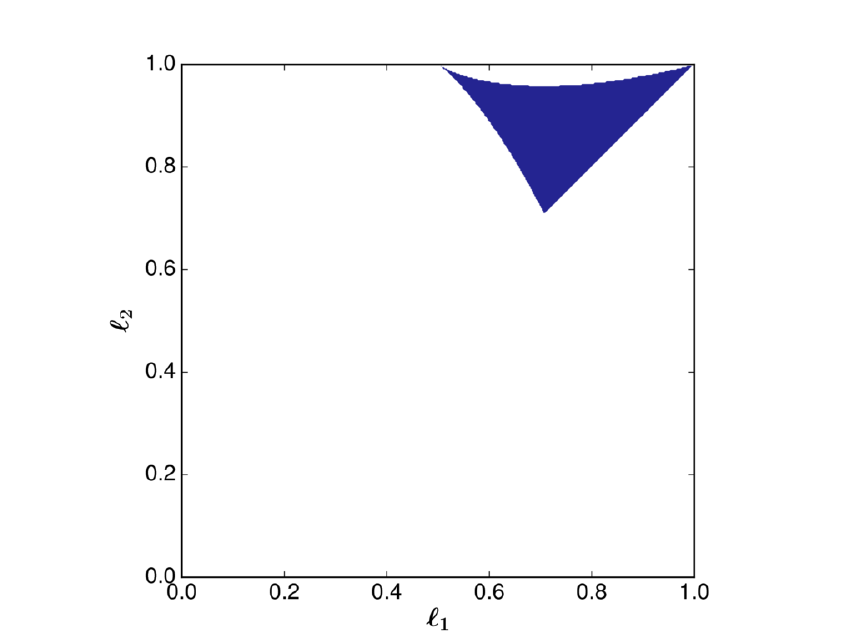}
\caption{m=3}
\end{subfigure}
\hfill
\begin{subfigure}{ 0.48 \linewidth}
\centering
\includegraphics[width=0.55 \textwidth ,trim={3cm 0cm 2cm 0cm}]{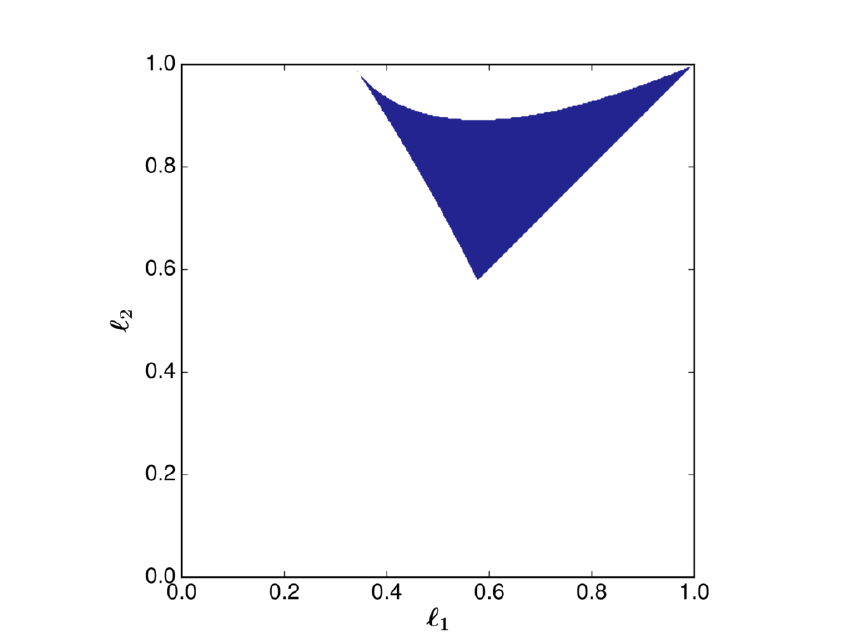}
\caption{m=4}
\end{subfigure}

\begin{subfigure}{ 0.48 \linewidth}
\centering
\includegraphics[width=0.55 \textwidth ,trim={3cm 0cm 2cm 0cm}]{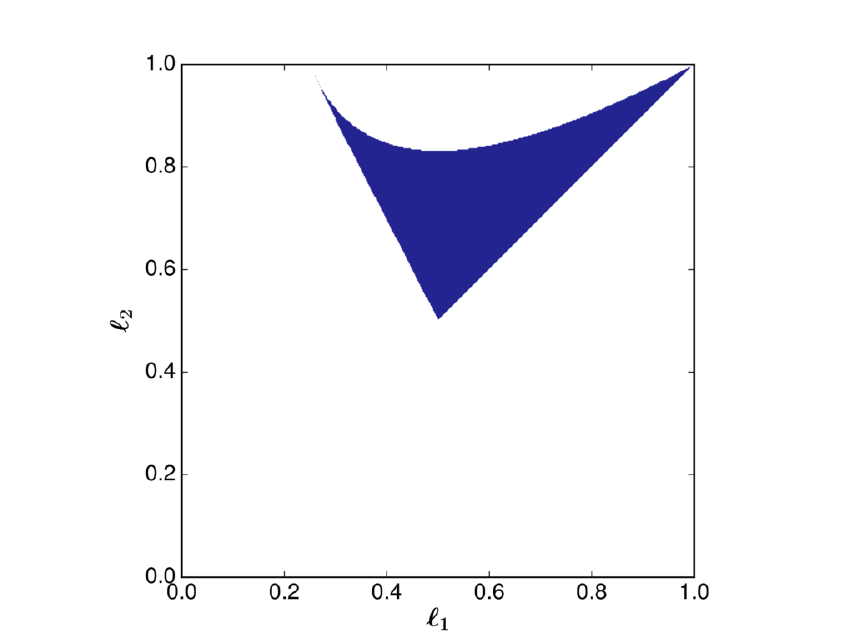}
\caption{m=5}
\end{subfigure}
\hfill
\begin{subfigure}{ 0.48 \linewidth}
\centering
\includegraphics[width=0.55 \textwidth ,trim={3cm 0cm 2cm 0cm}]{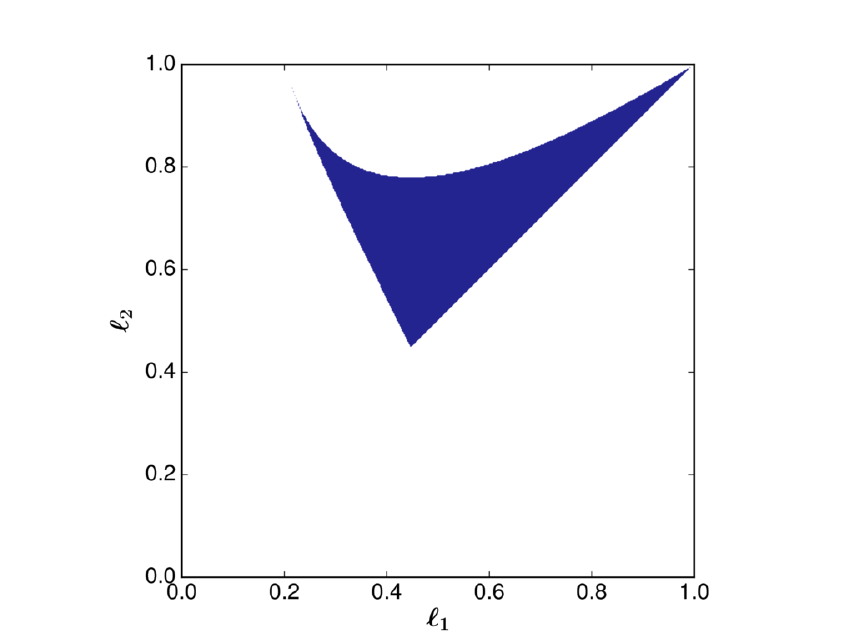}
\caption{m=6}
\end{subfigure}
\caption{\emph{Instability region in $(\ell_1,\,\ell_2)$ plane for $m=3,4,5,6\,$}.}
\label{fig_inst_reg}
\end{figure}

\subsection{Structure of normal modes}

\noindent
{\bf Stable modes:} When inequality~(\ref{eqn_instability_rel}) is not satisfied the two normal mode eigenfrequencies $\omega_m^\pm\,$, given by equation~(\ref{eqn_eigenvalue}), are both real with corresponding pattern speeds $\lambda_{\rm P}^\pm = \omega_m^\pm/m\,$. The DF of the normal modes is:
\beq
f_1^\pm(\ell, g, t; m) \;=\; 
{\rm Re}\Big{\{} A_{m1}^\pm\exp{[{\rm i}m(g - \lambda_{\rm P}^\pm t)]}\,\delta(\ell - \ell_1) 
+\; A_{m2}^\pm\exp{[{\rm i}m(g  - \lambda_{\rm P}^\pm t)]}\,\delta(\ell - \ell_2)\Big{\}}\,.
\label{eqn_stable_mode_f}
\eeq
The four complex amplitudes, $A_{m1}^\pm$ and $A_{m2}^\pm\,$, are related by equation~(\ref{eqn_eigenfunc}), which implies that $(A_{m2}/A_{m1})^\pm$ are real whenever $\omega_m^\pm\,$ are real. When the ratio is positive/negative, the normal mode is an in-phase/out-of-phase combination of the two sinusoidal edge disturbances. Moreover the product $(A_{m2}/A_{m1})^+\,(A_{m2}/A_{m1})^- \,=\, 1\,$, which implies (i) If the $+$ mode is an in-phase (or out-of-phase) combination of the two edge disturbances so is the $-$ mode, and vice versa; (ii) If disturbance at one of the edges makes a dominant contribution to the $+$ mode, then the other edge disturbance makes a dominant contribution to the $-$ mode. 
To summarize, a stable $\pm$ mode is either an in-phase or out-of-phase superposition of the edge disturbances, with generally unequal amplitudes. The 
pattern speeds, $\lambda_{\rm P}^\pm$, of the $\pm$ modes are generally
unequal.   

\bigskip
\noindent
{\bf Unstable modes:} When inequality~(\ref{eqn_instability_rel}) is satisfied the two normal mode eigenfrequencies $\omega_m^\pm\,$ given by equation~(\ref{eqn_eigenvalue}), are complex conjugates of each other.
We write $\omega_m^\pm = m\lambda_{\rm P} \pm {\rm i}\, \omega_{\rm I}$, where 
$\lambda_{\rm P}$ is the pattern speed and $\omega_{\rm I} > 0$ can be thought as the growth rate of the `$+$' mode, or as the damping rate 
of the `$-$' mode; we will refer to $\omega_{\rm I}$ as the growth rate. 
Equation~(\ref{eqn_eigenvalue}) gives:
\begin{subequations}
\begin{align}
& \lambda_{\rm P} \;=\; \frac{\Omega_1}{2} = - \frac{\ell_1}{1-\ell_1^2} \label{eqn_pattern_precession} \\[1em] 
& \omega_{\rm I} \;=\; \sqrt{ \frac{1}{\Delta \ell^2} \left( \frac{1-\ell_2^2}{1-\ell_1^2} \right)^m - \left( \frac{1}{\Delta \ell} - \frac{m \, \ell_1}{1-\ell_1^2} \right)^2 }.
\label{eqn_growth rate} 
 \end{align}
\end{subequations}
The pattern speed is negative and depends only on $\ell_1\,$. On the other
hand the growth rate depends on all of $(\ell_1, \ell_2, m)$.

Equations~(\ref{eqn_eigenvalue}) and (\ref{eqn_eigenfunc}) imply that 
whenever $\omega_m^\pm$ are complex conjugates, $(A_{m2}/A_{m1})^\pm$ are also complex conjugates. Moreover magnitude of the amplitude ratio, 
$\vert (A_{m2}/A_{m1})^\pm\vert = 1\,$, so we can write 
$(A_{m2}/A_{m1})^\pm = \exp{[\pm {\rm i}\, m \,\theta_m]}$, where 
\beq
\theta_m = \frac{1}{m} \, \cos^{-1}{ \left[ \left(\frac{1-\ell_1^2}{1-\ell_2^2} \right)^{m/2} \left(\frac{m \, \ell_1 \, \Delta \ell}{1-\ell_1^2} -1 \right)  \right]}\,,
\label{eqn_phase_shift_unstable_mode}
\eeq
where $\theta_m$ is the relative phase shift between the two edge disturbances 
composing a normal mode. Then the DF of the growing and damping normal modes of a given $m$ is given by the following superposition of the two edge disturbances:
\begin{align}
f_1^\pm(\ell, g, t; m) \;=\; \exp{[\pm\omega_{\rm I} \, t]}\,
{\rm Re}\Big{\{} &A_{m}^\pm\exp{[{\rm i}m(g - \lambda_{\rm P} t)]}\,\delta(\ell - \ell_1)\nonumber\\[1em] 
+\; &A_{m}^\pm\exp{[{\rm i}m(g \pm \theta_m - \lambda_{\rm P} t)]}\,\delta(\ell - \ell_2)\Big{\}}\,,
\label{eqn_unstable_mode_f}
\end{align}
where $A_m^\pm$ is a complex amplitude that is common to both edge disturbances.
In contrast to a stable mode, an unstable $\pm$ mode is a superposition of 
the edge disturbances with a relative phase shift but equal amplitudes, and a 
pattern speed $\lambda_{\rm P} = \Omega_1/2$ which is the same for both 
$\pm$ modes. 

In order to get an idea of the dependence of the growth rate as a function of the parameters, $(\ell_1, \ell_2, m)$ we plot in Figure~\ref{fig_grate_m} the growth rate as a function of $m$ for different values of $\Delta\ell$ and $\ell_2\,$. For fixed $\ell_2= 0.9$ and three different values of $\Delta\ell$, we see that bands with smaller $\Delta\ell$ are unstable over a larger range of $m$, with higher maximum growth rates occurring at larger $m$. For fixed $\Delta\ell = 0.1$ and three different values of $\ell_2$, the maximum growth rates are similar but occur at
smaller $m$ for larger $\ell_2\,$. 

We note that \texttt{waterbag\_1\_s0} has unstable modes for $m=3,4\,$ with the $m=3$ mode having the higher growth rate, $\omega_{\rm I} \sim 0.72 \, \ts^{-1} \simeq 2.4\,{\rm Gyr}^{-1}$; this is consistent with the initial growth of the $m=3$ mode in Figure~\ref{fig_wb1_DF_plane} and \ref{fig_wb1_mode_amp_grates}. In the next section we present a more detailed comparison of numerical experiments with linear theory.

\begin{figure}
\begin{subfigure}{ 0.5 \linewidth}
\centering
\includegraphics[width=0.55 \textwidth ,trim={2cm  0cm  2cm 0cm}]{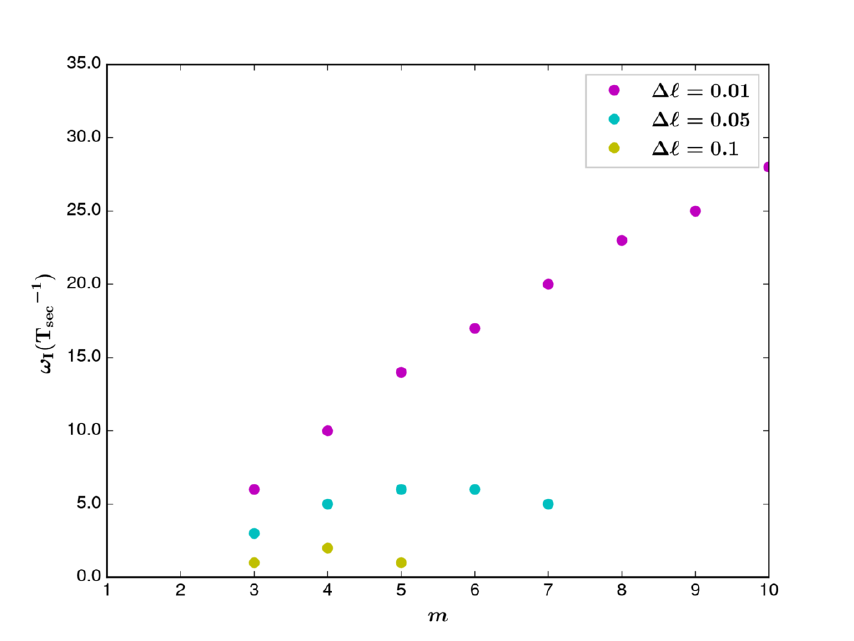}
\caption{$\ell_2 = 0.9$}
\end{subfigure}
\hfill
\begin{subfigure}{ 0.5 \linewidth}
\centering
\includegraphics[width=0.55 \textwidth ,trim={2cm  0cm  2cm 0cm}]{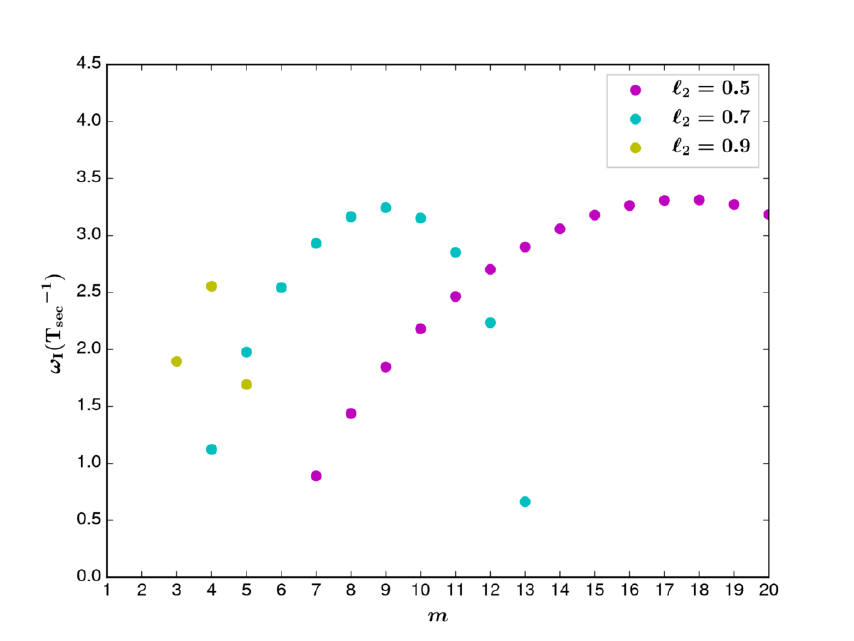}
\caption{$ \Delta \ell = 0.1$}
\end{subfigure}
\caption{Growth rate $\omega_{\rm I}$ variation with $m$: a). Left panel corresponds to waterbags with fixed $\ell_2 = 0.9$ b). Right panel for waterbags of fixed thickness $\Delta \ell = 0.1$ }
\label{fig_grate_m}
\end{figure}

\section{Evolution of instabilities}
\label{sec_simu_detailed}

\begin{table}
\centering
\footnotesize
 \captionsetup{font=footnotesize}
\begin{tabular}{|c| c| c| c| c|}
  \hline
  System Name                    & $\ell_1$ & $\ell_2$ & $T_{\rm end}$ & Stable~?\\[1ex]
  \hline
  \texttt{waterbag\_1\_s0}           & 0.7  & 0.9    & 2.5          & no  \\
  \texttt{waterbag\_2\_s0}           & 0.1  & 0.9    & 9.4          & yes \\
  \texttt{waterbag\_3\_s0}           & 0.8  & 0.9    & 10.0         & no  \\
  \texttt{waterbag\_4\_s0}           & 0.85 & 0.9    & 6.17         & no  \\
  \texttt{waterbag\_5\_s0}           & 0.7  & 0.97   & 8.79         & yes \\
  \hline
  \texttt{waterbag\_$\ell$1\_0.8\_$\ell$2\_0.81  } & 0.8  & 0.81 & 1.8            & no\\
  \texttt{waterbag\_$\ell$1\_0.8\_$\ell$2\_0.82 } & 0.8  & 0.82 & 10.0           & no\\
  \texttt{waterbag\_$\ell$1\_0.8\_$\ell$2\_0.83 } & 0.8  & 0.83 & 12.5           & no\\
  \texttt{waterbag\_$\ell$1\_0.8\_$\ell$2\_0.84 } & 0.8  & 0.84 & 13.3           & no\\
  \texttt{waterbag\_$\ell$1\_0.8\_$\ell$2\_0.85 } & 0.8  & 0.85 & 1.65           & no\\
  \texttt{waterbag\_$\ell$1\_0.8\_$\ell$2\_0.86 } & 0.8  & 0.86 & 34.2           & no\\
  \texttt{waterbag\_$\ell$1\_0.8\_$\ell$2\_0.87 } & 0.8  & 0.87 & 0.28           & no\\
  \texttt{waterbag\_$\ell$1\_0.8\_$\ell$2\_0.88 } & 0.8  & 0.88 & 5.9            & no\\
  \texttt{waterbag\_$\ell$1\_0.8\_$\ell$2\_0.89 } & 0.8  & 0.89 & 5.9            & no\\
  \texttt{waterbag\_$\ell$1\_0.8\_$\ell$2\_0.90 } & 0.8  & 0.90 & 41.2           & no\\
  \texttt{waterbag\_$\ell$1\_0.8\_$\ell$2\_0.91 } & 0.8  & 0.91 & 20.0           & no\\
  \texttt{waterbag\_$\ell$1\_0.8\_$\ell$2\_0.92 } & 0.8  & 0.92 & 10.8           & no\\
  \texttt{waterbag\_$\ell$1\_0.8\_$\ell$2\_0.93 } & 0.8  & 0.93 & 6.4            & no\\
  \texttt{waterbag\_$\ell$1\_0.8\_$\ell$2\_0.94 } & 0.8  & 0.94 & 44.0           & no\\
  \texttt{waterbag\_$\ell$1\_0.8\_$\ell$2\_0.95 } & 0.8  & 0.95 & 38.7           & no\\
  \texttt{waterbag\_$\ell$1\_0.8\_$\ell$2\_0.96 } & 0.8  & 0.96 & 18.4           & no\\
  \texttt{waterbag\_$\ell$1\_0.8\_$\ell$2\_0.97 } & 0.8  & 0.97 & 5.1            & no\\
  \texttt{waterbag\_$\ell$1\_0.8\_$\ell$2\_0.98 } & 0.8  & 0.98 & 211            & yes\\
  \texttt{waterbag\_$\ell$1\_0.8\_$\ell$2\_0.99 } & 0.8  & 0.99 & 16.3           & yes\\
  \texttt{waterbag\_$\ell$1\_0.8\_$\ell$2\_1.00 } & 0.8  & 1.00 & 19.0           & yes\\
  \hline
\end{tabular}
\caption{\emph{List of all numerical simulations}. The upper 
five cases correspond to Set~I and the lower ones to Set~II. The 
total duration of each simulation, $T_{\rm end}$, is given in units of 
Gyr; it is of order a few secular times and differs from case to case.} 
\label{tbl:simulations}
\end{table}

We ran a suite of numerical simulations of waterbag bands, with parameters listed in the Table~\ref{tbl:simulations}. The primary goal is to put the linear theory of the previous section to stringent tests, and is explored through the upper (Set~I) and lower (Set~II) groups shown in Table~\ref{tbl:simulations}:
\begin{itemize}
\item
{\bf Set~I} consists of five cases, of which two --- the unstable band 
\texttt{waterbag\_1\_s0} and the stable band \texttt{waterbag\_2\_s0} --- have already been discussed. 

\item
{\bf Set~II} is a detailed test of the linear theory prediction of the transition from instability to stability of a band with fixed $\ell_1 = 0.8$, as $\ell_2$ is varied
over a range of values. 
\end{itemize}

Then we give a taste of the long-term evolution of an unstable band, 
that goes well beyond the applicability of linear theory. Here the 
point of interest is in the collisionless relaxation to a state with a 
wide spread in eccentricities.

\subsection{Set~I}
\label{sec_set1_bags}

\begin{table}
\centering
\footnotesize
 \captionsetup{font=footnotesize}
  \begin{tabular}{|c|c|c|c|c|}
  \cline{3-5}
   \multicolumn{2}{c|}{  }             &  \multicolumn{3}{c|}{Fastest growing mode} \\
   \hline
System name  & Unstable $m$  &   $m_0$ & $\omega_{\rm I, max}$(Gyr$^{-1}$) &  $\lambda_{\rm P0}$(rad Gyr$^{-1}$)  \\
 \hline
\texttt{waterbag\_1\_s0}  & 3,4\,    &\,3     &2.4     &-4.57\\
\texttt{waterbag\_3\_s0}  & 3,4,5    &\,4     &8.5     &-7.41\\
\texttt{waterbag\_4\_s0}  & 3 - 7    &\,6     &20.6    &-10.21\\  
\hline
\end{tabular}
 \caption{\emph{Theoretical predictions for the unstable bands of Set~I}.}
 \label{tbl:theo-predict}
\end{table}
 
\begin{table}
\centering
\footnotesize
 \captionsetup{font=footnotesize}
\begin{tabular}{|c| c| c| c|}

            \cline{2-4}
   \multicolumn{1}{c|}{}          & \multicolumn{3}{c|}{Fastest growing mode}                        \\
\hline           
System name  &  $m_0$ (Theory) & $m_0$ (Simulations)    & Agreement  \\
\hline
\texttt{waterbag\_1\_s0}    & 3      & 3      & yes      \\
\texttt{waterbag\_3\_s0}    & 4      & 4      & yes$^*$   \\
\texttt{waterbag\_4\_s0}    & 6      & 6      & yes$^*$   \\
\hline
\end{tabular}
 \caption{\emph{Comparison between linear theory and simulations for the unstable bands of Set~I}. $^*$ There is good agreement for \texttt{waterbag\_3\_s0} for $t < 0.2$~Gyr, and for \texttt{waterbag\_4\_s0} for 
$0.05 < t < 0.15$~Gyr.}
\label{tbl:bunch1_mode_comparison}
\end{table}

\begin{figure}

\begin{subfigure}{0.3\linewidth}
\includegraphics[width=1.25 \textwidth ]{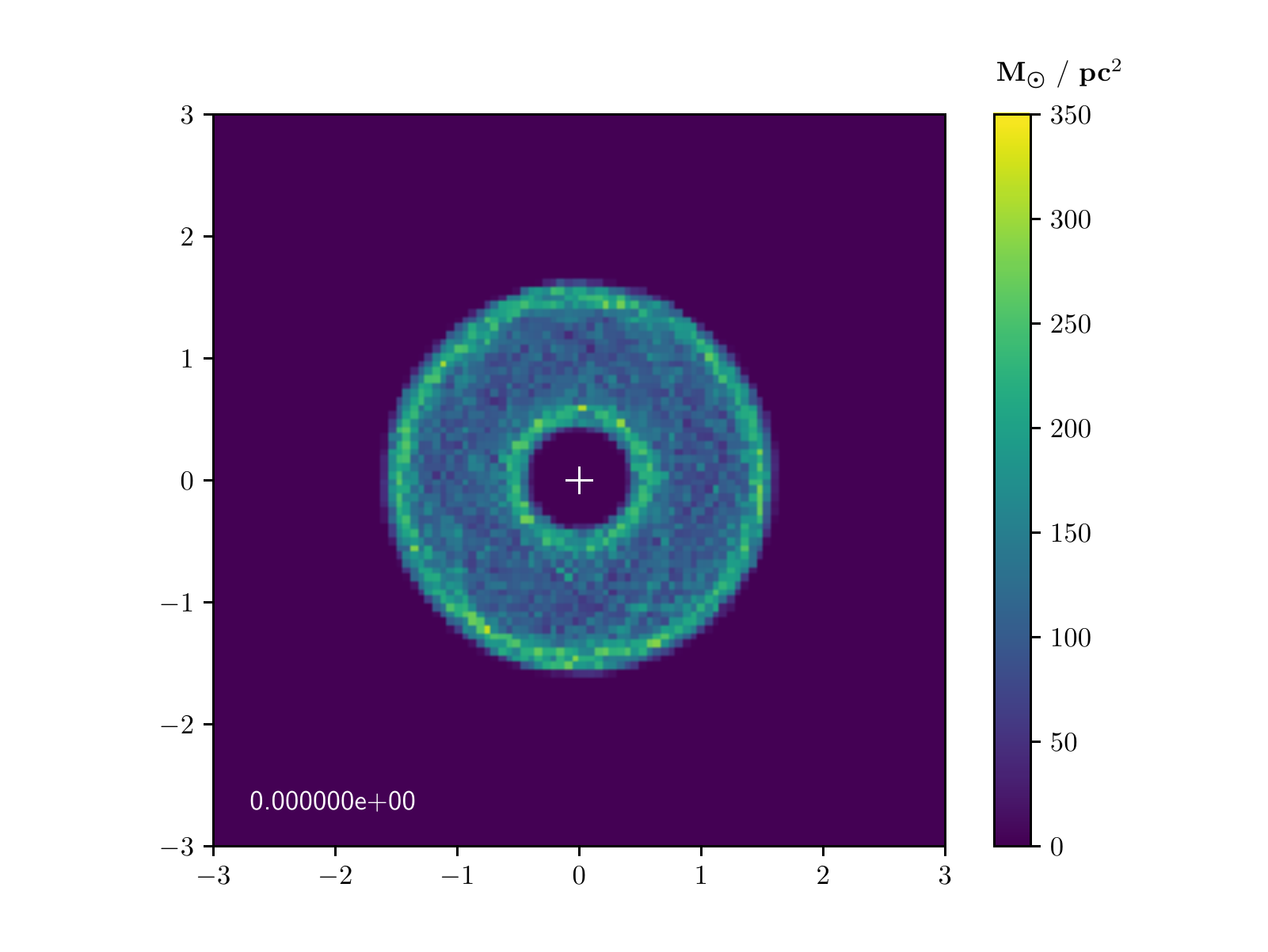}
\end{subfigure}
\hfill
\begin{subfigure}{0.3\linewidth}
\includegraphics[width=1.25 \textwidth ]{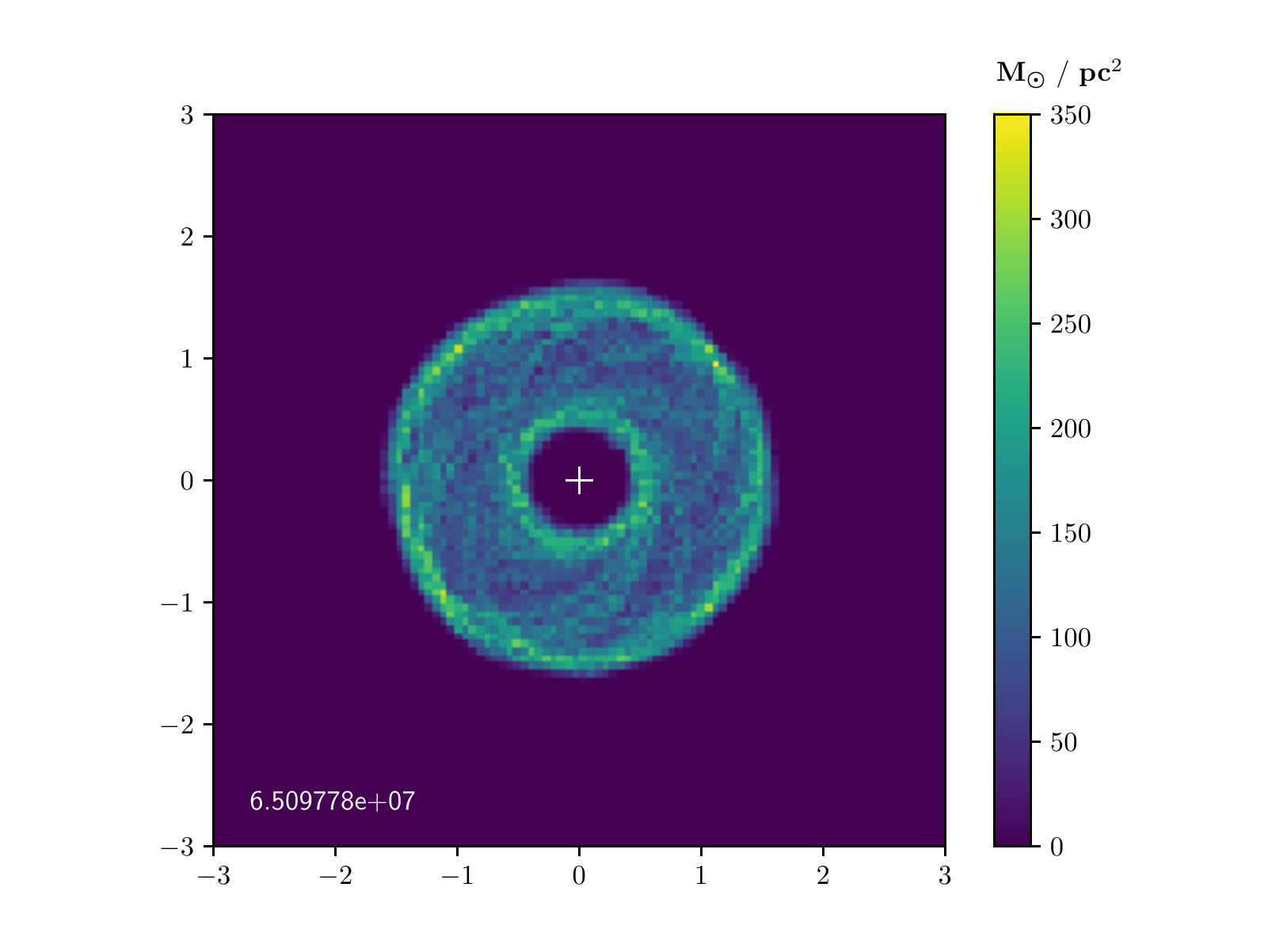}
\end{subfigure}
\hfill
\begin{subfigure}{0.3\linewidth}
\includegraphics[width=1.25 \textwidth ]{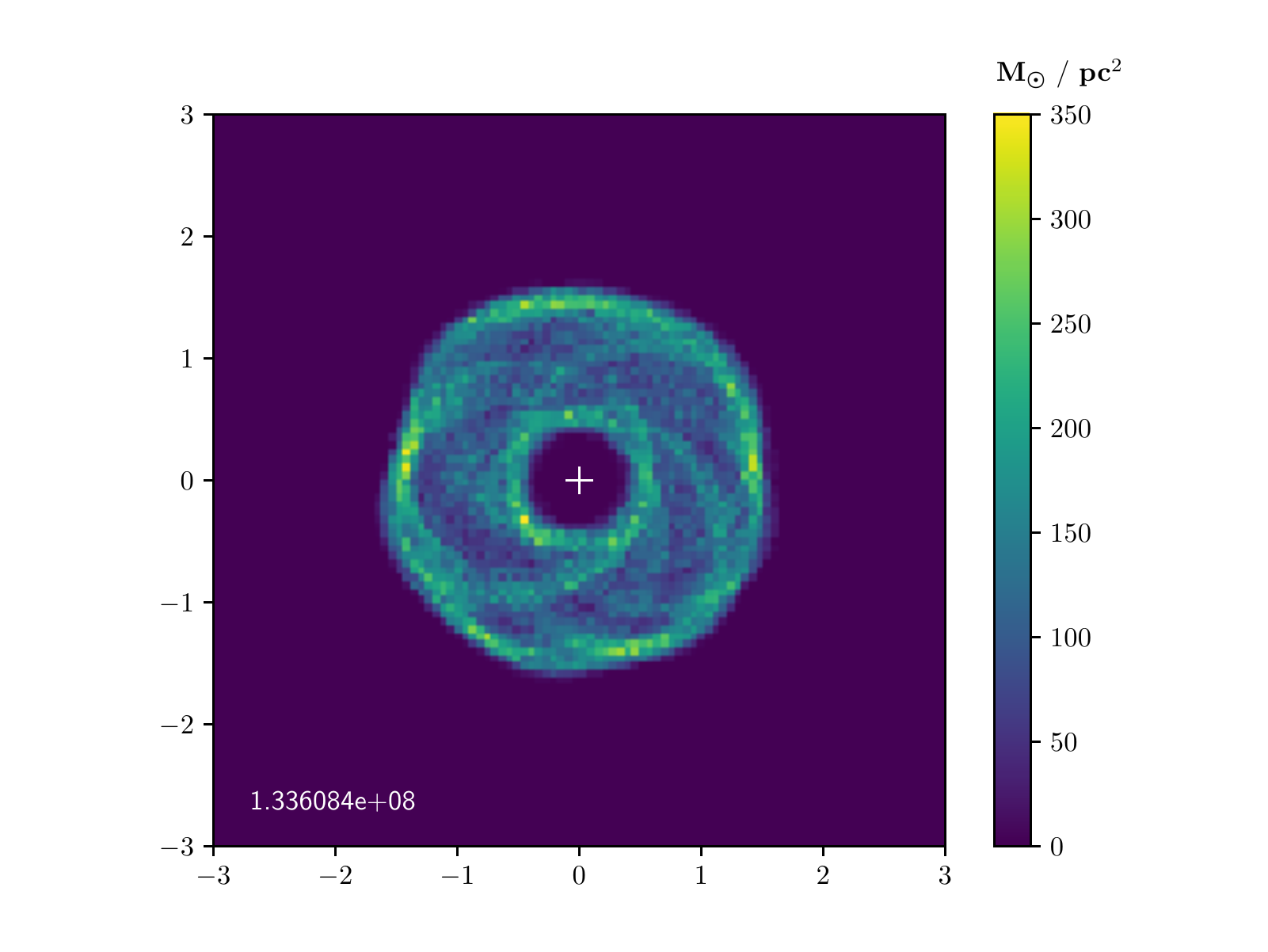}
\end{subfigure}
\hfill

\begin{subfigure}{0.3\linewidth}
\includegraphics[width=1.25 \textwidth ]{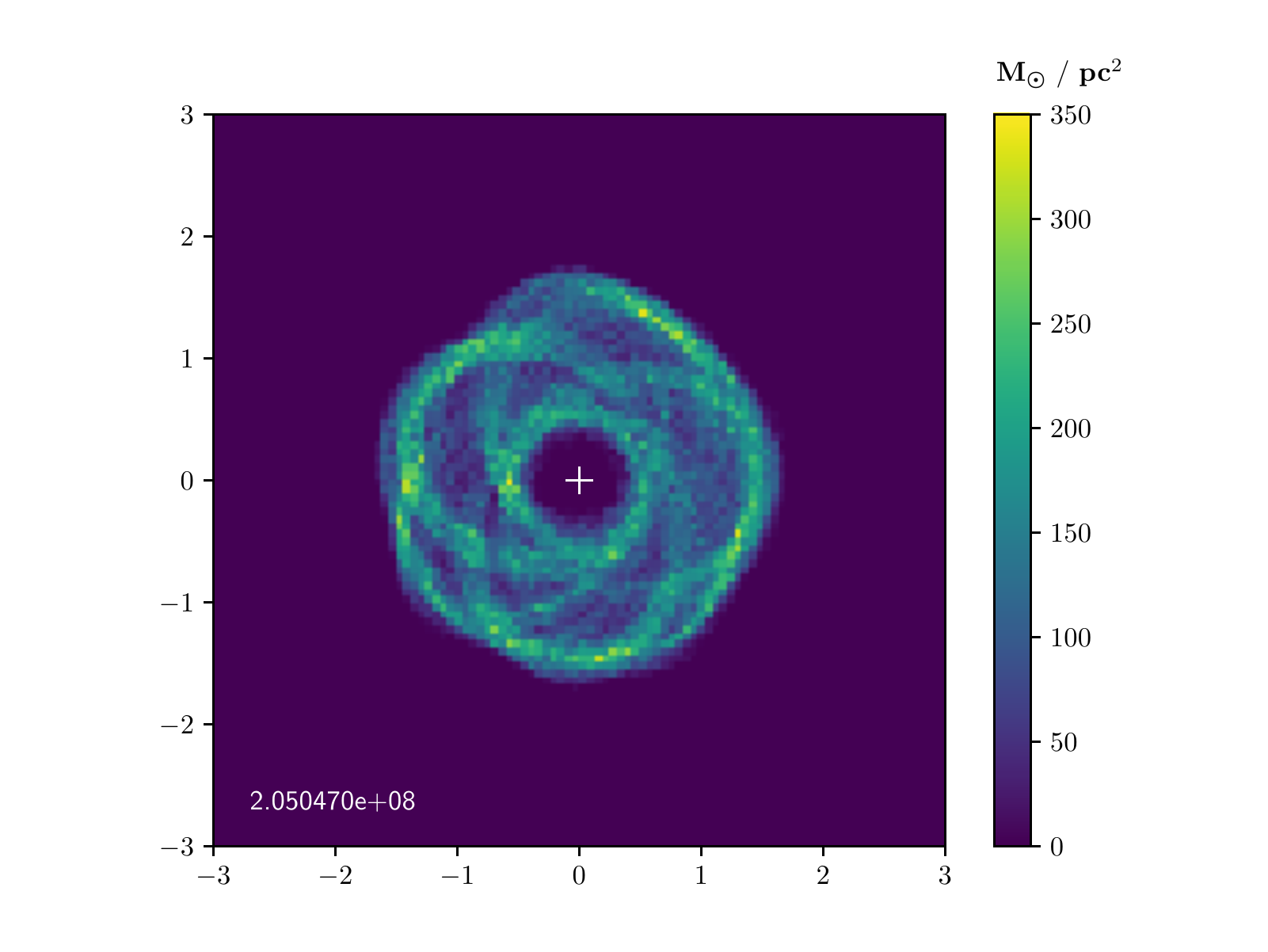}
\end{subfigure}
\hfill
\begin{subfigure}{0.3\linewidth}
\includegraphics[width=1.25 \textwidth ]{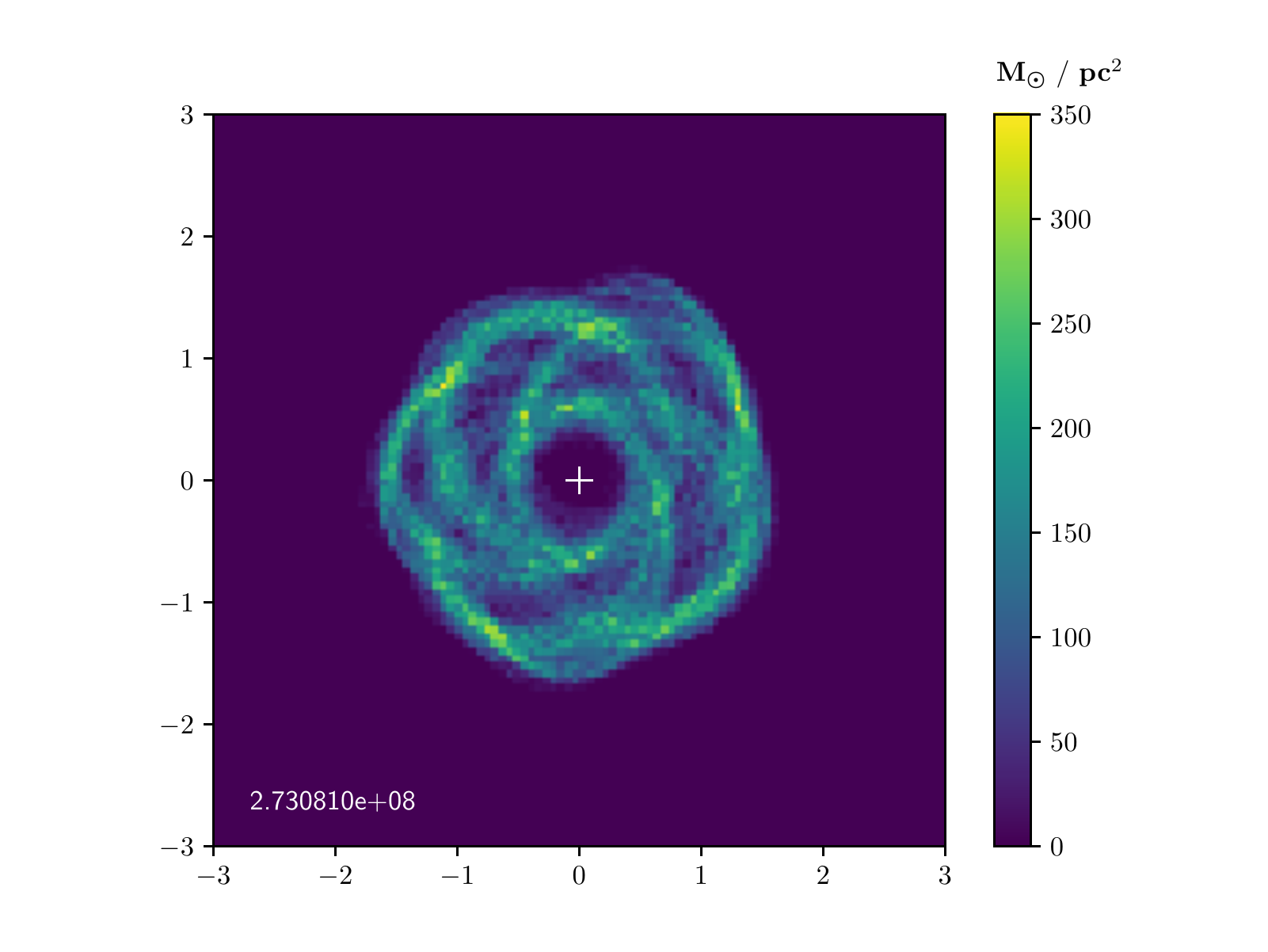}
\end{subfigure}
\hfill
\begin{subfigure}{0.3\linewidth}
\includegraphics[width=1.25 \textwidth ]{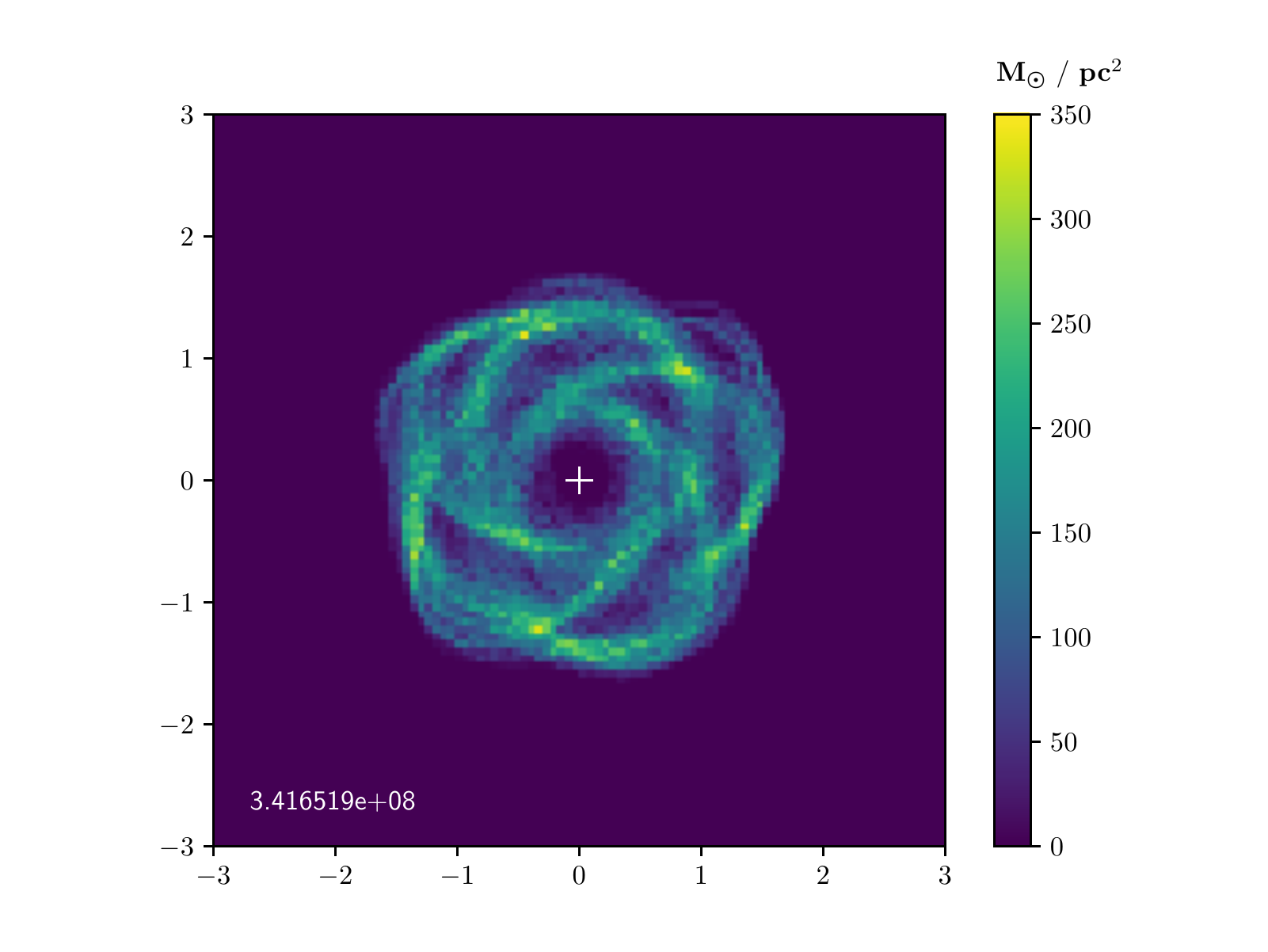}
\end{subfigure}

\begin{subfigure}{0.3\linewidth}
\includegraphics[width=1.0 \textwidth ]{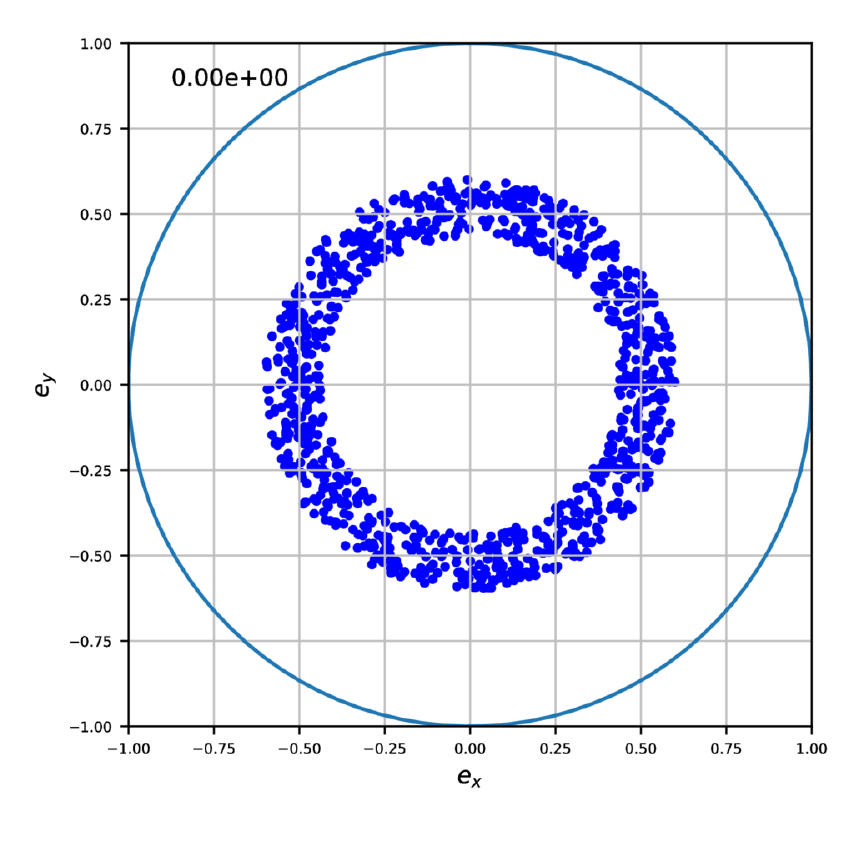}
\end{subfigure}
\hfill
\begin{subfigure}{0.3\linewidth}
\includegraphics[width=1.0 \textwidth ]{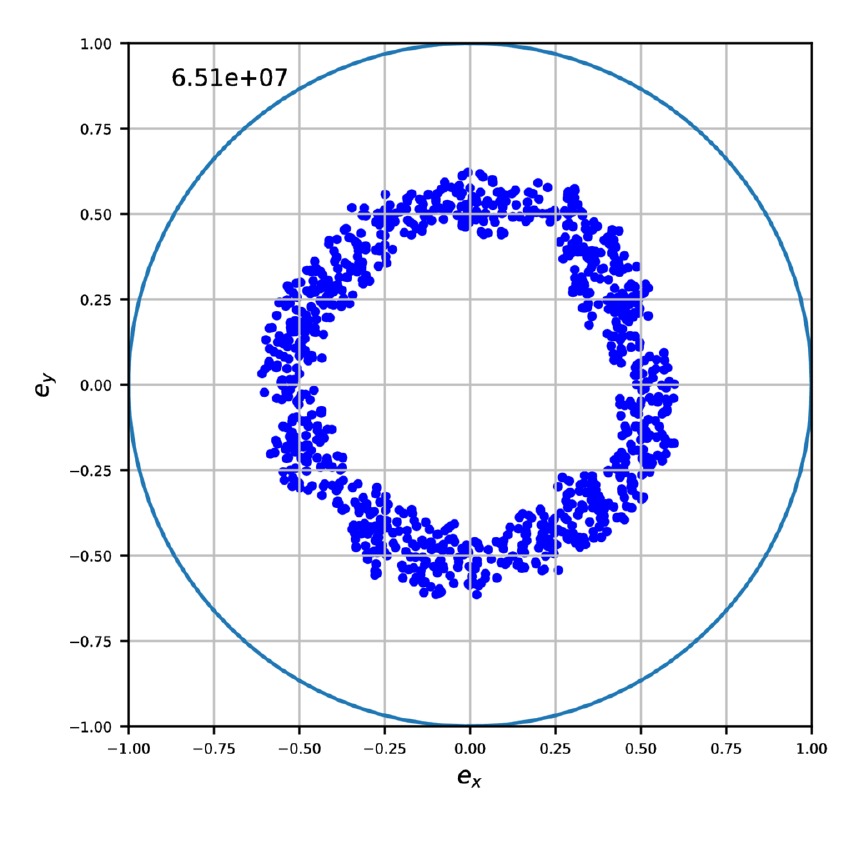}
\end{subfigure}
\hfill
\begin{subfigure}{0.3\linewidth}
\includegraphics[width=1.0 \textwidth ]{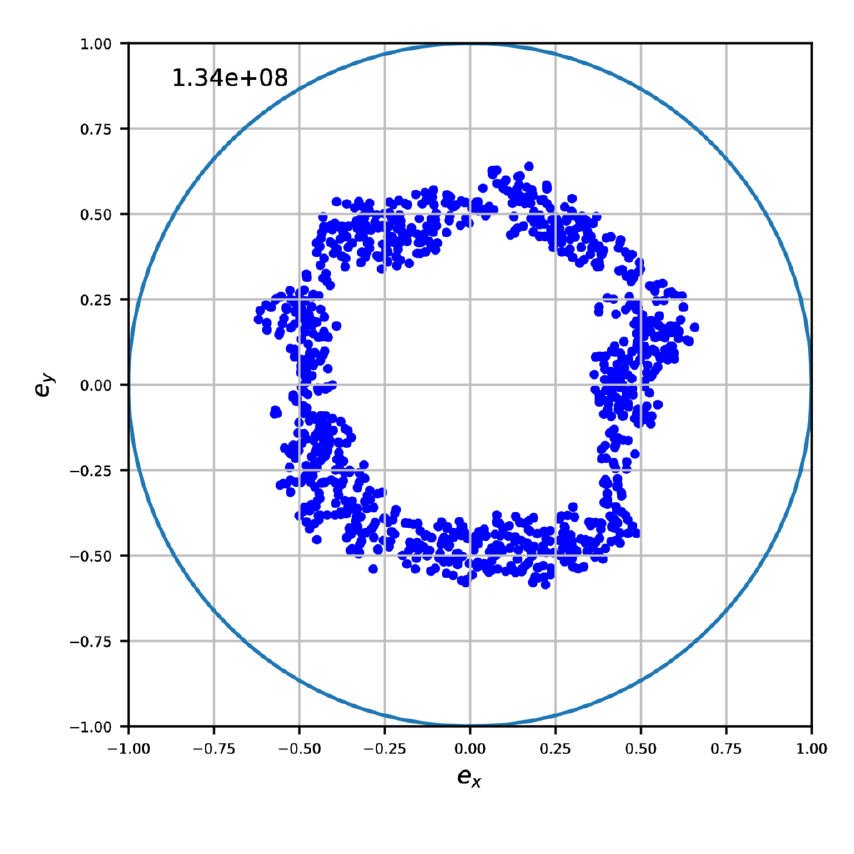}
\end{subfigure}
\hfill

\begin{subfigure}{0.3\linewidth}
\includegraphics[width=1.0 \textwidth ]{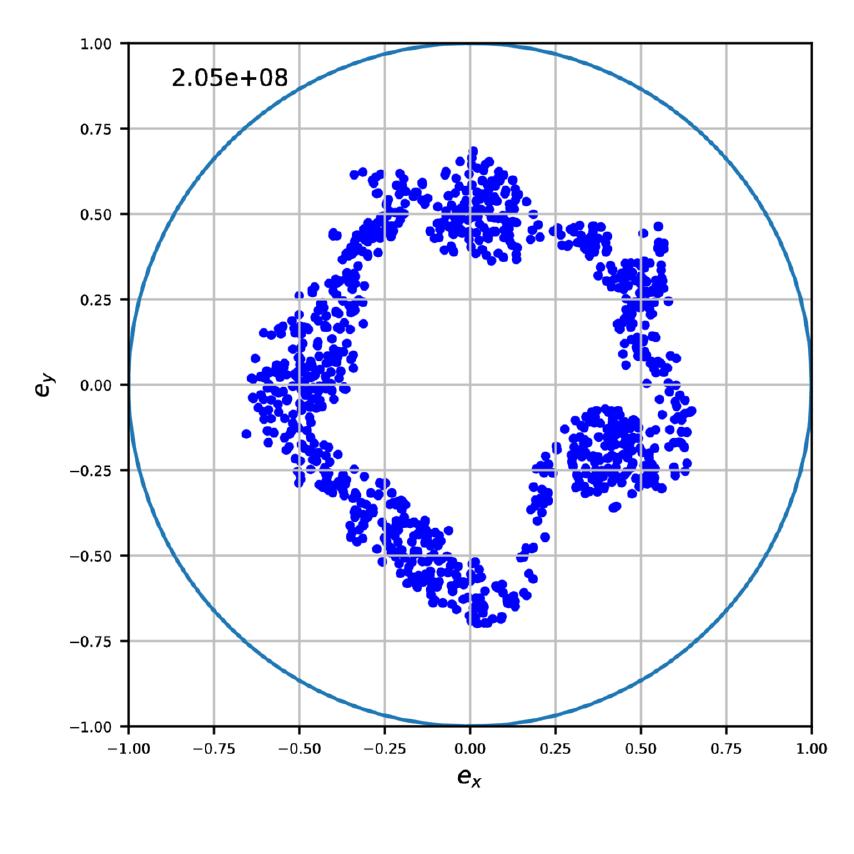}
\end{subfigure}
\hfill
\begin{subfigure}{0.3\linewidth}
\includegraphics[width=1.0 \textwidth ]{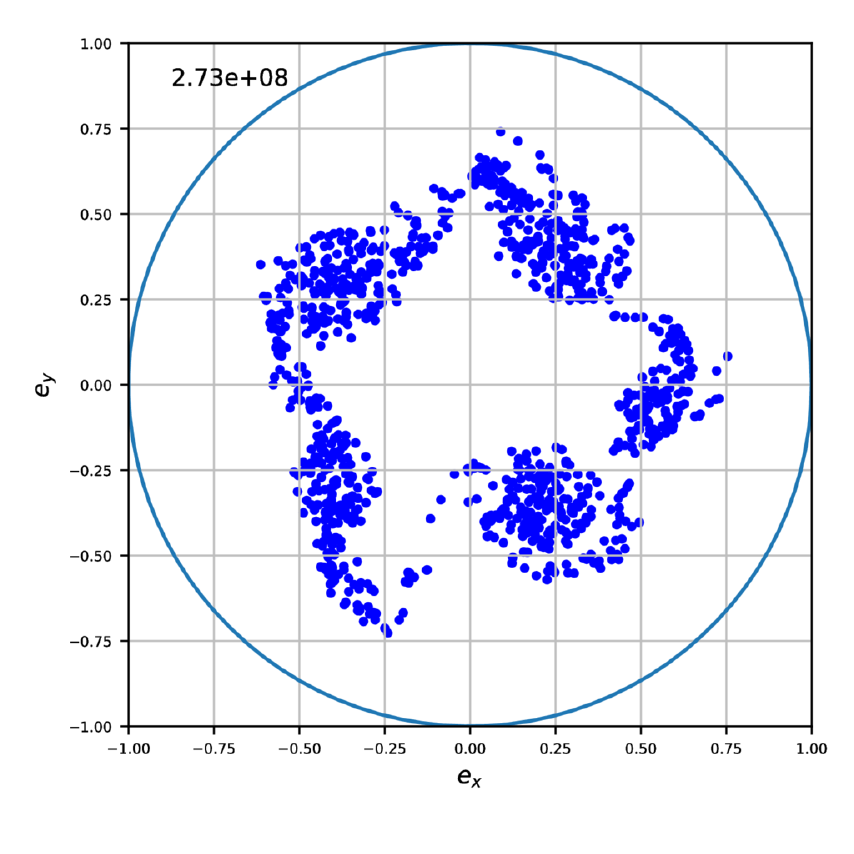}
\end{subfigure}
\hfill
\begin{subfigure}{0.3\linewidth}
\includegraphics[width=1.0 \textwidth ]{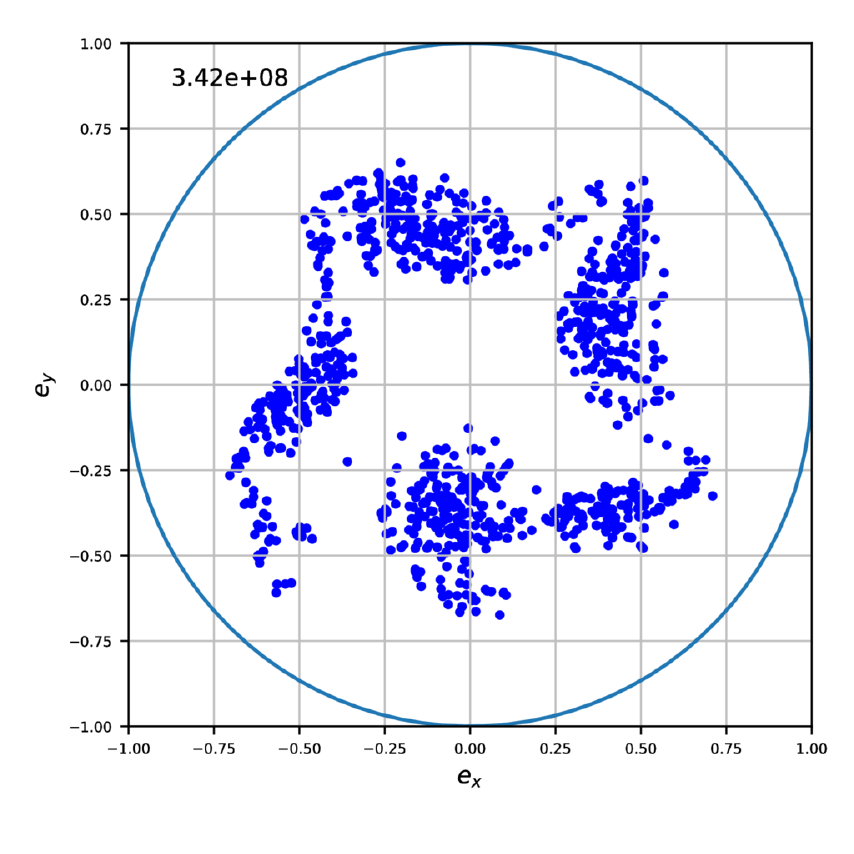}
\end{subfigure}
  \caption{\emph{Similar to Fig.~\ref{fig_wb1_DF_plane}, but for} \texttt{waterbag\_3\_s0}. A $m=4$ pattern emerges by $\sim 0.06$~Gyr.}
\label{fig_wb3_DF_plane}

\end{figure}

\begin{figure}

\begin{subfigure}{0.3\linewidth}
\includegraphics[width=1.25 \textwidth ]{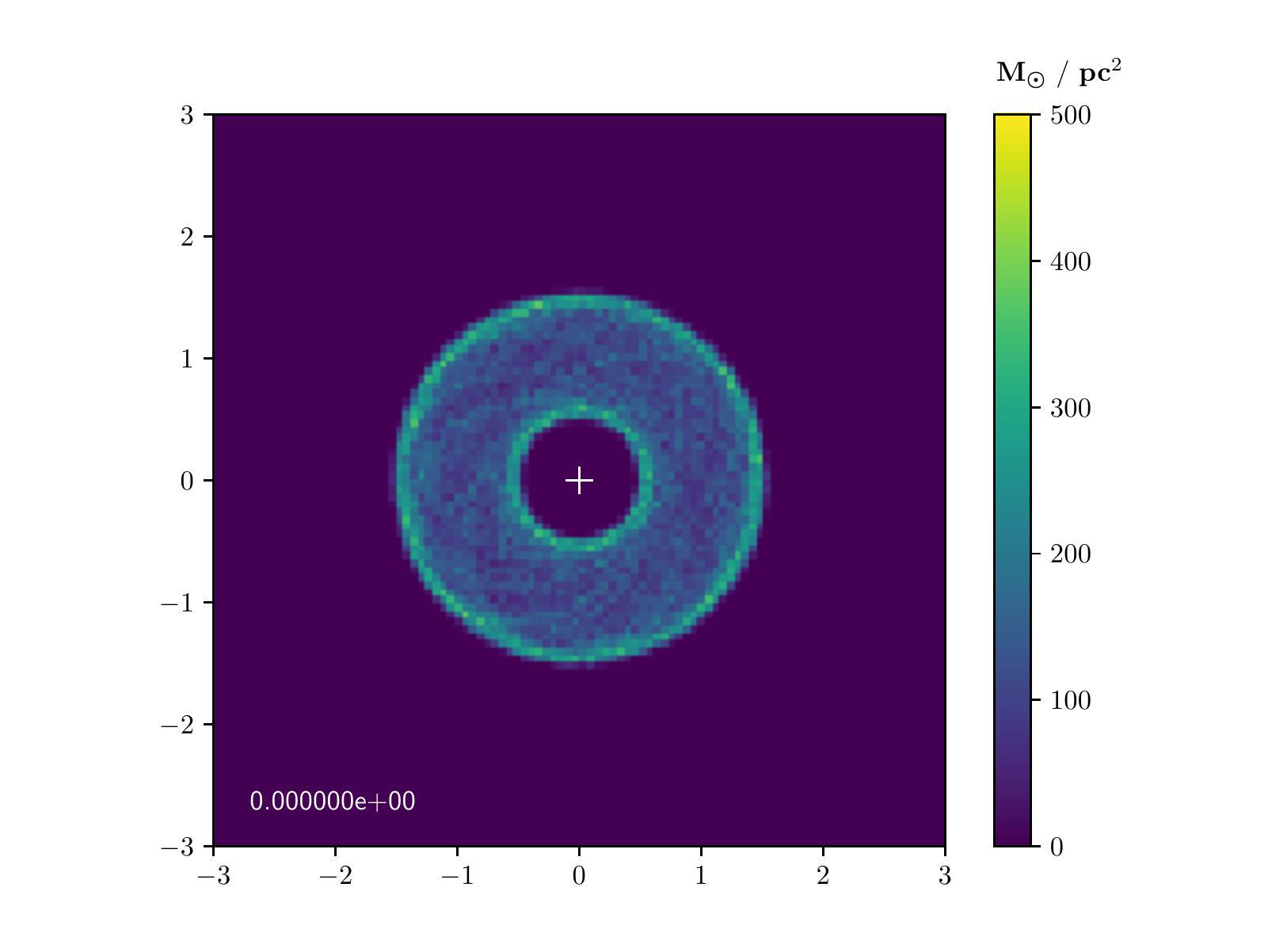}
\end{subfigure}
\hfill
\begin{subfigure}{0.3\linewidth}
\includegraphics[width=1.25 \textwidth ]{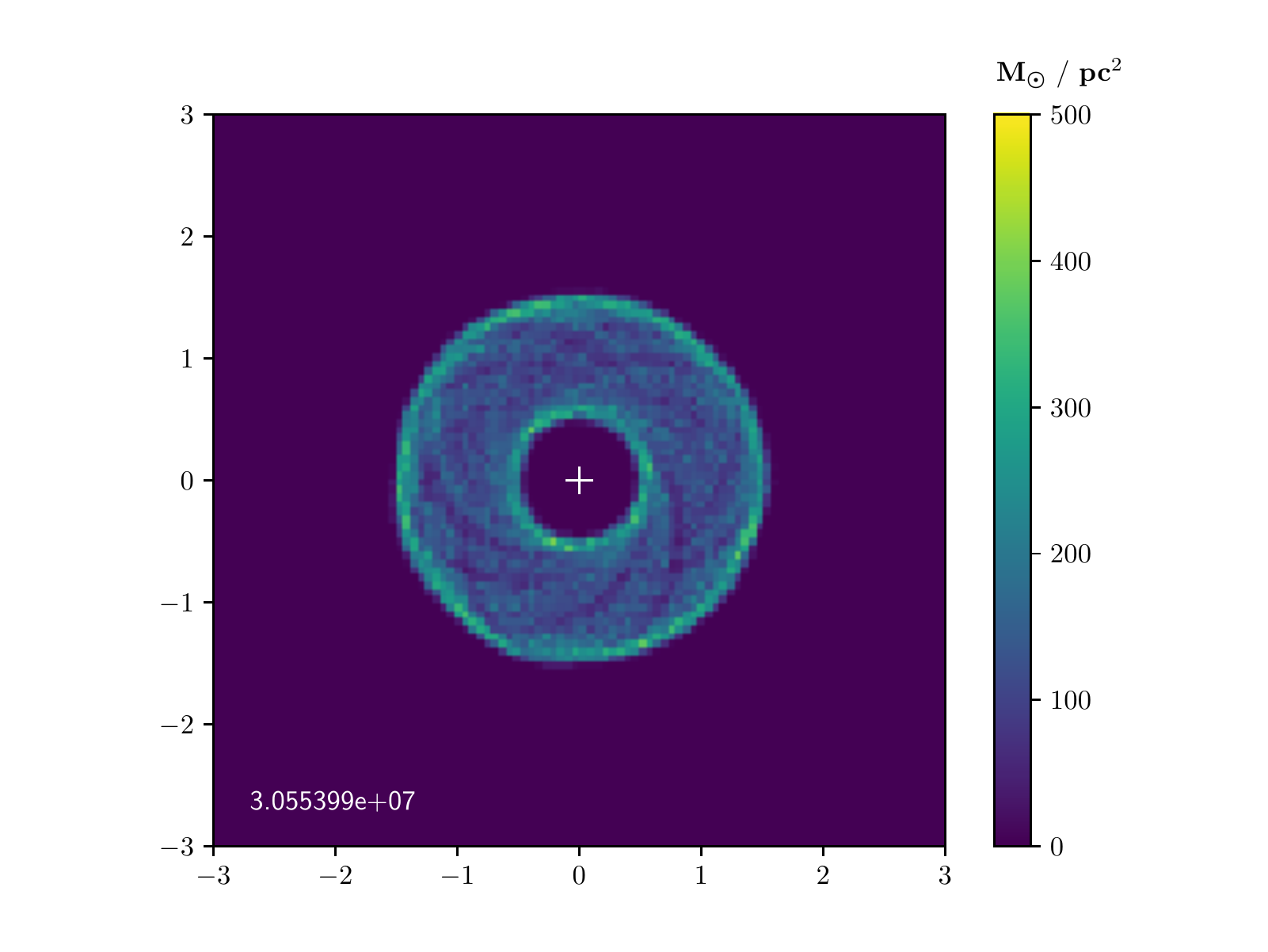}
\end{subfigure}
\hfill
\begin{subfigure}{0.3\linewidth}
\includegraphics[width=1.25 \textwidth ]{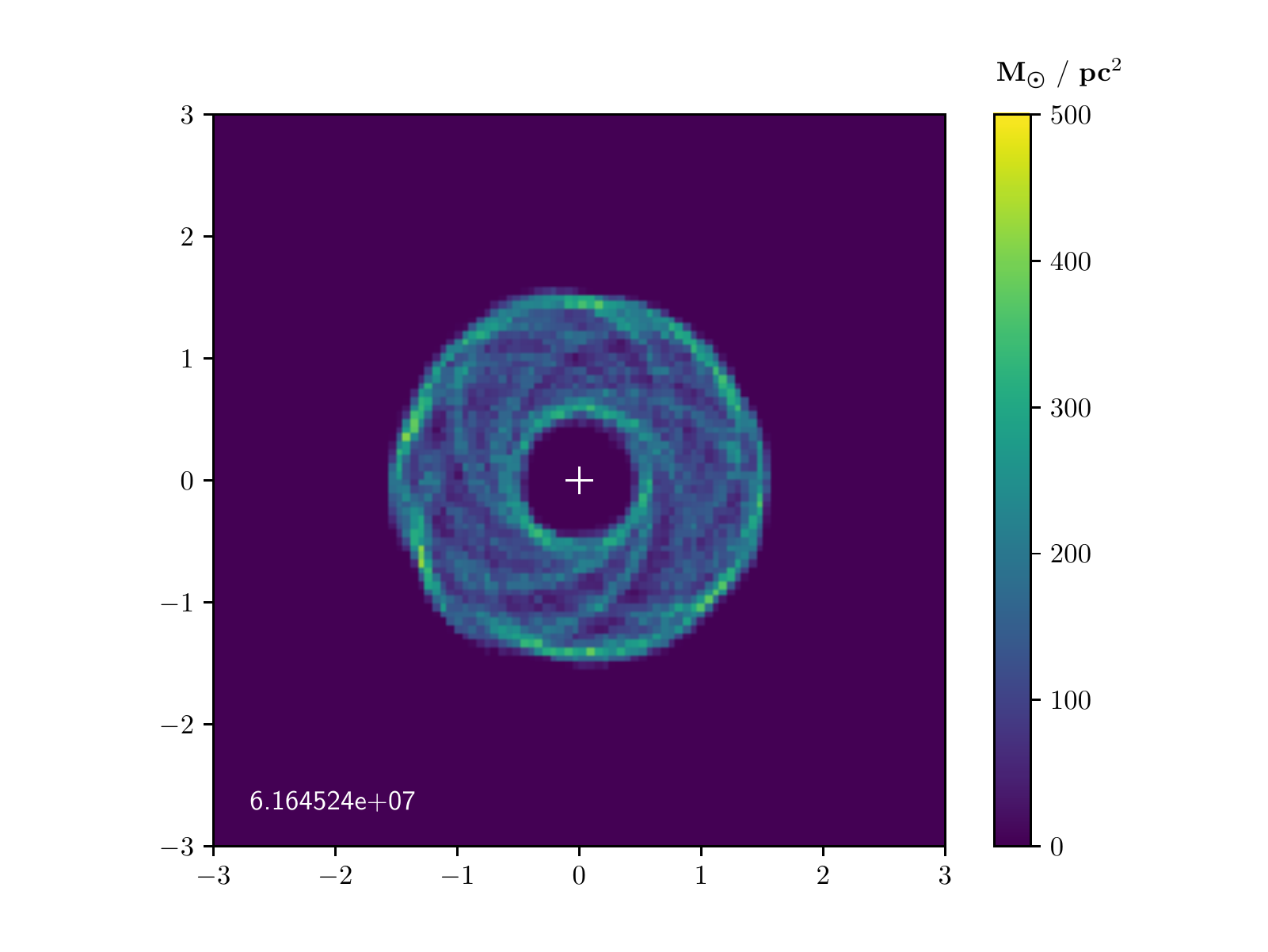}
\end{subfigure}
\hfill

\begin{subfigure}{0.3\linewidth}
\includegraphics[width=1.25 \textwidth ]{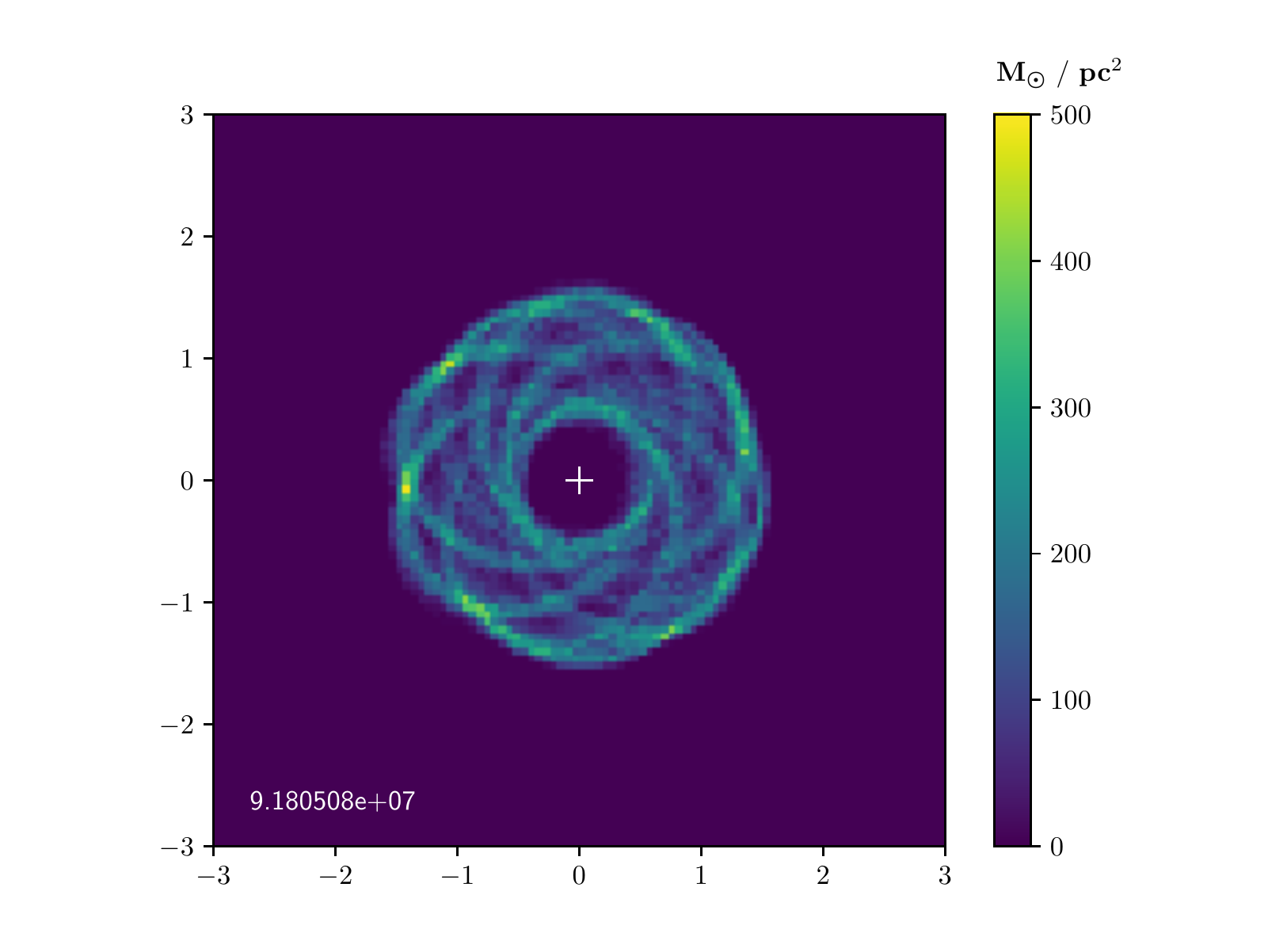}
\end{subfigure}
\hfill
\begin{subfigure}{0.3\linewidth}
\includegraphics[width=1.25 \textwidth ]{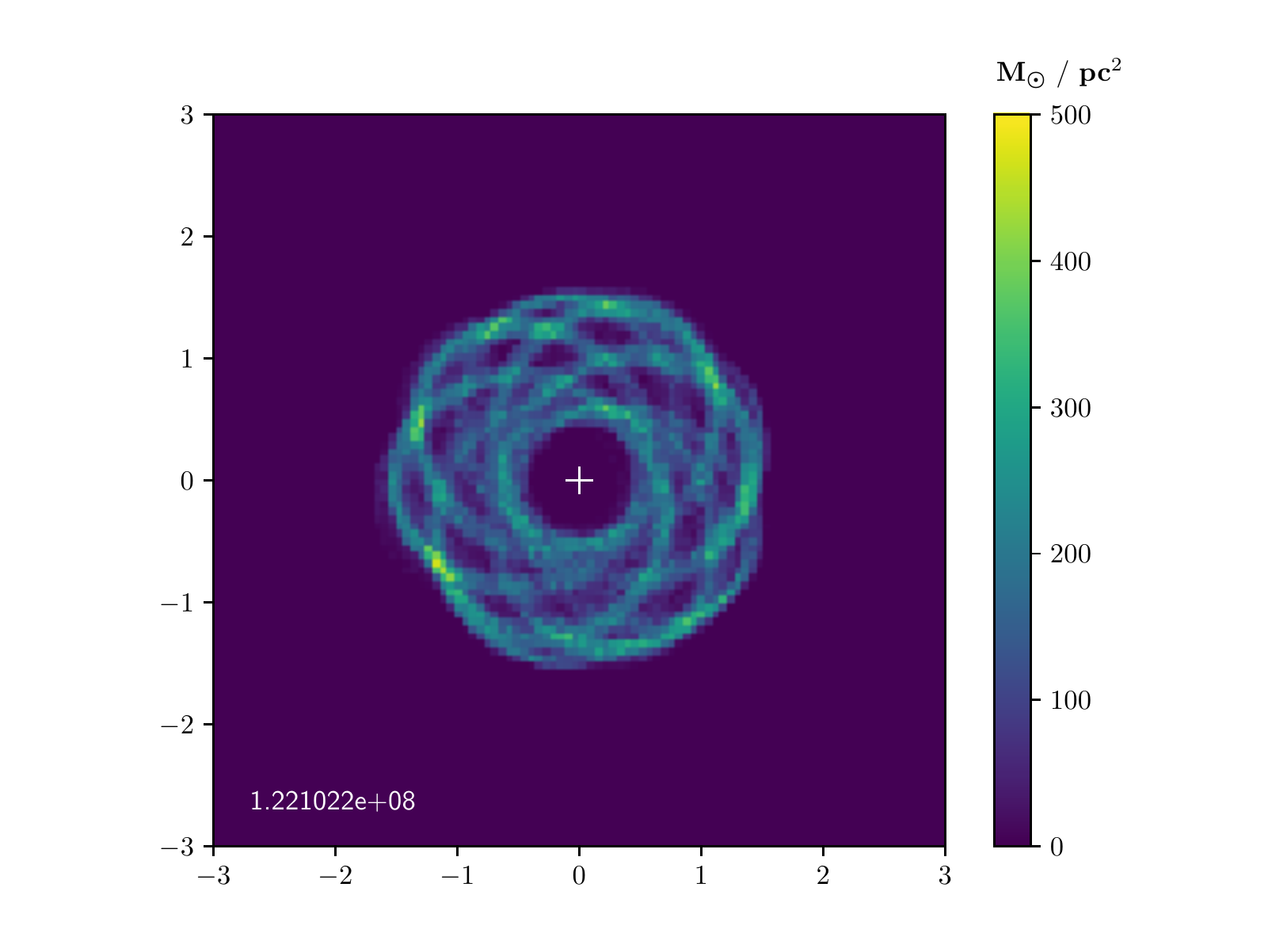}
\end{subfigure}
\hfill
\begin{subfigure}{0.3\linewidth}
\includegraphics[width=1.25 \textwidth ]{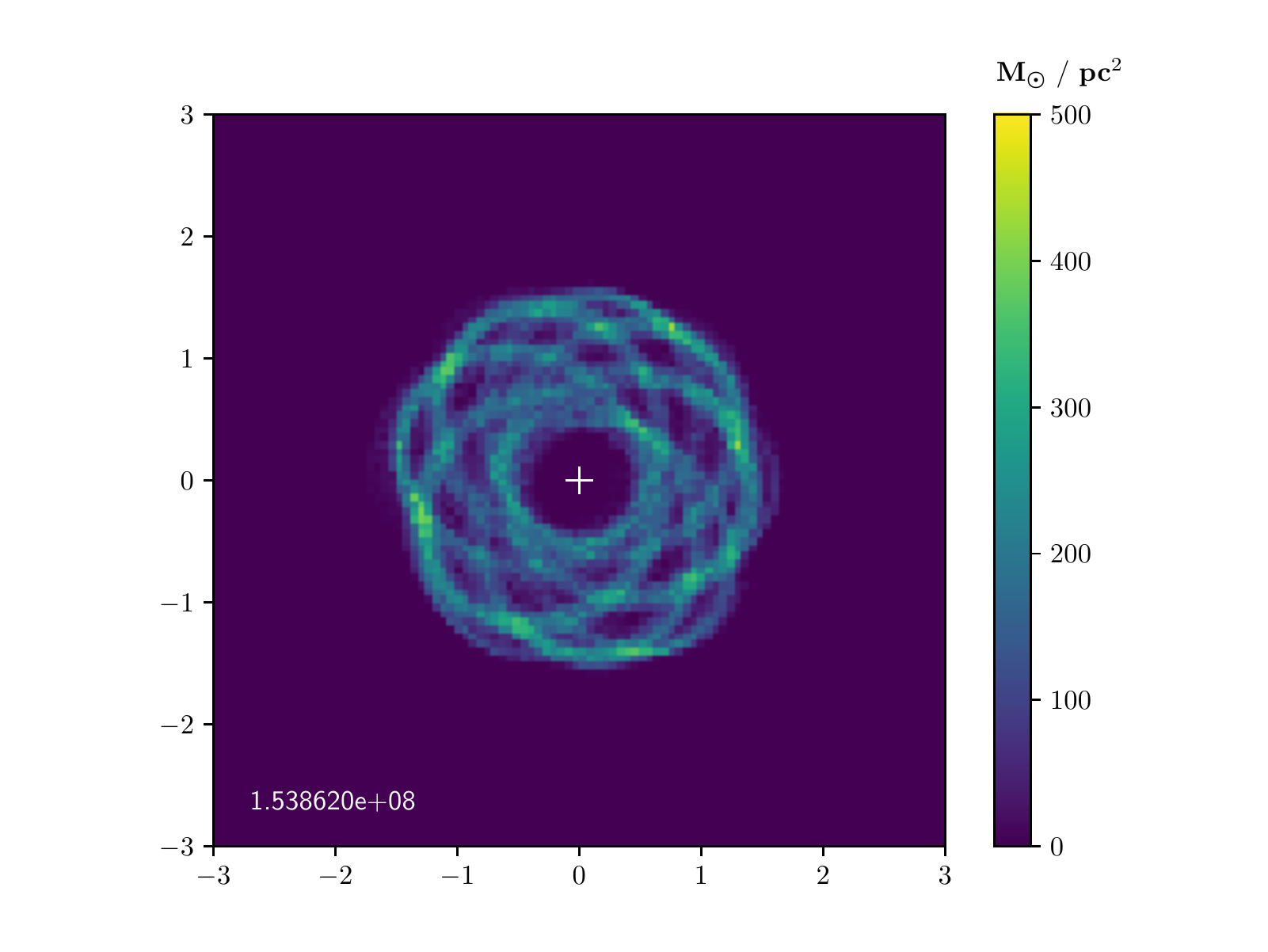}
\end{subfigure}

\begin{subfigure}{0.3\linewidth}
\includegraphics[width=1.0 \textwidth ]{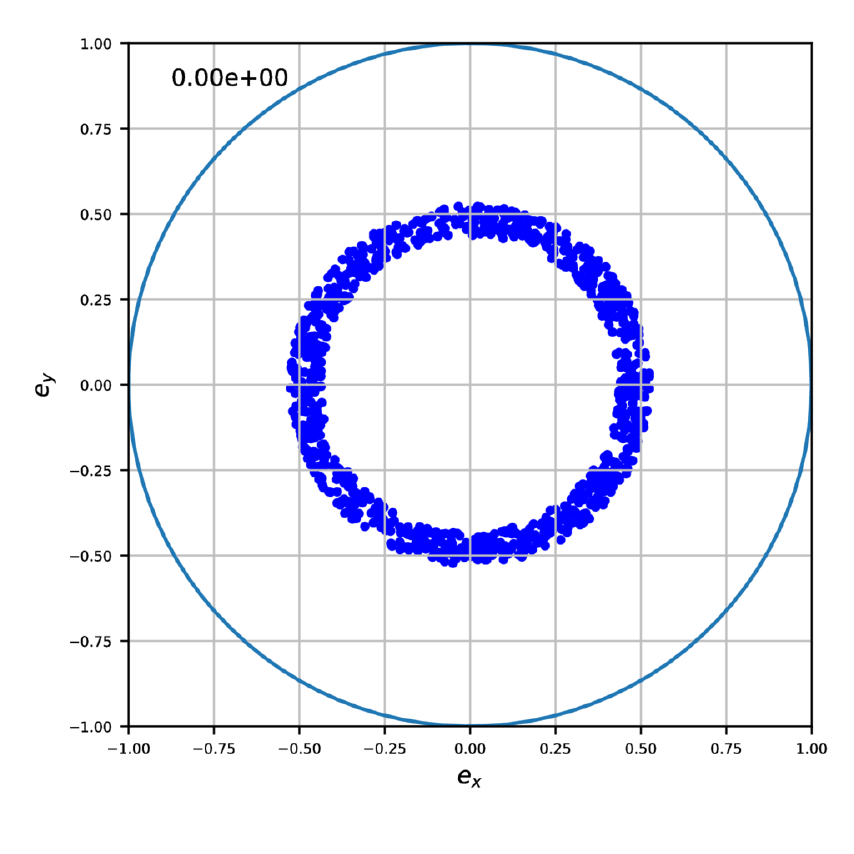}
\end{subfigure}
\hfill
\begin{subfigure}{0.3\linewidth}
\includegraphics[width=1.0 \textwidth ]{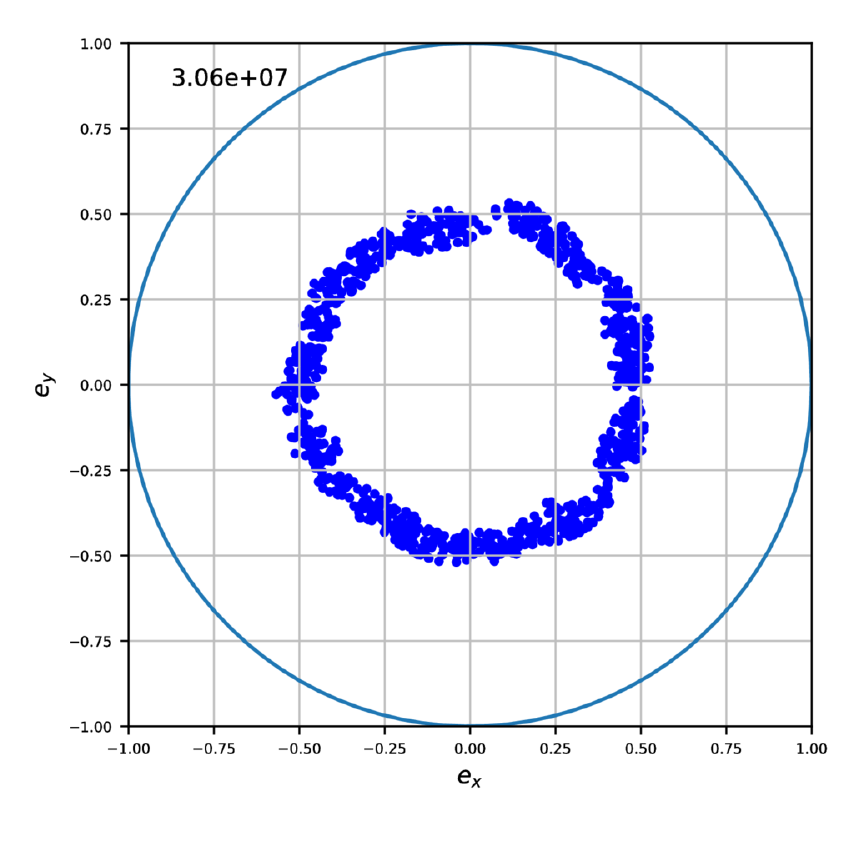}
\end{subfigure}
\hfill
\begin{subfigure}{0.3\linewidth}
\includegraphics[width=1.0 \textwidth ]{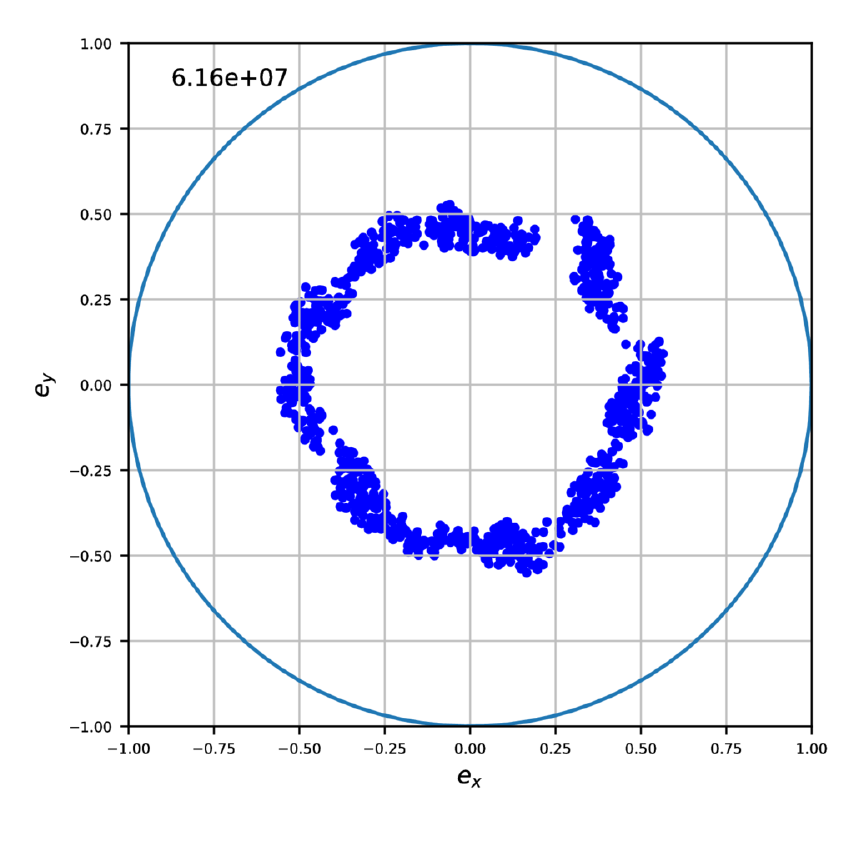}
\end{subfigure}
\hfill

\begin{subfigure}{0.3\linewidth}
\includegraphics[width=1.0 \textwidth ]{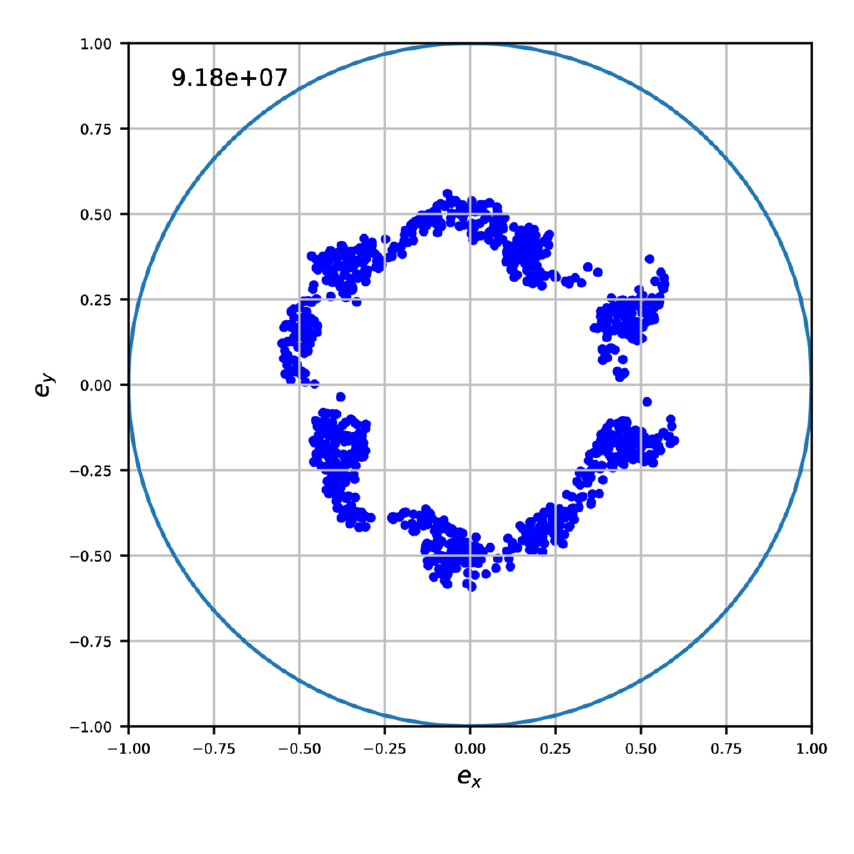}
\end{subfigure}
\hfill
\begin{subfigure}{0.3\linewidth}
\includegraphics[width=1.0 \textwidth ]{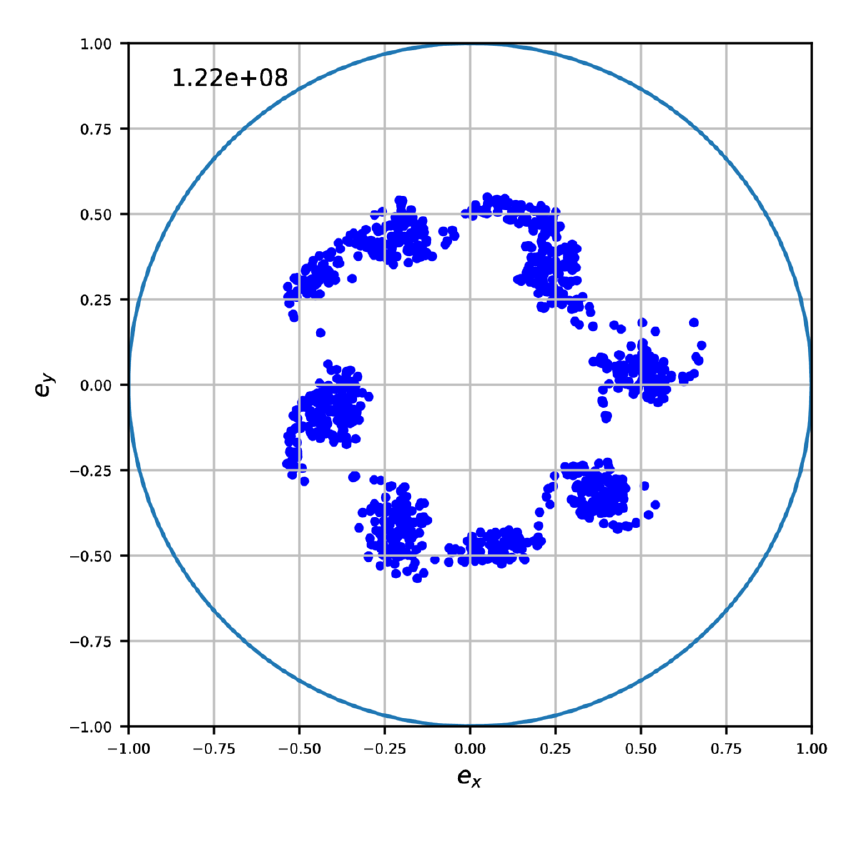}
\end{subfigure}
\hfill
\begin{subfigure}{0.3\linewidth}
\includegraphics[width=1.0 \textwidth ]{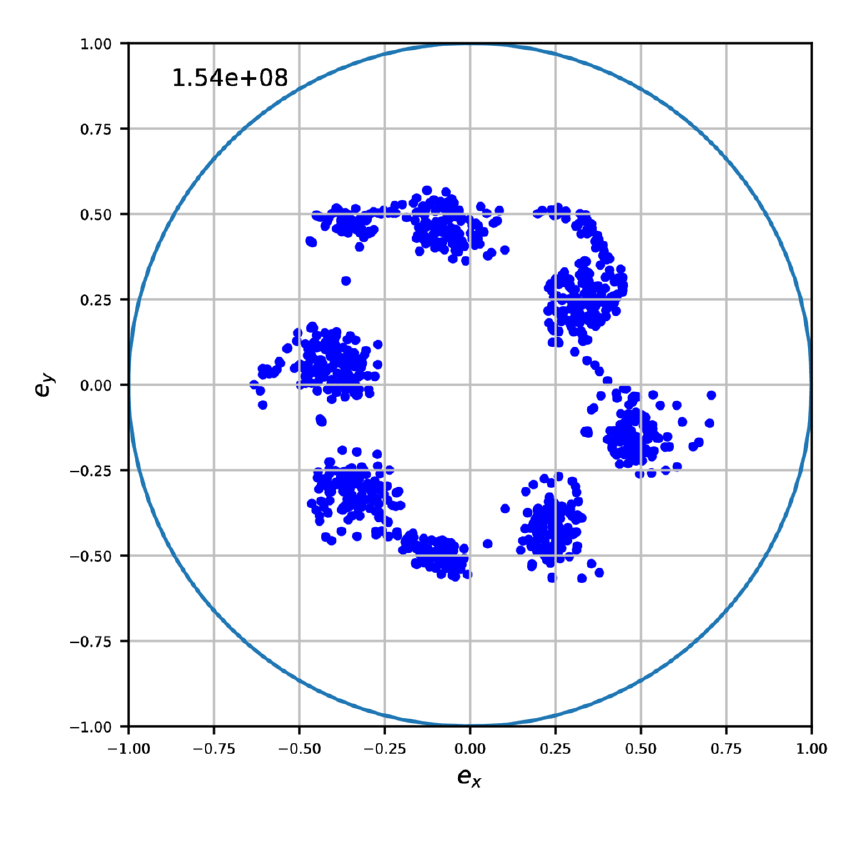}
\end{subfigure}
 \caption{\emph{Similar to Fig.~\ref{fig_wb1_DF_plane}, but for} \texttt{waterbag\_4\_s0}. A $m=6$ pattern emerges by $\sim 0.03$~Gyr.} 
\label{fig_wb4_DF_plane}

\end{figure}

Of the five cases in Set~I, \texttt{waterbag\_1\_s0} and \texttt{waterbag\_2\_s0} have been discussed earlier. \texttt{waterbag\_5\_s0} is stable according to linear theory, and the simulation results confirmed this, showing stable evolution similar to \texttt{waterbag\_2\_s0}. We now consider two new unstable bands, \texttt{waterbag\_3\_s0} and \texttt{waterbag\_4\_s0}. In Table~\ref{tbl:theo-predict} we list the predictions of linear theory for these two bands, including also \texttt{waterbag\_1\_s0} whose instability was discussed earlier. For each band all its unstable modes are identified, and the growth rate and pattern speed of the most unstable mode $(m_0)$ are computed using equations~(\ref{eqn_growth rate}) and (\ref{eqn_pattern_precession}).

\begin{figure}
\begin{subfigure}{0.5\linewidth}%
\centering
\includegraphics[width=1.0 \textwidth]{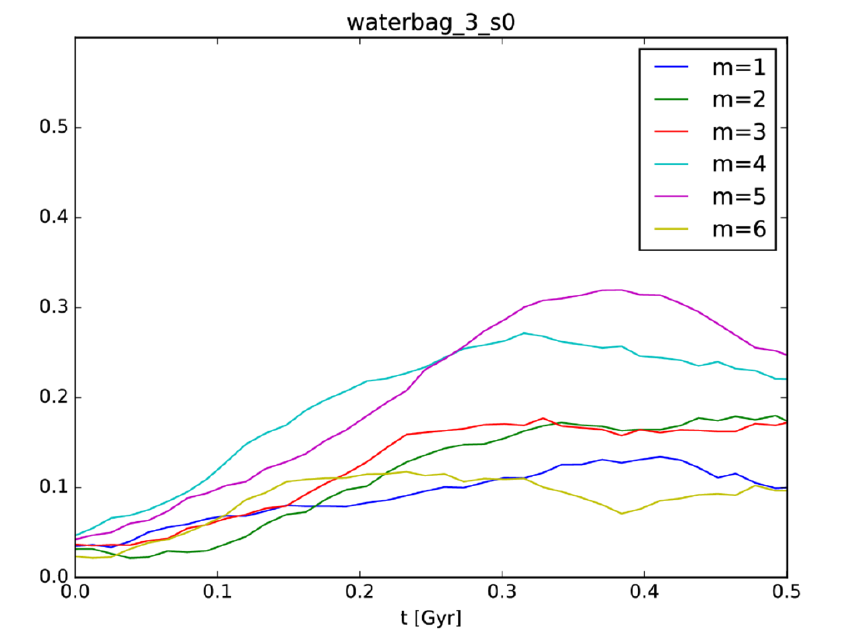}
  \caption{\texttt{waterbag\_3\_s0} }
\label{fig_wb3_modes}
\end{subfigure}
\hfill
\begin{subfigure}{0.5\linewidth}
\centering
\includegraphics[width=1.0 \textwidth]{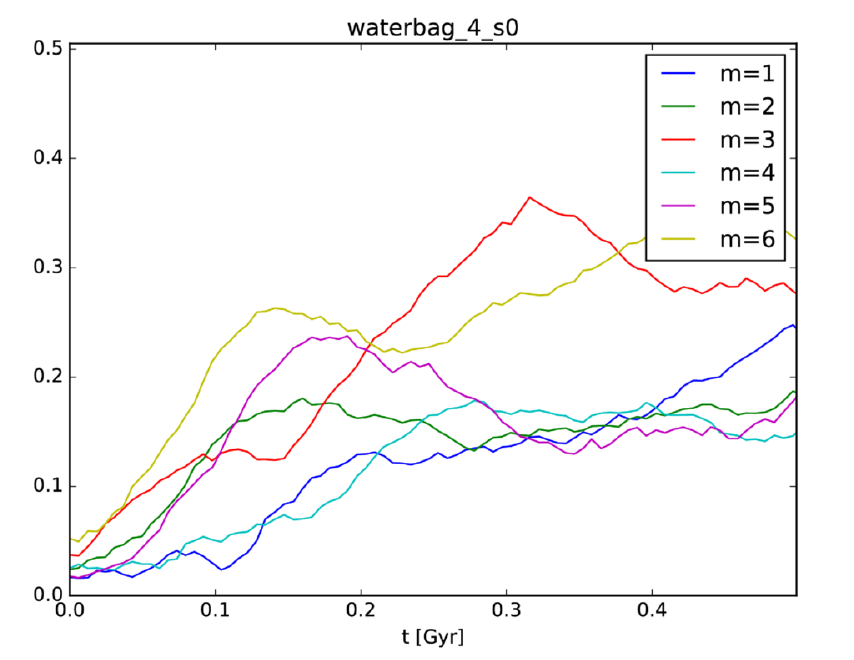}
  \caption{\texttt{waterbag\_4\_s0}}
\label{fig_wb4_modes}
\end{subfigure}
  \caption{\emph{Evolution of mode amplitudes $a_m$}. (a) \texttt{waterbag\_3\_s0}, (b) \texttt{waterbag\_4\_s0}.}
\label{fig_wb3&4_modes}
\end{figure}

\medskip
\noindent
{\bf Simulations of \texttt{waterbag\_3\_s0}:} From Figure~\ref{fig_wb3_DF_plane} we see that a $m=4$ pattern emerges by $\sim 0.06$~Gyr, which is in agreement with linear theory. Non-linear interactions, mainly with the unstable $m=5$ mode, lead to distortions of the pattern. This 
can be seen clearly in Figure~\ref{fig_wb3_modes} which plots the mode 
amplitudes $a_m$ versus time: the $m=4$ mode has the maximum amplitude 
until $\sim 0.2$~Gyr, after which the $m=5$ mode begins to dominate.  

\medskip
\noindent
{\bf Simulations of \texttt{waterbag\_4\_s0}:} From Figure~\ref{fig_wb4_DF_plane} we see that a $m=6$ pattern emerges by $\sim 0.03$~Gyr, which is in agreement with linear theory. Non-linear interactions with other unstable
modes lead to distortions of the pattern. This can be seen clearly in Figure~\ref{fig_wb4_modes} which plots the mode amplitudes $a_m$ versus time: the $m=6$ mode dominates until $\sim 0.2$~Gyr, after which there seems to be
non-linear interactions among many modes.  

\medskip\noindent
Table~\ref{tbl:bunch1_mode_comparison} shows the general agreement between 
linear theory and simulations. 

\subsection{Set~II}
\label{sec_set2_bags}

\begin{figure}

\begin{subfigure}{0.3\linewidth}
\includegraphics[width=1.25 \textwidth ]{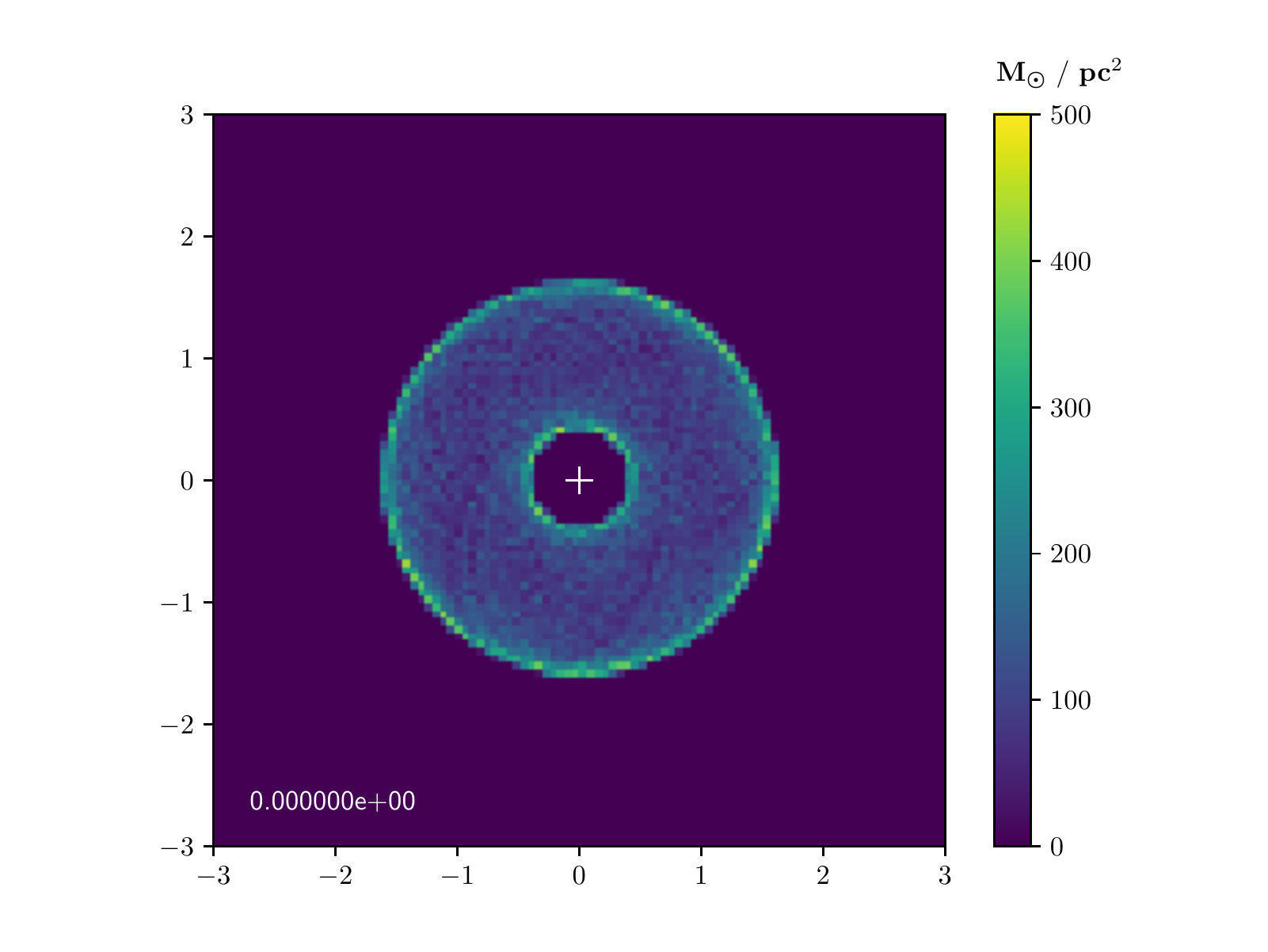}
\end{subfigure}
\hfill
\begin{subfigure}{0.3\linewidth}
\includegraphics[width=1.25 \textwidth ]{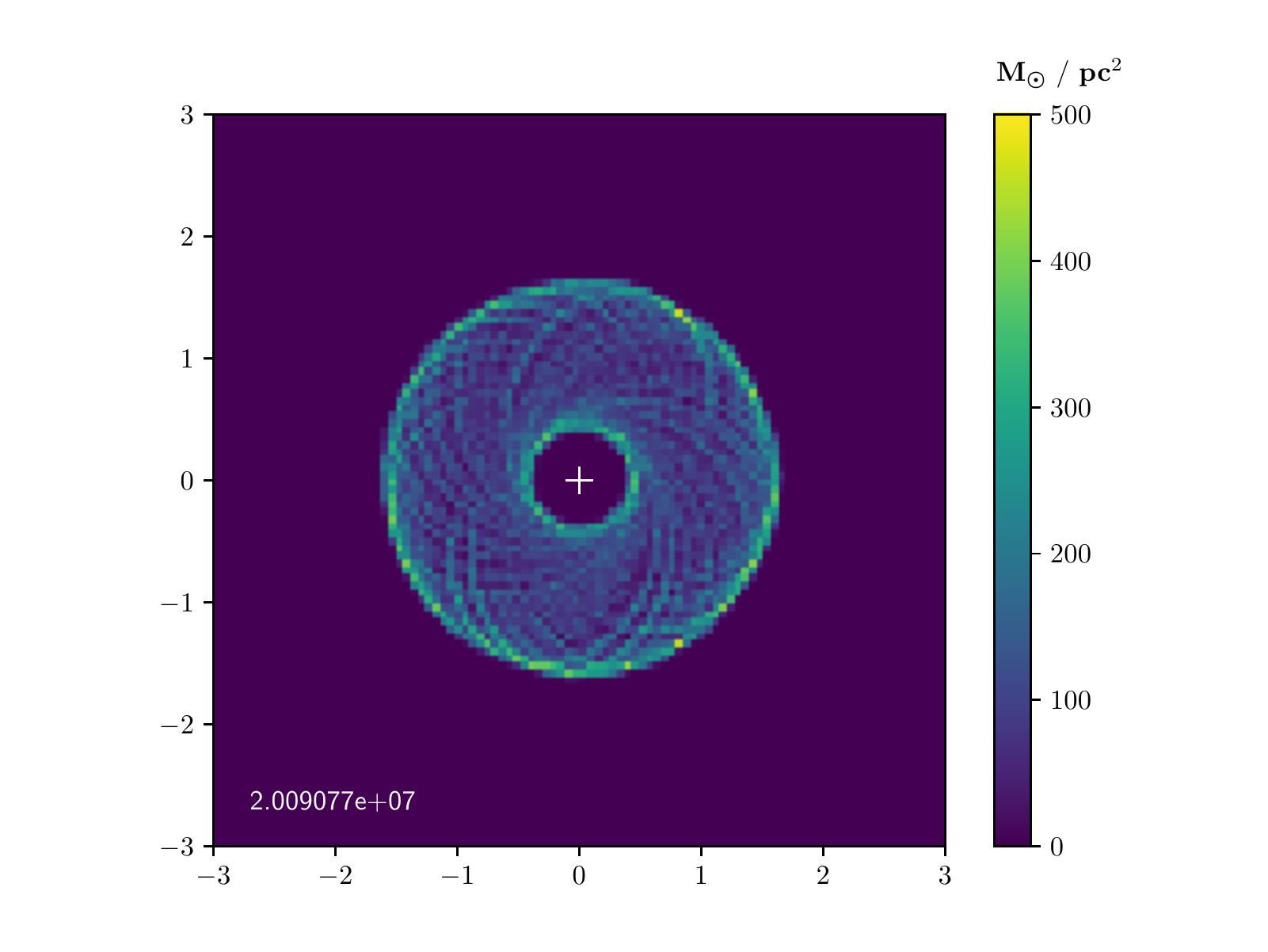}
\end{subfigure}
\hfill
\begin{subfigure}{0.3\linewidth}
\includegraphics[width=1.25 \textwidth ]{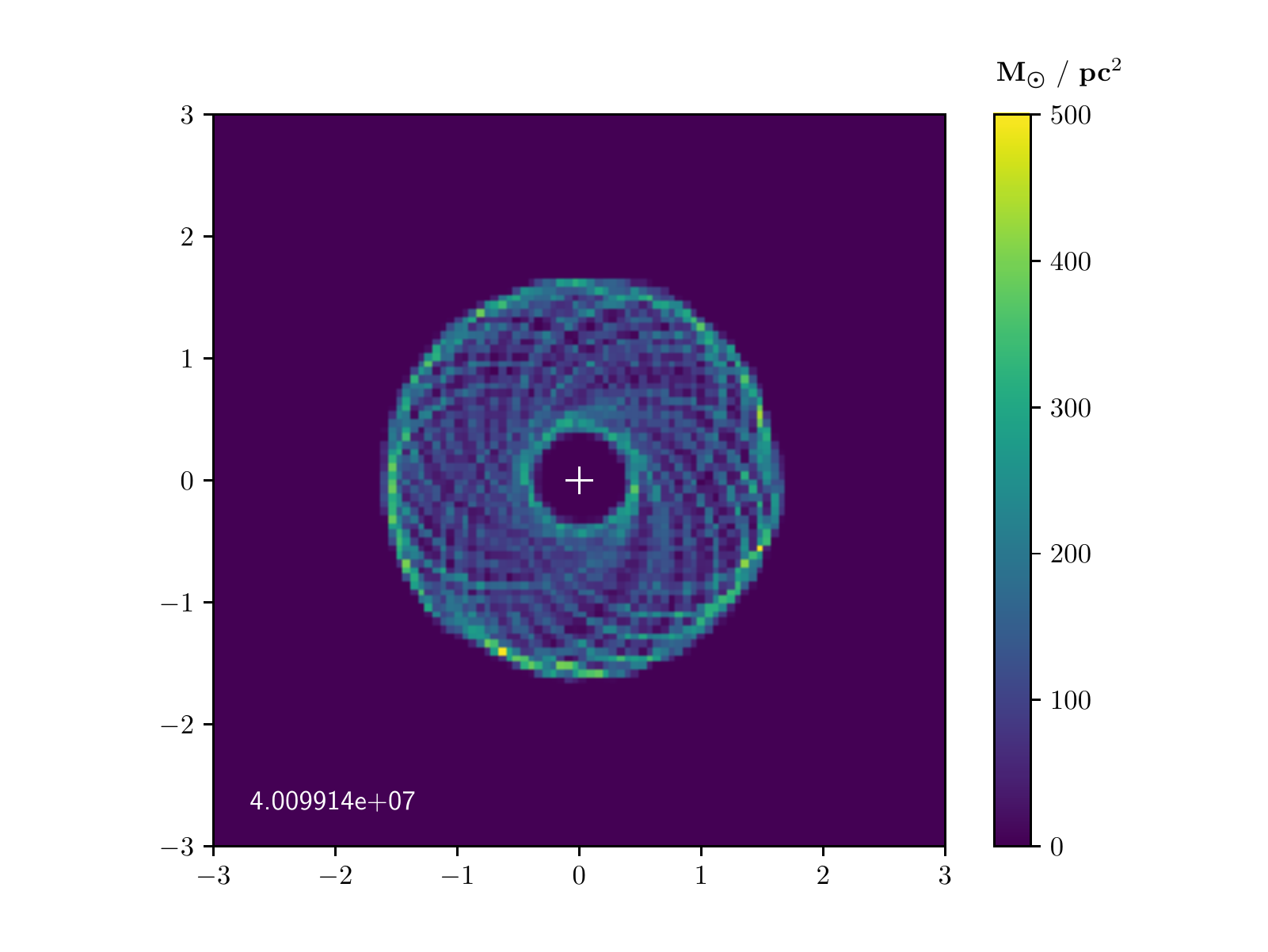}
\end{subfigure}
\hfill

\begin{subfigure}{0.3\linewidth}
\includegraphics[width=1.25 \textwidth ]{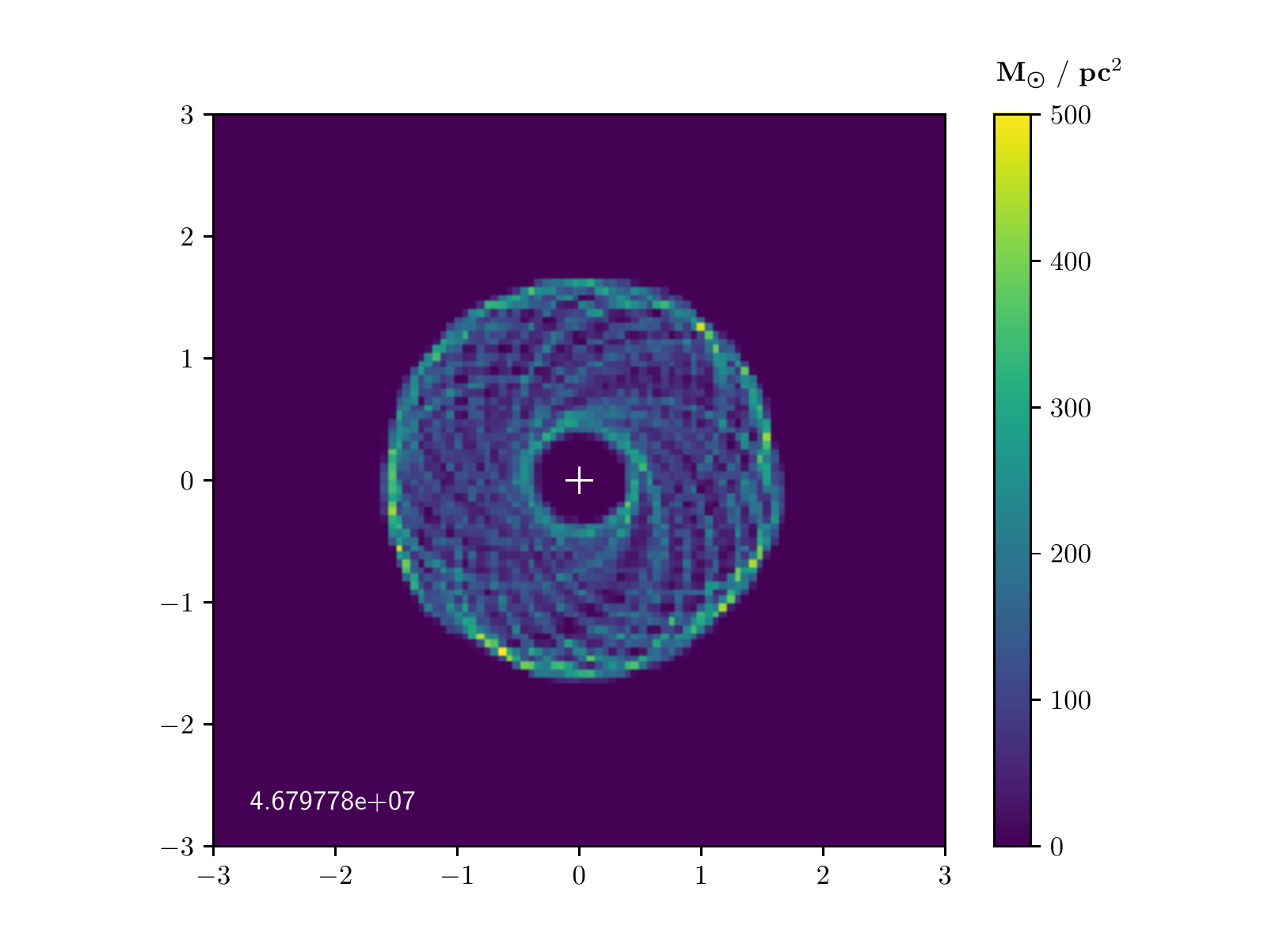}
\end{subfigure}
\hfill
\begin{subfigure}{0.3\linewidth}
\includegraphics[width=1.25 \textwidth ]{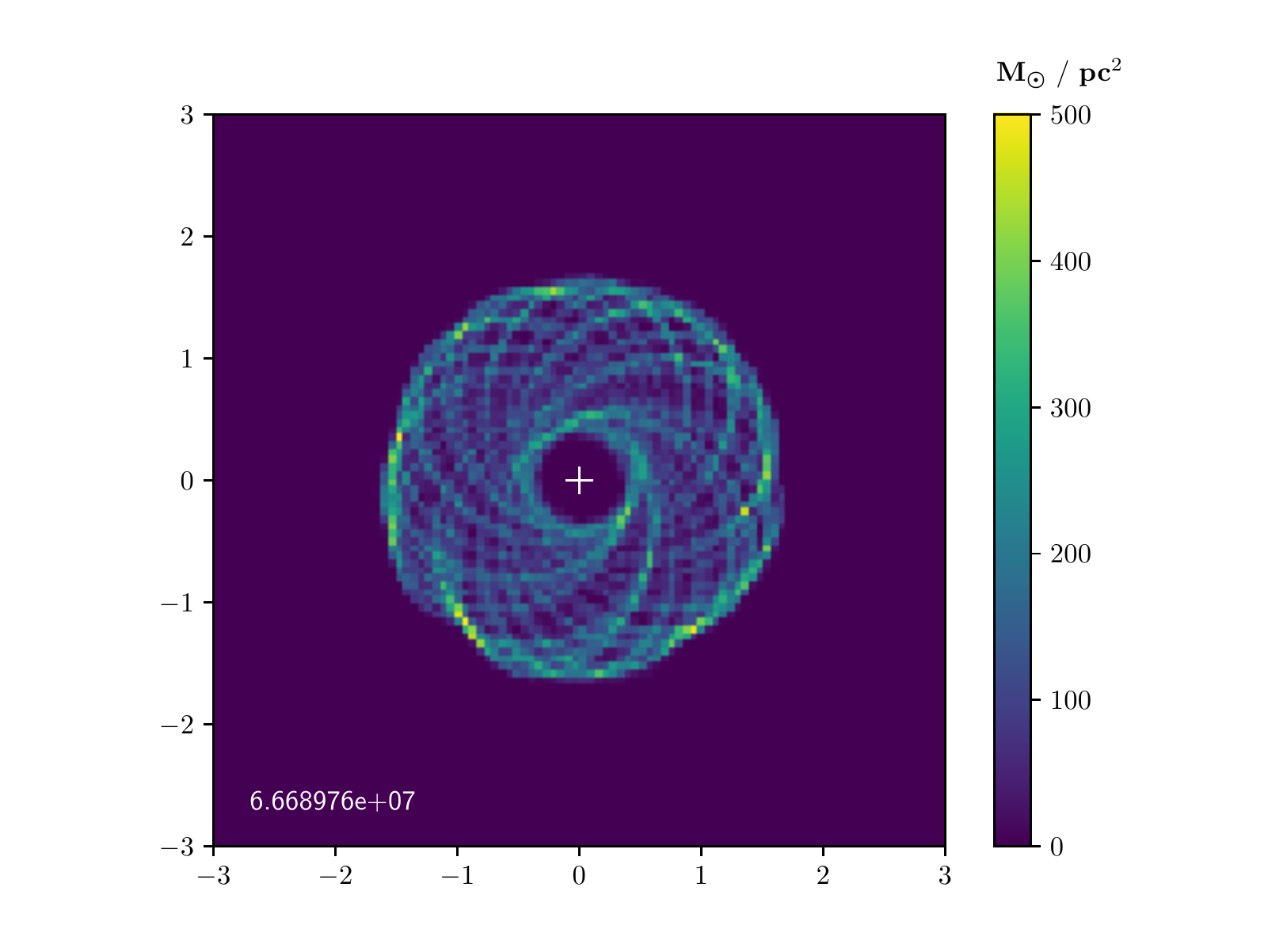}
\end{subfigure}
\hfill
\begin{subfigure}{0.3\linewidth}
\includegraphics[width=1.25 \textwidth ]{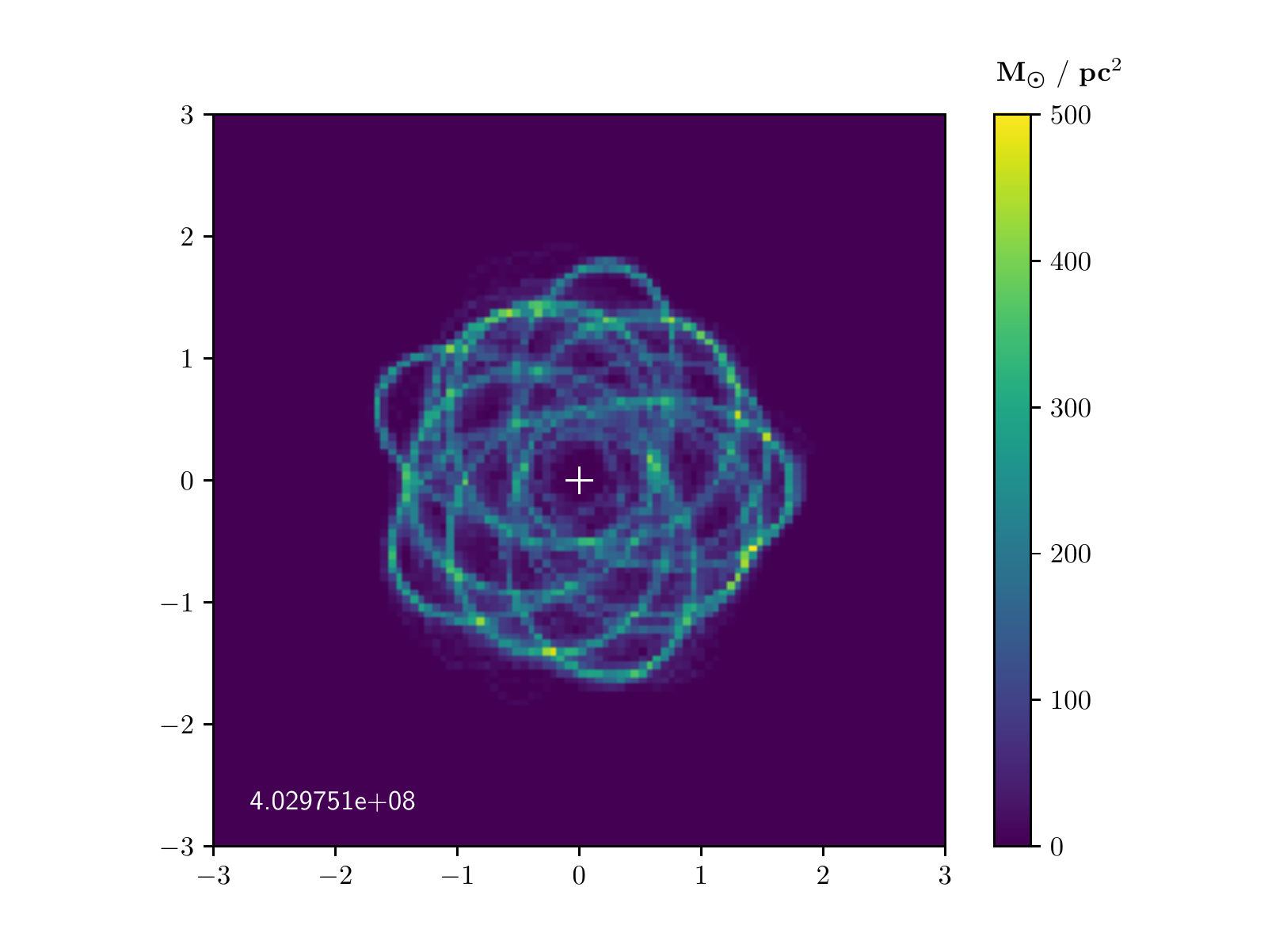}
\end{subfigure}

\begin{subfigure}{0.3\linewidth}
\includegraphics[width=1.0 \textwidth ]{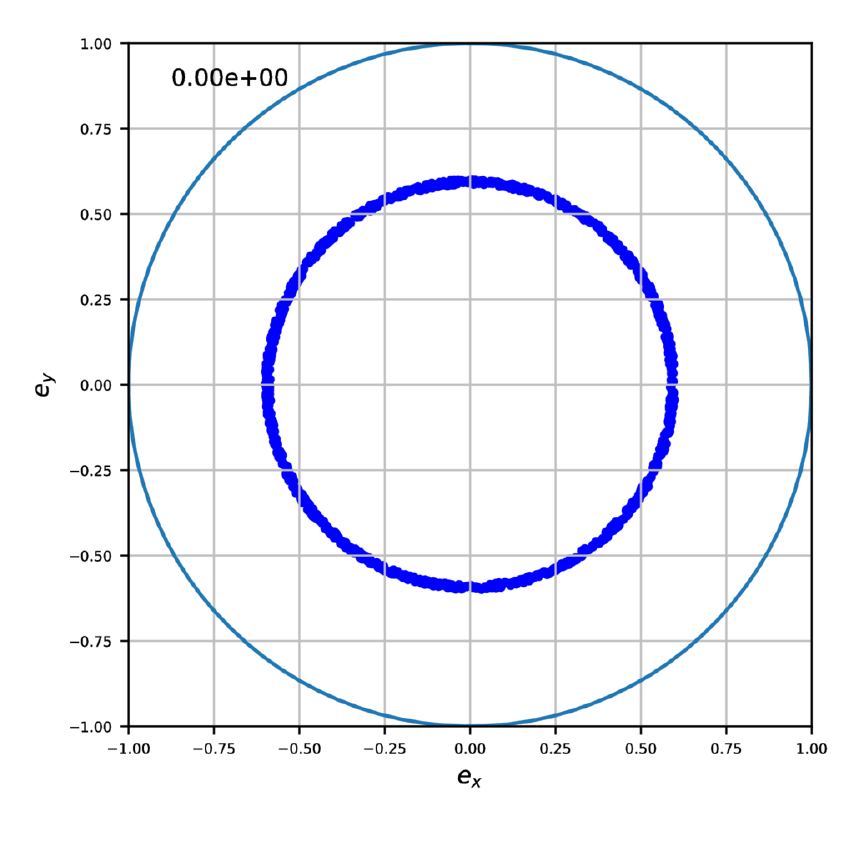}
\end{subfigure}
\hfill
\begin{subfigure}{0.3\linewidth}
\includegraphics[width=1.0 \textwidth ]{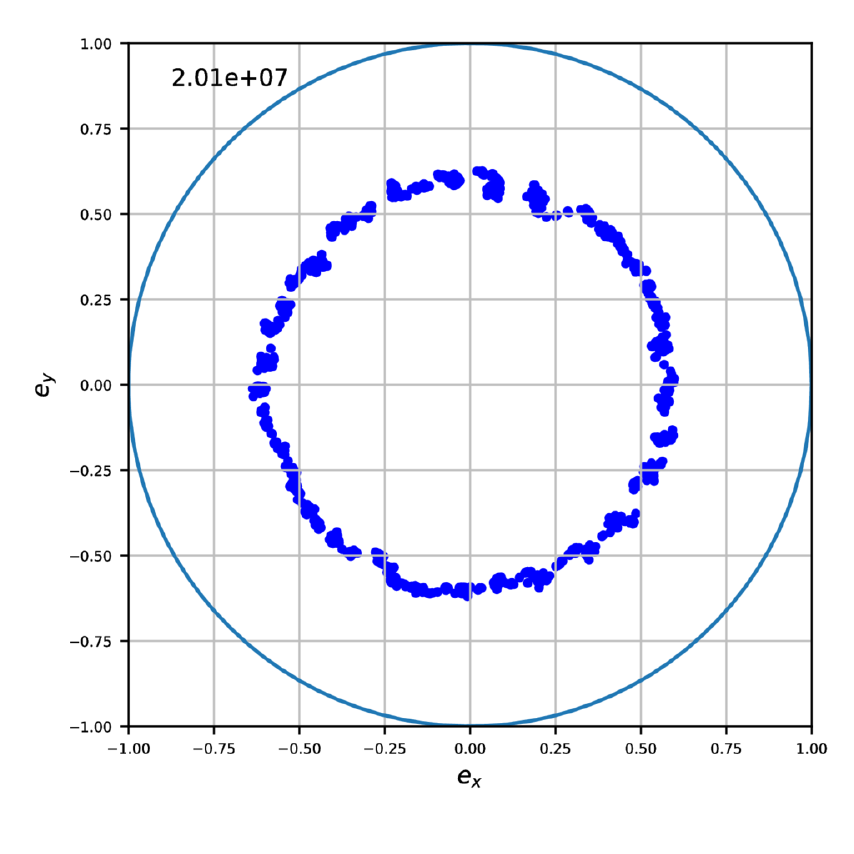}
\end{subfigure}
\hfill
\begin{subfigure}{0.3\linewidth}
\includegraphics[width=1.0 \textwidth ]{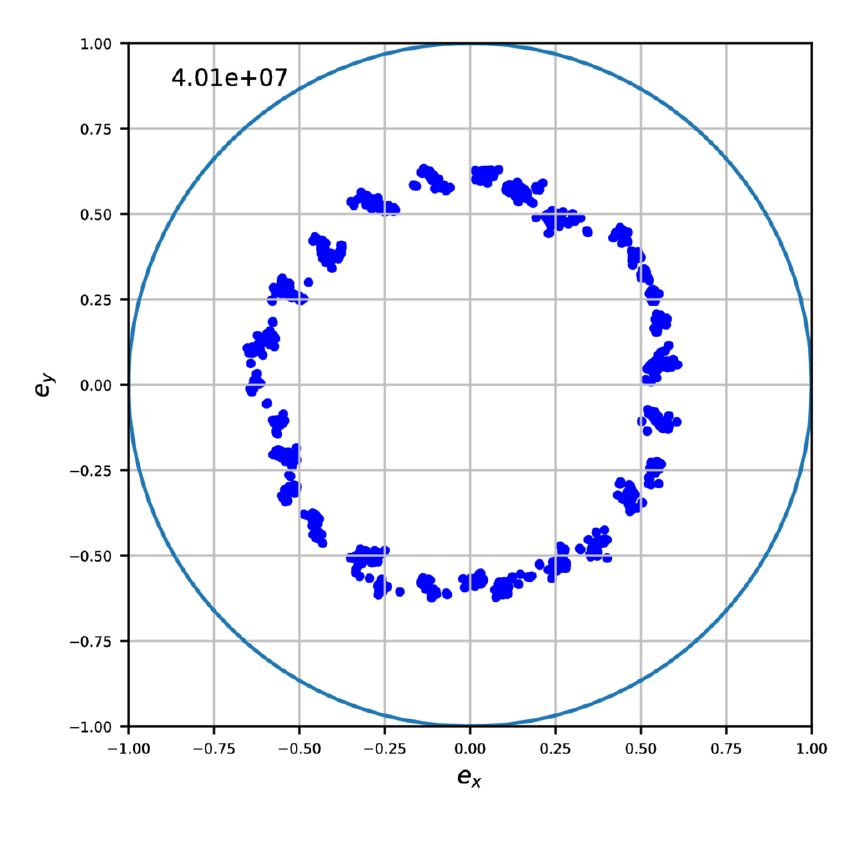}
\end{subfigure}
\hfill

\begin{subfigure}{0.3\linewidth}
\includegraphics[width=1.0 \textwidth ]{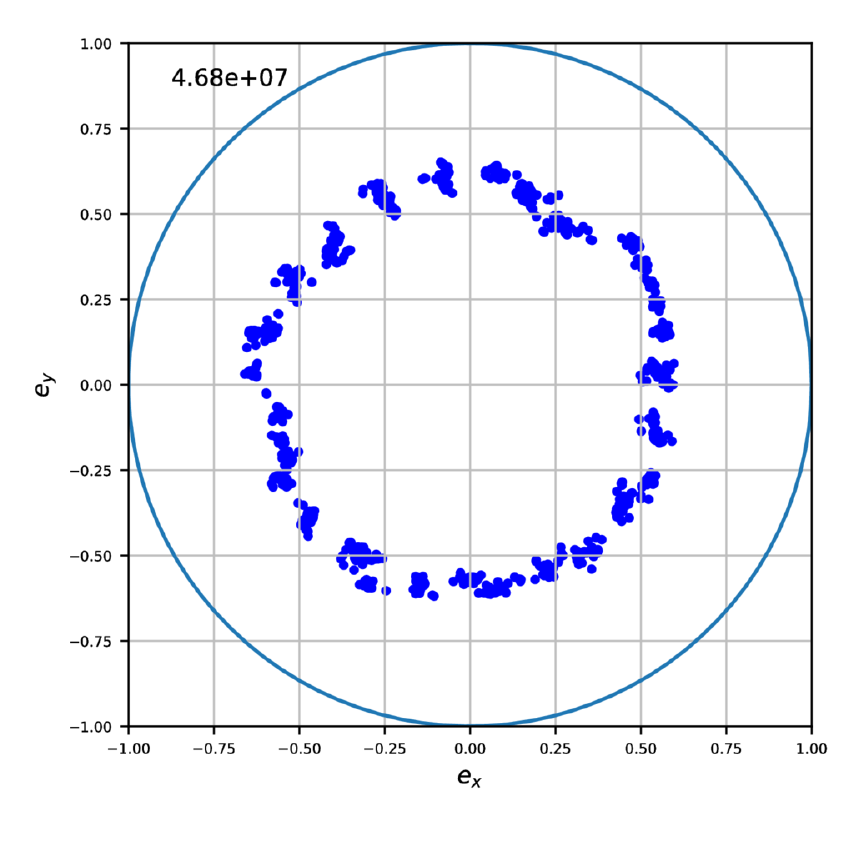}
\end{subfigure}
\hfill
\begin{subfigure}{0.3\linewidth}
\includegraphics[width=1.0 \textwidth ]{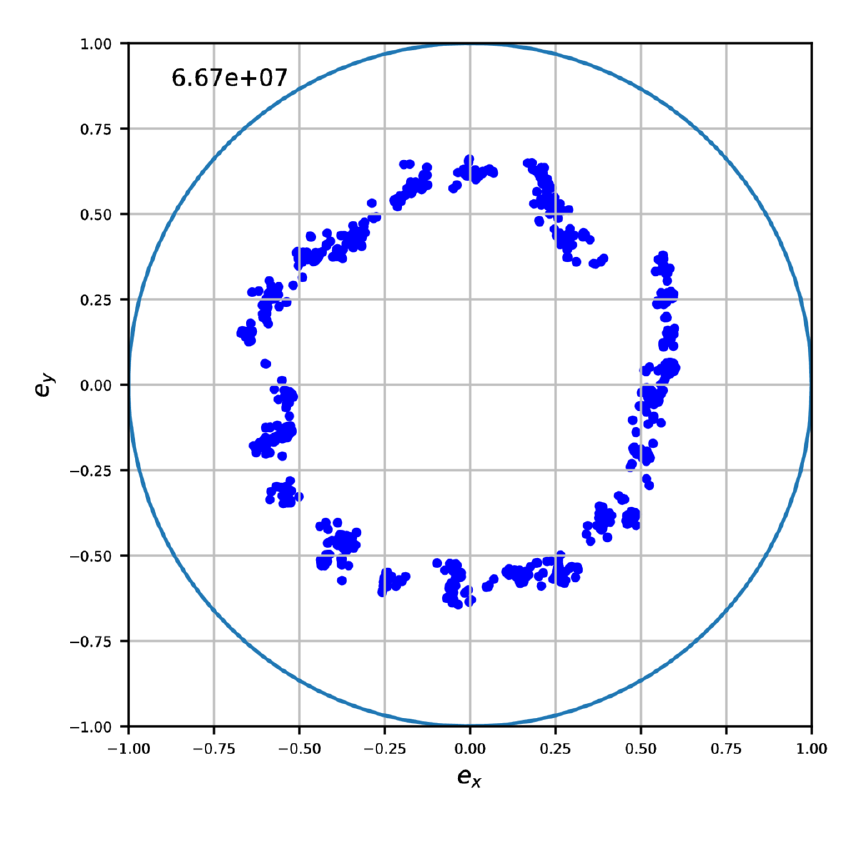}
\end{subfigure}
\hfill
\begin{subfigure}{0.3\linewidth}
\includegraphics[width=1.0 \textwidth ]{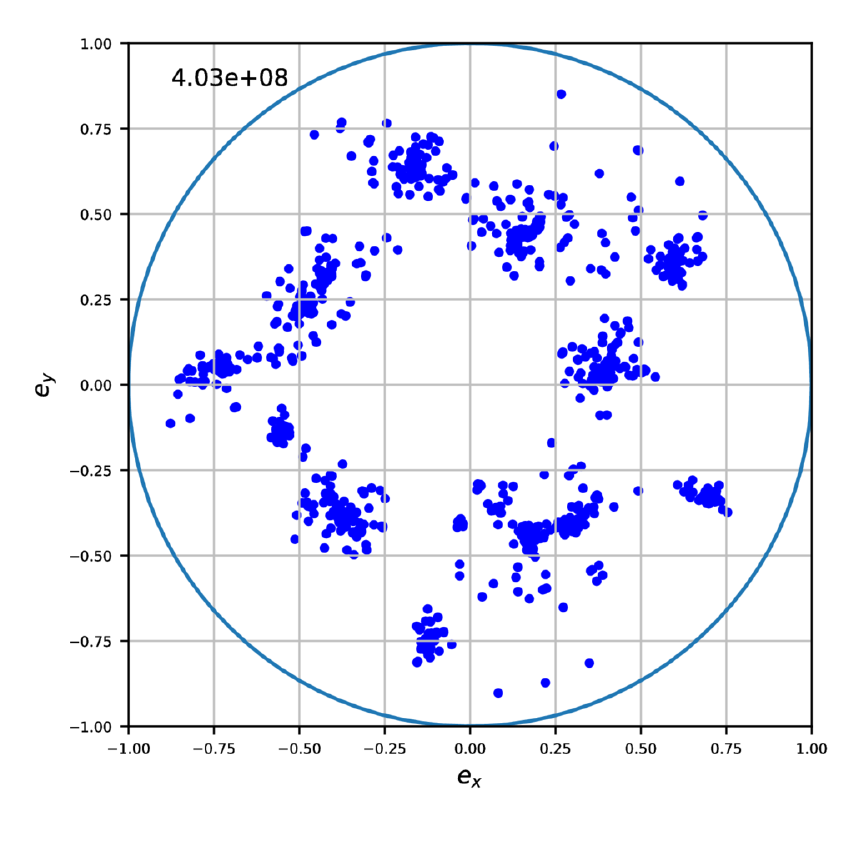}
\end{subfigure}
 \caption{\emph{Similar to Fig.~\ref{fig_wb1_DF_plane}, but for} 
{\texttt{waterbag\_$\ell$1\_0.8\_$\ell$2\_0.81  }}. An high $m$ pattern emerges by $\sim 0.02$~Gyr.}
\label{fig_wb81_DF_plane}

\end{figure}

The narrowest band in Table~\ref{tbl:simulations} is {\texttt{waterbag\_$\ell$1\_0.8\_$\ell$2\_0.81  }}, with 
$\Delta\ell = 0.01\,$. According to linear theory this band is unstable to 
a wide range of modes with $m = 3 - 57$, with $m = 36$ having the fastest growth rate. Figure~\ref{fig_wb81_DF_plane} shows the evolution of this 
narrow band, whose initial evolution shows an instability dominated by
$m \sim 36$ mode, in agreement with linear theory. 

Linear theory also predicts a transition from instability to stability
when the lower boundary is held fixed at $\ell_1 = 0.8$ and the band  
is made broader by increasing $\ell_2$. This transition occurs at $\ell_2 = \ell_{\rm crit} \simeq 0.963\,$: bands with $\ell_2 < \ell_{\rm crit}$ are unstable to various modes whereas broader bands with $\ell_{\rm crit} <\ell_2 < 1 $ are stable for all $m$. In order to test this precise prediction, we
ran a total of 20 simulations increasing $\ell_2$ in steps of $0.01$, from $0.81$ to $1$, and looked for signs of instabilities. From the last column
of Table~\ref{tbl:simulations} we see that the simulations confirm linear theory, with the small difference that the transition seems to happen when 
$\ell_2$ crosses $0.97$, instead of the predicted value of $0.963$. 

\subsection{Collisionless relaxation}
\label{sec_relax of a bag}

As instabilities unfold and non-linear interactions between modes 
dominate, what can we expect of evolution over long times? We have 
earlier in this section followed the short-time evolution of the unstable band \texttt{waterbag\_3\_s0}, with its initial growth of a dominant $m=4$
mode over $\sim 0.06$~Gyr, followed by the rise of a $m=5$ mode around
$\sim 0.2$~Gyr lasting until at least $\sim 0.34$~Gyr. What happens after this? Here we follow the evolution for $\sim 4$~Gyr. 

Figure~\ref{fig_lopsided} shows both the initial and final states   
of \texttt{waterbag\_3\_s0}. When compared with the intermediate states of Figure~\ref{fig_wb3_DF_plane}, the final state appears more axisymmetric. 
The final state also has a wider range of eccentricities than the initial state. It consists of a nearly circular high density ring, surrounded by a lower-density halo of particles with a wide range of eccentricites. The strong non-axisymmetric instabilities that plagued the initial state seem to have saturated, leaving behind a relaxed, coarse-grained state that is 
approximately axisymmetric and steady in time. The secular precessional timescale for the initial state is $\ts \sim 0.8$ Gyr, so the total duration of the run, $4$~Gyr is about $5 \, \ts$. This is too short a duration for a collisional process like resonant relaxation to be effective. Hence what we have witnessed must be collisionless relaxation, where non-axisymmetric instabilities provide the pathway for transition from one axisymmetric state to another.

\begin{figure}
   \hspace{-0.4cm}
   \begin{subfigure}{0.5\linewidth}
     \includegraphics[width=1.0 \textwidth ]{fig_wb3/wb3_sig_t0.pdf}
   \end{subfigure}
   \hspace{0.1cm}
   \begin{subfigure}{0.5\linewidth}
   \includegraphics[width=1.0 \textwidth ]{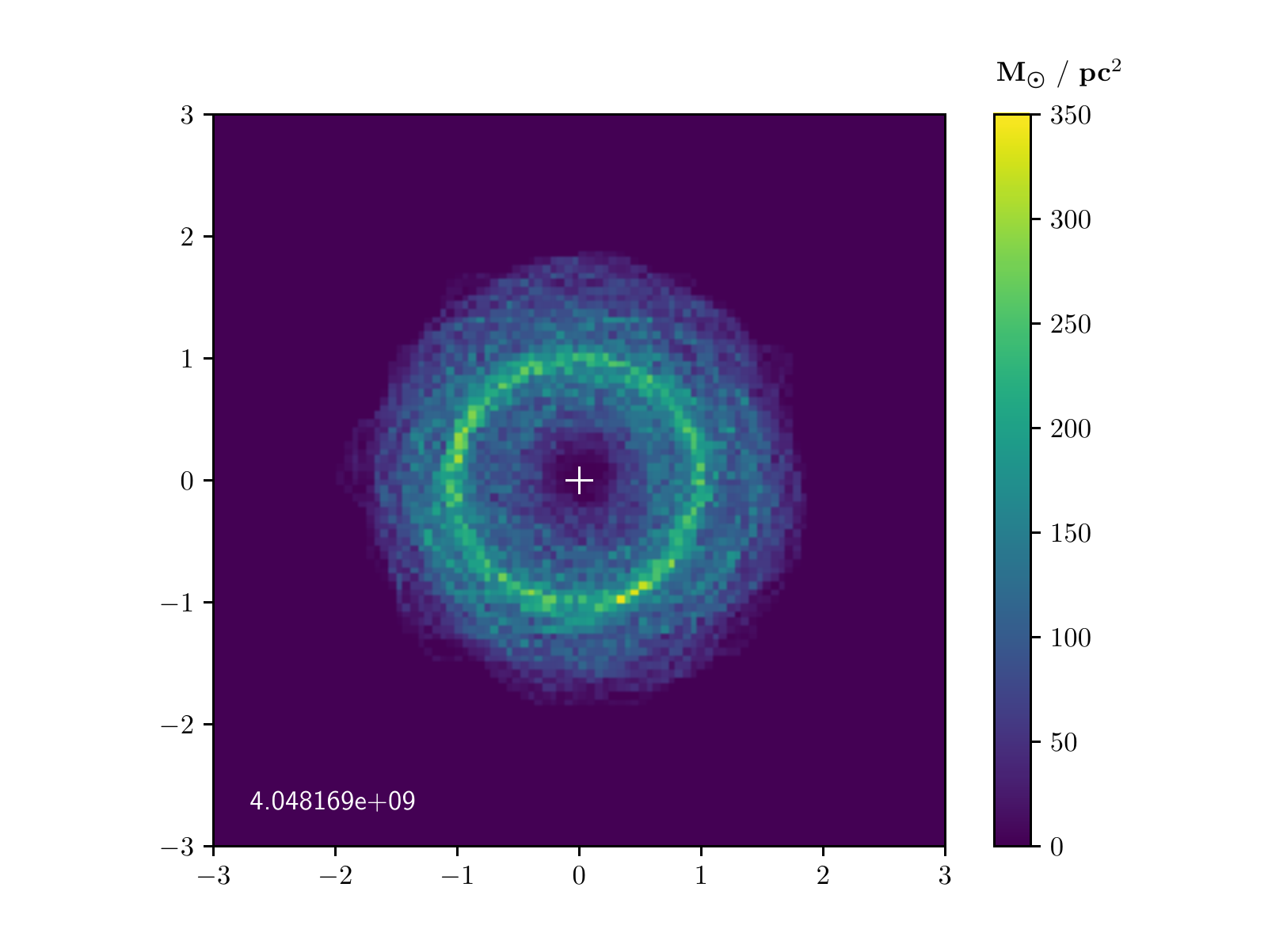}
   \end{subfigure}
   
   \begin{subfigure}{0.4\linewidth}
   \includegraphics[width=1.0 \textwidth ]{fig_wb3/wb3_t0.pdf}
   \caption{Initial state}
   \end{subfigure}
   \hspace{2cm}
   \begin{subfigure}{0.4\linewidth}
   \includegraphics[width=1.0 \textwidth ]{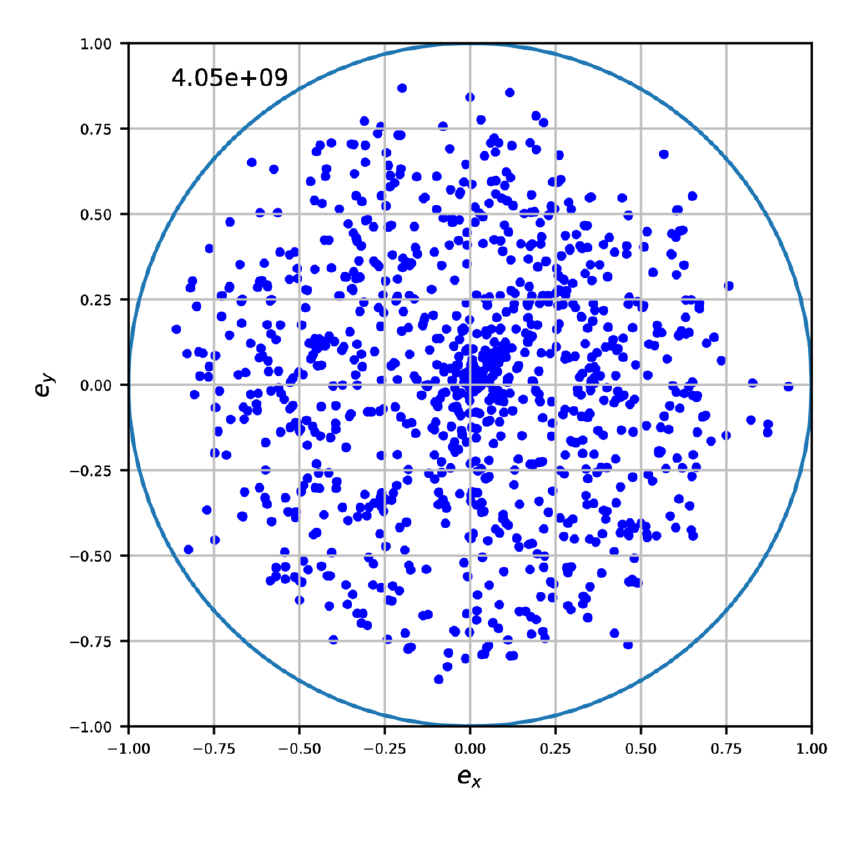}
   \caption{Relaxed state at $4.05$ Gyr}
   \end{subfigure}
 \caption{\emph{Collisionless relaxation of} \texttt{waterbag\_3\_s0}.}
\label{fig_lopsided}
\end{figure}

 \FloatBarrier

\section{Conclusions}
\label{sec_conclusions}

Mono-energetic waterbags are the simplest models of low mass stellar discs around a MBH. We studied, analytically and numerically, the stability of initial states that are prograde and axisymmetric. These waterbags have a DF, $f_0(\ell)$, which is constant when $0 \leq \ell_1 \leq \ell \leq \ell_2 \leq 1$, and zero when $\ell$ is outside this range. There are two types of waterbags, polarcaps with $\ell_2 = 1$ and bands with $\ell_2 < 1$. The linear stability problem can be solved simply: for each $m$ the growth rates of instabilities, pattern speeds of stable and unstable modes and the complete normal mode structure have been determined explicitly as functions of $(\ell_1, \ell_2)$, the waterbag parameters. 
\begin{itemize}
\item
Polarcaps have one stable normal mode for each $m$, with the noteworthy feature that the $m=1$ mode always has positive pattern speed. For a polarcap consisting of orbits with eccentricities $e < 0.94$, only the $m=1$ mode has a positive pattern speed. 

\item
Bands have two normal modes for each $m$, and can be either stable or unstable. Very narrow bands (with $\ell_1\simeq\ell_2$) are unstable to modes with a wide range in $m$, whereas broad bands approaching a polarcap (with $\ell_2 \simeq 1$) are stable. 
\end{itemize}

The evolution of instabilities was also explored through numerical simulations, which can explore both linear and non-linear regimes. 
A variety of numerical experiments were performed by which we demonstrated
good agreement with linear theory. Long-time integration showed the 
growth of instabilities of different $m$, that interacted with each
other non-linearly, then saturated and later relaxed collisionlessly 
into a quasi-steady state, which has a wider range of eccentric orbits than the initial state. This suggests secular non-axisymmetric instabilities could 
provide pathways for stars to exchange angular momentum via the mean self-gravitational field, and spread out in eccentricities. 

It is straightforward to extend our study to include external gravitational 
sources (such as nuclear density cusps or distant perturbers) and general
relativity, as described in ST1. But one clearly needs to go well beyond our simple models in order to study real systems, like the disc of young stars at the Galactic centre. We need to consider more general DFs and include orbits with a range of semi-major axes and inclinations. But self-gravitational dynamics poses difficult problems and secular dynamics is still in its infancy, so we need to build the tools step by step; describing the collisionless relaxation of even an unstable band remains a challenge for dynamists.


\appendix
\section{Surface probability density}
\label{app:sec: surface_density_waterbags}
The surface probability density function is obtained by integrating 
the disc DF over velocity space:
\beq
\Sigma(\bfr) = \int \rmd\bfu\, \hat{f}(\bfr,\bfu) 
\label{eqn_surf-den}
\eeq
where the DF $\hat{f}(\bfr,\bfu)$ is written a function of $\bfr$ and $\bfu$,
which are the position vector and velocity of a star, respectively, in the MBH's rest frame. For a razor-thin disc, the four dimensional phase volume, 
$\rmd\bfr\,\rmd\bfu = I\rmd w\,\rmd I\,\rmd g\,\rmd \ell$. Hence the DF of
an axisymmetric  monoenergetic disc (not necessarily a waterbag) is 
related to the DF, $f_0(\ell)$, of Section~\ref{sec_lin_axi}, as follows:
\beq
\hat{f}(\bfr,\bfu) \;=\; \frac{f_0(\ell)}{4 \pi^2 I_0} \,\delta(I-I_0)\,.
\label{eqn_realspace_DF}
\eeq
Then
\beq
\Sigma_0(r) \;=\; \frac{1}{4 \pi^2 I_0} \int \rmd u \, \rmd \phi \, u \, f_0(\ell) \, \delta{(I-I_0)}\,, 
\label{eqn_sig_prelim}
\eeq
where $u$ is the speed, $\phi$ is the angle between $\bfu$ and $\bfr$, 
and $I_0 = \sqrt{GM_\bullet a_0\,}\,$. Since the discs we consider have only prograde orbits, $\ell > 0$ which implies that $0\leq \phi \leq \pi$. 
We now express the (scaled) Delaunay variables, $\{\ell,I \}$, in terms of 
$\{u,\phi \}$:
\begin{subequations}
\begin{align}
I &\;=\; \left(\frac{2}{GM_{\bullet}r} - \frac{u^2}{(GM_{\bullet})^2}  \right)^{-1/2}\,,
\label{eqn_I_u_relation}\\[1ex] 
\ell &\;=\; L/I \;=\; I^{-1}ru\sin{\phi}\,. 
\label{eqn_l_phi_relation}
\end{align}
\end{subequations}
Hence
\begin{align}
\delta(I-I_0) &\;=\; \frac{\delta(u-u_0)}{ \displaystyle{ \left|\rmd I / \rmd u \right|_{u_0}  } } \;=\; \left(\frac{G M_\bullet}{a_0^3}\right)^{1/2}\,\frac{\delta(u-u_0)}{u_0}\,,
\label{eqn_delta-simplified}\\
\mbox{where}\qquad\qquad\qquad\qquad &\nonumber\\
u_0(r) &\;=\;
\begin{cases} 
\;\displaystyle{\sqrt{GM_\bullet\left(\frac{2}{r} \,-\, \frac{1}{a_0}\right)}}\,,\qquad\qquad\mbox{for $r \leq 2a_0$}\\[1ex] 
\qquad\qquad\quad 0\,,\qquad\qquad\qquad\quad\mbox{\;for $r > 2a_0$.}               
\end{cases}
\label{eqn_u0} 
\end{align}
is the speed at radius $r$, of an orbit with semi-major axis $a_0$. 
Substituting equation~(\ref{eqn_delta-simplified}) in (\ref{eqn_sig_prelim}) and using equations~(\ref{eqn_l_phi_relation}) and (\ref{eqn_u0}), the surface density for 
a general monoenergetic DF: 
\begin{align}
\Sigma_0(r) &\;=\; \frac{1}{4 \pi^2 a_0^2}\,\int \rmd \phi\, 
f_0\!\left(\ell_0(r)\sin{\phi}\right)\,,
\label{eqn_sigma_int_u_phi}\\[1em] 
\mbox{where}\qquad\qquad\ell_0(r) &\;=\; \frac{ru_0(r)}{I_0} \;=\;
\begin{cases} 
\;\displaystyle{\sqrt{\frac{2r}{a_0} \,-\, \frac{r^2}{a_0^2}\,}}\,,
\qquad\qquad\,\mbox{for $r \leq 2a_0$}\\[1ex] 
\qquad\quad 0\,,\qquad\qquad\qquad\mbox{\;for $r > 2a_0$.}               
\end{cases}
\label{eqn_l0_def}
\end{align}
 
For the waterbag DF of equation~(\ref{eqn_waterbag-def}), $f_0(\ell) = 1/\Delta\ell = 1/(\ell_2 -\ell_1) = \mbox{constant}$ for $0 \le \ell_1 <\ell_2 \le 1$ and is zero outside this range. This implies that $\Sigma_0(r)$ is non zero only when $\left|r - a_0\right| \leq a_0e_1$. Within this range of radii, 
\beq
\Sigma_0(r)\;=\; \frac{1}{4 \pi^2 a_0^2\Delta\ell}\,\Delta\phi(r)\,,
\label{eqn_sig_delta}
\eeq
where $\Delta\phi(r)$ is the range in $\phi$ for which 
\beq
\frac{\ell_1}{\ell_0(r)} \;\leq\;   \sin{\phi} \;\leq\; \frac{\ell_2}{\ell_0(r)}\,.
\label{eqn_phi_inequality}
\eeq
All we need to do now is to determine $\Delta\phi(r)$. There are two cases to consider:
\begin{itemize}
\item[{\bf 1.}] $\;\ell_2 \leq \ell_0(r)\,$:
Using equation~(\ref{eqn_l0_def}), this condition is equivalent to 
$\left|r - a_0\right| \leq a_0e_2$. Then $\Delta\phi(r) \,=\, 2\left(\phi_2 - \phi_1\right)$, where $\phi_1(r) = \sin^{-1}\left[\ell_1/\ell_0(r)\right]$ and $\phi_2(r) = \sin^{-1}\left[\ell_2/\ell_0(r)\right]$.

\item[{\bf 2.}] $\;\ell_1 \leq \ell_0(r) \leq \ell_2\,$: Using equation~(\ref{eqn_l0_def}), this condition is equivalent to 
$a_0e_2 \leq \left|r - a_0\right| \leq a_0e_1 $. Then $\Delta\phi(r) \,=\, 2\left(\pi/2 - \phi_1\right)$.
\end{itemize}
Substituting these expressions for $\Delta\phi(r)$ in 
equation~(\ref{eqn_sig_delta}), we obtain  equation~(\ref{eqn_surface density-final}) for the surface probability density of an axisymmetric mono-energetic waterbag.
\end{document}